\documentclass[reprint,twocolumn,aps,superscriptaddress,longbibliography]{revtex4-2}
\usepackage{amsmath,amssymb,amsfonts,mathrsfs}
\usepackage{bm}
\usepackage{graphicx}
\usepackage{hyperref}
\usepackage{amsbsy}
\usepackage{dcolumn}
\usepackage{lipsum}
\usepackage{tikz}
\usetikzlibrary{matrix, arrows.meta, decorations.pathreplacing}
\usepackage{mathdots}
\usepackage[ruled, linesnumbered]{algorithm2e}
\hypersetup{
	colorlinks=true,
	citecolor=red,  
	linkcolor=blue  
}

\begin{document}

	\title{Frequency collisions in parametrically modulated superconducting circuits} 
	
	
	\author{Zhuang Ma}
	\affiliation{National Laboratory of Solid State Microstructures, School of Physics, Nanjing University, Nanjing 210093, China}
	\affiliation{Shishan Laboratory, Suzhou Campus of Nanjing University, Suzhou 215000, China}
	\affiliation{Jiangsu Key Laboratory of Quantum Information Science and Technology, Nanjing University, Suzhou 215163, China}
	\author{Peng Zhao}
	\email{shangniguo@sina.com}
	\affiliation{Quantum Science Center of Guangdong-Hong Kong-Macao Greater Bay Area, Shenzhen 518045, China}
	
	\author{Xinsheng Tan}
	\email{tanxs@nju.edu.cn}
	\affiliation{National Laboratory of Solid State Microstructures, School of Physics, Nanjing University, Nanjing 210093, China}
	\affiliation{Shishan Laboratory, Suzhou Campus of Nanjing University, Suzhou 215000, China}
	\affiliation{Jiangsu Key Laboratory of Quantum Information Science and Technology, Nanjing University, Suzhou 215163, China}
	\affiliation{Synergetic Innovation Center of Quantum Information and Quantum Physics, University of Science and Technology of China, Hefei, Anhui 230026, China}
	\affiliation{Hefei National Laboratory, Hefei 230088, China}
	
	\author{Yang Yu}
	\email{yuyang@nju.edu.cn}
	\affiliation{National Laboratory of Solid State Microstructures, School of Physics, Nanjing University, Nanjing 210093, China}
	\affiliation{Shishan Laboratory, Suzhou Campus of Nanjing University, Suzhou 215000, China}
	\affiliation{Jiangsu Key Laboratory of Quantum Information Science and Technology, Nanjing University, Suzhou 215163, China}
	\affiliation{Synergetic Innovation Center of Quantum Information and Quantum Physics, University of Science and Technology of China, Hefei, Anhui 230026, China}
	\affiliation{Hefei National Laboratory, Hefei 230088, China}
	\date{\today}
	
	\begin{abstract}
	
	Superconducting circuits are a leading platform for scalable quantum computing, where parametric modulation is a widely used technique for implementing high-fidelity multiqubit operations. A critical challenge, however, is that this modulation can induce a dense landscape of parasitic couplings, leading to detrimental frequency collisions that constrain processor performance. In this work, we develop a comprehensive numerical framework, grounded in Floquet theory, to systematically analyze and mitigate these collisions. Our approach integrates this numerical analysis with newly derived analytical models for both qubit-modulated and coupler-modulated schemes, allowing us to characterize the major parasitic sideband interactions captured by the model and their distinct error budgets. This analysis forms the basis of a constraint-based optimization methodology designed to identify parameter configurations that satisfy the derived physical constraints, thereby avoiding detrimental parasitic interactions. We illustrate the utility of this framework with applications to analog quantum simulation and gate design. Our work provides a predictive tool for co-engineering device parameters and control protocols, enabling the systematic suppression of crosstalk and paving the way for large-scale, high-performance quantum processors.
		
	\end{abstract}
	
	\pacs{}
	
	\maketitle 
	

\section{Introduction}

Superconducting circuits have emerged as a leading platform for quantum information processing, demonstrating milestones such as large-scale quantum simulations \cite{Liu2025} and quantum error correction below the surface-code threshold \cite{Acharya2025}. The performance and scalability of superconducting circuits are critically determined by the co-design of their architecture, fabrication, and control strategies. Consequently, significant research has focused on developing circuit architectures and optimized control protocols to perform high-fidelity operations in increasingly complex multi-qubit systems \cite{Yan2018, Sete2021, Klimov2024}.

Fixed- and tunable-frequency qubits represent two competing paradigms, each with distinct advantages. Fixed-frequency qubits offer simplicity and reduced flux noise. In these systems, two-qubit entanglement is typically achieved via microwave-activated cross-resonance gates \cite{Paraoanu2006, Rigetti2010, Groot2010, Chow2011, Heya2024}. In contrast, tunable-frequency qubits provide greater operational flexibility at the expense of increased complexity and susceptibility to flux noise. These qubits, often integrated with tunable couplers, have become a prevalent platform for implementing entangled gates based on baseband \cite{Strauch2003} or sideband (parametric) \cite{McKay2016, Caldwell2018, Reagor2018} flux control.

The frequency tunability of these qubits allows for the activation of interactions via an external drive, a technique known as parametric modulation. This technique typically involves applying a time-periodic flux pulse to a frequency-tunable qubit or coupler, dynamically bridging the energy gap of interacting components to realize a coupling \cite{McKay2016, Didier2018, Caldwell2018}. Monochromatic parametric drives are inherently periodic, which grants them a high degree of robustness against flux noise, pulse distortions, and crosstalk \cite{Ganzhorn2020, Abrams2020, Sete2024, Ma2025}. This combination of flexibility and robustness has made parametric modulation a powerful tool for both analog quantum simulation and high-fidelity entangled gates \cite{McKay2016, Caldwell2018, Reagor2018, Abrams2020, Chu2020, Ganzhorn2020, Li2021, Sete2021a, Li2022, Ma2023, Zheng2022, Zhang2024a, Zhang2025b}.

However, as the scale and density of quantum processors increase, a fundamental challenge emerges: spectral crowding. The limited frequency bandwidth becomes densely populated with the spectrum of qubits, couplers, and their various parametrically induced sidebands. This crowded spectrum makes the system highly susceptible to frequency collisions---a critical form of crosstalk where a drive intended for a target interaction inevitably activates parasitic, off-resonant transitions elsewhere in the circuit \cite{Brink2018, Malekakhlagh2020, Zhao2023}. These collisions, along with other frequency-dependent errors such as those from two-level-system (TLS) defects \cite{Mueller2019, Klimov2024}, represent a major bottleneck for scaling up quantum processors.

Mitigating this spectral complexity has become a central focus of research. For fixed-frequency architectures, where frequency allocation is permanent, optimizing qubit placement to enhance collective yield is crucial for scaling. This has led to the development of methods such as mixed-integer programming \cite{Morvan2022}, Floquet analysis \cite{Heya2024}, frequency-aware analytical placement frameworks \cite{Zhang2025}, improved optimization techniques \cite{Zhang2025a}, and the use of chiplet architectures \cite{Smith2022}. For tunable-frequency architectures, the challenge expands to encompass the dynamic choreography of qubit frequency trajectories to avoid detrimental frequency-dependent errors during operation. This non-convex, high-constraint optimization problem has been tackled with a variety of powerful techniques, including the Snake optimizer \cite{Klimov2020, Klimov2024}, graph theory \cite{Kelly2018}, context-aware coupler reconfiguration \cite{Hour2024}, frequency-aware
compilation \cite{Ding2020}, automatic frequency allocation \cite{Wang2024b}, and neural network approaches \cite{Ai2025, Lu2025}. Combined with fabrication, including post-fabrication tuning \cite{Hertzberg2021, Kim2022, Pappas2024}, these strategies can effectively improve overall solution quality.

While these high-level optimization strategies are powerful, their efficacy is ultimately limited by the quality of the underlying physical model used to identify and quantify the frequency-dependent errors they seek to avoid. Although specific aspects of the crosstalk and error budgets for parametric gates have been previously analyzed \cite{Ganzhorn2020, Sete2024, Krauss2025}, and simulations have demonstrated the potential for high performance at scale \cite{Osman2023}, a systematic, physics-based framework for predicting the systematic landscape of frequency collisions has been lacking. Such a framework is essential for generating the precise, physics-informed constraints needed to guide high-level optimizers and for understanding the fundamental performance limits of a given circuit architecture.

In this work, we address this gap by developing a comprehensive numerical framework based on Floquet theory---the natural mathematical language for time-periodic systems \cite{Floquet1883}---to systematically analyze and map the landscape of frequency collisions in multi-qubit superconducting circuits under parametric drives. Our framework combines a full numerical treatment with newly derived analytical models for both qubit-modulated and coupler-modulated interactions, which we validate against simulations to reveal the distinct error budgets associated with each scheme. Building on recent work that has applied Floquet analysis to frequency allocation challenges \cite{Heya2024, Paolo2022}, our approach provides a powerful tool for identifying the error sources from parasitic interactions. The resulting analysis serves as the foundation for a constraint-based optimization algorithm designed to find optimal circuit and control configurations. This provides a systematic methodology for co-designing high-fidelity parametric operations while mitigating errors from frequency collisions and other unwanted frequency-dependent effects. We note that while this method does not capture all possible frequency collisions, it provides an intuitive and efficient tool for identifying and estimating the major parasitic resonances in parametrically modulated superconducting circuits.

The remainder of this article is organized as follows. In Sec. \ref{sec:parametric_modulation}, we analytically model the qubit-modulated and coupler-modulated couplings and introduce the Floquet formalism used for their analysis. In Sec. \ref{sec:frequency_collisions_in_parametric_modulation}, we analyze the constraints arising from frequency collisions for both qubit-modulated and coupler-modulated schemes and present an illustrative algorithm to solve the corresponding optimization problem. In Sec. \ref{sec:application_on_lattices}, we illustrate applications of our method to analog quantum simulation and entangled gates. Finally, in Sec. \ref{sec:summary}, we summarize our main results and discuss potential directions for future work.

	 \begin{figure*}[htbp]
	\centering
	\includegraphics{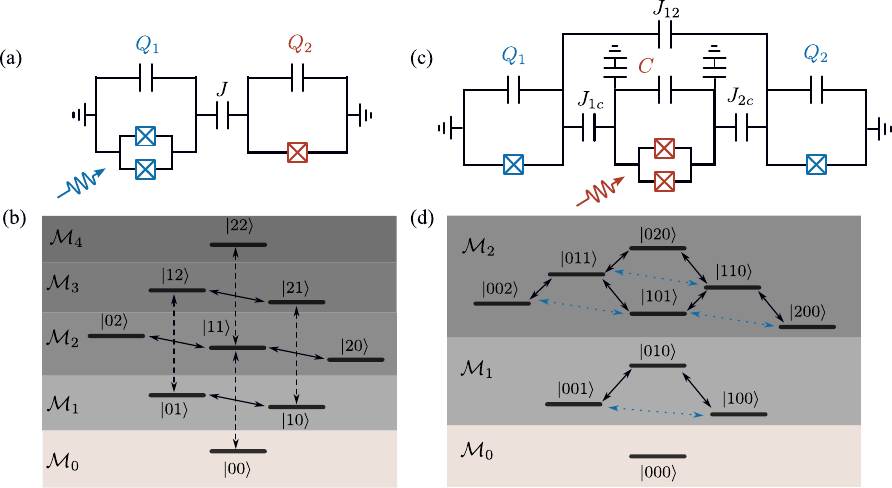}
	\caption{Two-qubit systems for qubit-modulated and coupler-modulated parametric interactions.
	(a) Schematic of two capacitively coupled transmon qubits with the static coupling strength $J$. A fixed-frequency qubit ($Q_2$) is coupled to a tunable transmon ($Q_1$). A parametric flux pulse is applied to $Q_1$ to modulate its frequency and induce a parametric interaction.	
	(b) Energy level diagram of the two-qubit system. Solid and dashed arrows indicate co-rotating and counterrotating transitions, respectively. The system is labeled using excitation manifolds $\mathcal{M}_{0,1,2,3,4}$ and the subsequent demos focus on the single-excitation manifold $\mathcal{M}_1$, i.e., the $\{|10\rangle,| 01\rangle\}$ subspace.
	(c) Schematic of the circuit architecture, featuring two transmon qubits ($Q_1, Q_2$) capacitively coupled to a central tunable transmon-type coupler ($C$) with static strengths $J_{1c}$ and $J_{2c}$, respectively. A direct static coupling of strength $J_{12}$ also exists between the two qubits. A parametric flux pulse is applied to the coupler to mediate parametric interactions.
	(d) Energy level diagram showing the lowest three excitation manifolds of the system: the ground state manifold $\mathcal{M}_0$, the single-excitation manifold $\mathcal{M}_1$, and the double-excitation manifold $\mathcal{M}_2$. Black solid arrows represent the strong, static qubit-coupler strengths (with strengths $\propto J_{1c},J_{2c}$), while the blue dotted arrow indicates the weak direct qubit-qubit strengths (with strength $\propto J_{12}$). counterrotating terms are omitted for clarity.
}
	\label{fig:TF_FTF_demo}
\end{figure*}

	\section{Parametric modulation}
	\label{sec:parametric_modulation}
	
In the circuit architecture containing frequency-tunable transmon qubits \cite{Koch2007}, interactions can be dynamically activated using parametric modulation. This technique is broadly categorized into two paradigms: qubit-modulated \cite{Didier2018, Caldwell2018, Reagor2018} and coupler-modulated \cite{McKay2016, Roth2017} schemes. In the former, a parametric pulse is applied to a tunable qubit that has a static coupling to the adjacent qubit, inducing an effective, tunable interaction. In the latter, a tunable coupler placed between two qubits is modulated, mediating a tunable interaction. We denote $|Q_1Q_2\rangle$ and $|Q_1C Q_2\rangle$ as the states of the qubit-qubit and the qubit-coupler-qubit systems, in Figs. \ref{fig:TF_FTF_demo}(a) and \ref{fig:TF_FTF_demo}(c), respectively. We make the simplifying assumption that the tunable qubit and coupler frequencies depend linearly on the applied external flux (see Appendix \ref{app:full_circuit} for the nonlinear full-circuit discussion) and adopt Hamiltonian parameters listed in Tables \ref{tab:paras_qq} and \ref{tab:paras_qcq}.  We also adopt the convention of setting $\hbar=1$.

\begin{table}[!ht]
	\centering
	\caption{Hamiltonian parameters of the qubit-qubit system in Fig. \ref{fig:TF_FTF_demo} (a).}
\begin{ruledtabular}
	\begin{tabular}{lcc}
		& $Q_1$ & $Q_2$ \\
		\hline
		$\omega_{1,2}/2\pi$ (GHz) & 4.85 & 5.00 \\

		$\alpha_{1,2}/2\pi$ (MHz) & -220 & -260 \\

		$J/2\pi$ (MHz) & \multicolumn{2}{c}{5}\\
	\end{tabular}
\end{ruledtabular}
	\label{tab:paras_qq}
\end{table}

\begin{table}[!ht]
	\centering
	\caption{Hamiltonian parameters of the qubit-coupler-qubit system in Fig. \ref{fig:TF_FTF_demo} (c).}
\begin{ruledtabular}
	\begin{tabular}{lllllll}
		& \multicolumn{2}{c}{$Q_1$} & \multicolumn{2}{c}{$C$ } & \multicolumn{2}{c}{$Q_2$}  \\
		\hline
		$\omega_{1,c,2}/2\pi$ (GHz) & \multicolumn{2}{c}{5.801} & \multicolumn{2}{c}{6.990} & \multicolumn{2}{c}{5.921} \\

		$\alpha_{1,c,2}/2\pi$ (MHz) & \multicolumn{2}{c}{-205} & \multicolumn{2}{c}{-105}  & \multicolumn{2}{c}{-300} \\

		$J_{1c,2c}/2\pi$ (MHz) & 
		\multicolumn{3}{c}{\makebox[1.5cm][c]{100}} & 
		\multicolumn{3}{c}{\makebox[1cm][l]{100}} \\

		$J_{12}/2\pi$ (MHz) & \multicolumn{6}{c}{5} \\

	\end{tabular}
\end{ruledtabular}
	\label{tab:paras_qcq}
\end{table}
	
	\subsection{Qubit-modulated parametric coupling}
	\label{sec:qubit_modulate_parametric_coupling}
We consider a system of two transmon qubits, $Q_1$ and $Q_2$, with a static, direct coupling strength $J$ shown in Fig. \ref{fig:TF_FTF_demo} (a). The Hamiltonian $	\mathcal{H}_{qq} = 	\mathcal{H}_{qq}^{0} + \mathcal{V}_{qq}$ in the laboratory frame is 
\begin{equation}
	\begin{split}
 \mathcal{H}_{qq}^{0} &=\sum_{i=1,2} \left(\omega_i b_i^{\dagger} b_i+ \frac{\alpha_i}{2} b_i^{\dagger} b_i^{\dagger} b_ib_i\right), \\
	 \mathcal{V}_{qq} &= J (b_1 + b_1^{\dagger} ) (b_2 + b_2^{\dagger} ),
	\end{split}
	\label{eq:hamiltonian_lab}
\end{equation}
where $\omega_i$, $\alpha_i$, and $b_i$ ($b^\dagger_i$) are the frequency, anharmonicity, and annihilation (creation) operators, respectively, for qubit $i$.

An effective, tunable interaction is induced by sinusoidally modulating the frequency of $Q_1$, as depicted in Fig. \ref{fig:TF_FTF_demo} (a). The modulation is described by $\omega_1(t) = \bar{\omega}_1 + \epsilon_p \cos(\omega_{p}t + \phi_{p})$. After transforming to an interaction picture with respect to $\mathcal{H}_{qq}^0$, the effective Hamiltonian can be derived \cite{Ma2025} as
\begin{equation}
	\begin{split}
		H_{\mathrm{eff}}/J
		= &b_1b_2 e^{i[-F_1-A_1 (b_1^{\dagger} b_1-I) -F_2-A_2(b_2^{\dagger} b_2-I)]} \\
		&+b_1^\dagger b_2 e^{i[F_1+A_1 b_1^{\dagger} b_1 -F_2-A_2(b_2^{\dagger} b_2-I)]}\\
		&+b_1 b_2^\dagger e^{i[-F_1-A_1 (b_1^{\dagger} b_1-I) +F_2+A_2b_2^{\dagger} b_2]}\\
		&+b_1^\dagger b_2^\dagger e^{i(F_1+A_1 b_1^{\dagger} b_1 + F_2+A_2b_2^{\dagger} b_2)},
		\label{eq:hamiltonian_rot}
	\end{split}
\end{equation}
where $F_i (t) = \int_0^t \omega_i(\tau)d\tau b_i^{\dagger} b_i$ and $A_i (t) = \int_0^t \alpha_i(\tau)d\tau$. To simplify this expression, we truncate the Hilbert space of each qubit to its lowest three energy levels shown in Fig. \ref{fig:TF_FTF_demo} (b). Applying the Jacobi-Anger expansion $e^{i z \sin \theta}=\sum_{n=-\infty}^{\infty} J_{n}(z) e^{i n \theta}$, the Hamiltonian in Eq. \eqref{eq:hamiltonian_rot} becomes

\begin{equation}
	\begin{split}
		H_{\mathrm{eff}}/J = &\sum_{n=-\infty}^{\infty}  J_{n}\left(\frac{\epsilon_{p}}{\omega_{p}}\right) e^{i(n\omega_p t +\beta_n)}\\
		&\times \left\{e^{i\Delta t} |10\rangle \langle 01|\right. \\
		&+\sqrt{2}e^{i(\Delta +\bar\alpha_2)t} |11\rangle \langle 02| \\
		&+\sqrt{2}e^{i(\Delta -\bar\alpha_1)t} |20\rangle \langle 11| \\
		&+2e^{i(\Delta +\bar\alpha_2-\bar\alpha_1)t} |21\rangle \langle 12| \\
		&+e^{-i\Sigma t} |11\rangle \langle 00| \\
		&+\sqrt{2}e^{-i(\Sigma +\bar\alpha_2)t} |12\rangle \langle 01| \\
		&+\sqrt{2}e^{-i(\Sigma +\bar\alpha_1)t} |21\rangle \langle 10| \\
		&+\left.2e^{-i(\Sigma +\bar\alpha_1+\bar\alpha_2)t} |22\rangle \langle 11|\right\} +\text{H.c.},
	\end{split}
	\label{eq:Ham8}
\end{equation}
where $J_n$ is the $n$th Bessel function of the first kind. We have assumed the frequencies and anharmonicities of the undriven qubit ($Q_2$) and the static part of the driven qubit ($Q_1$) are constant, denoted by $\bar{\omega}_i$ and $\bar{\alpha}_i$. Here, $\Delta = \bar{\omega}_2 - \bar{\omega}_1$ (we assume $\Delta>0$) is the detuning, $\Sigma = \bar{\omega}_2 + \bar{\omega}_1$ is the sum frequency, and $\beta_n =  n(\phi_{p}+\pi) + \frac{\epsilon_{p}}{\omega_{p}}\sin(\phi_{p})$ is the drive-dependent phase. In the curly brace of Eq. \eqref{eq:Ham8}, the first and last four terms represent the co-rotating (number-conserving) and counterrotating (number-non-conserving) couplings, respectively. Each term in the summation over $n$ corresponds to a specific sideband. A desired interaction can be resonantly activated by choosing the drive frequency $\omega_p$ to satisfy a resonance condition, such as $|10\rangle \leftrightarrow | 01\rangle$ with $\Delta + n\omega_{p} = 0$, $|11\rangle \leftrightarrow | 02\rangle$ with $\Delta+\bar\alpha_2 + n\omega_{p} = 0$, $|20\rangle \leftrightarrow |11\rangle$ with $\Delta-\bar\alpha_1 + n\omega_{p} = 0$, and $|11\rangle \leftrightarrow | 00\rangle$ with $\Sigma - n\omega_{p} = 0$ \cite{Didier2018, Caldwell2018, Reagor2018}. The resulting effective coupling strength is given by
\begin{gather}
	g_{\text{eff}}^{(n)}= \sqrt{C}JJ_{n}\left(\frac{\epsilon_{p}}{\omega_{p}}\right),
	\label{eq:qubit_strength}
\end{gather}
where $C = \max(i_1,i_2) \max(j_1,j_2)$  is a coefficient determined by the specific energy levels involved in the transition $|i_1j_1\rangle \leftrightarrow | i_2j_2\rangle$. 

In the parametrically modulated system, the drive at frequency $\omega_p$ on $Q_1$ creates a series of harmonic sidebands, effectively replicating the original interaction at energy intervals of $\hbar \omega_p$ \cite{Ma2025}. The drive frequency, $\omega_p$, therefore sets the energy separation between these adjacent sidebands. The static coupling, $\sqrt{C} J$, on the other hand, determines the gap between sidebands and the $Q_2$ spectrum. The condition $\omega_p \gg \sqrt{C}J$ ensures the separation between the adjacent sidebands is much larger than the gap. This spectral separation is crucial, as it allows one to selectively address a single, resonant or near-resonant sideband interaction without simultaneously exciting other, off-resonant sidebands. This spectral picture corresponds to all terms oscillating rapidly on the characteristic timescale of the system's evolution, with the exception of the single resonant or near-resonant contribution in the summation of Eq. \eqref{eq:Ham8}.
The rapidly oscillating nonresonant terms can be neglected under the rotating-wave approximation (RWA), leaving only the slowly varying resonant contribution to dictate the evolution, i.e., the activated parametric coupling.

	\subsection{Coupler-modulated parametric coupling}
	\label{sec:coupler_modulated_parametric}
	
We now analyze the coupler-modulated parametric coupling, illustrated schematically in Fig. \ref{fig:TF_FTF_demo} (c). In this system, two transmon qubits ($Q_1, Q_2$) are coupled to a central tunable transmon-type coupler ($C$) with strengths $J_{1c}$ and $J_{2c}$, respectively, while also possessing a direct coupling $J_{12}$ \cite{Yan2018, Sete2021}. The corresponding energy level diagram, including the ground, single-, and double-excitation manifolds ($\mathcal{M}_{0,1,2}$), is shown in Fig. \ref{fig:TF_FTF_demo}(d). The Hamiltonian for this three-mode system $	\mathcal{H}_{qcq} = \mathcal{H}_{qcq}^{0} + \mathcal{V}_{qcq}$ is
	\begin{equation}
	\begin{split}
		 \mathcal{H}_{qcq}^{0} &=\sum_{i} \left(\omega_i b_i^{\dagger} b_i+ \frac{\alpha_i}{2} b_i^{\dagger} b_i^{\dagger} b_ib_i\right),  \\
		 \mathcal{V}_{qcq} &= \sum_{i< j}J_{ij} (b_i + b_i^{\dagger} ) (b_j + b_j^{\dagger} ),
		\label{eq:hamiltonian_coupler_lab}
	\end{split}
\end{equation}
	with $i,j\in\{1,2,c\}$, where $\omega_{i}$, $ \alpha_{i}$, and $b_{i} (b^\dagger_{i})$ are the frequency, anharmonicity, and annihilation (creation) operators, respectively, for mode $i$ (qubit or coupler). 
	
	In the dispersive regime, where $|\Delta_{ic}| \gg |J_{ic}| \gg |J_{12}|$ (with $\Delta_{ic}=\omega_{i}-\omega_c,i\in\{1,2\}$), and assuming the coupler remains in its ground state, the coupler's degrees of freedom can be perturbatively eliminated. This is achieved via a Schrieffer-Wolff (SW) transformation, $U=\exp\{\sum_{i=1,2}[{J_{i c}}/{\Delta_{ic}}({b}_{i}^{\dagger} {b}_{c}-{b}_{i} {b}_{c}^{\dagger})+{J_{i c}}/{\Sigma_{ic}}({b}_{i}^{\dagger} {b}_{c}^{\dagger}-{b}_{i} {b}_{c})]\}$ with $\Sigma_{ic}=\omega_{i}+\omega_c$ \cite{Bravyi2011, Yan2018, Sete2021a, Rasmussen2021}. To second order in terms of the perturbation parameters $ J_{i c}/\Delta_{ic}(\Sigma_{ic})$ and assuming small anharmonicities ($|\Delta_{ic}| \gg |\alpha_{i,c}|$), the transformation yields an effective qubit-qubit Hamiltonian:
	 \begin{gather}
	\begin{split}
	\tilde{\mathcal{H}}_{qq} = &\sum_{i=1,2} \left(\tilde{\omega}_i b_i^{\dagger} b_i + \frac{\tilde{{\alpha}}_i}{2} b_i^{\dagger} b_i^{\dagger} b_ib_i\right) \\
	&+ \tilde{J}_{12} (b_1b_2^\dagger +b_1^\dagger b_2),
	\label{eq:hamiltonian_decoupler_lab}
	\end{split}
\end{gather}
with $\tilde{\alpha}_i \approx \alpha_{i}$. Applying the SW transformation reveals how the underlying interactions renormalize the qubit frequencies and modify the effective qubit-qubit coupling strength, 
	\begin{gather}
		\tilde{\omega}_i  \approx \omega_i + J_{ic}^2\left(\frac{1}{\Delta_{ic}} - \frac{1}{\Sigma_{ic}} \right), \label{eq:lamb_shift} \\
		\tilde{J}_{12} \approx J_{12} +\frac{J_{1c}J_{2c}}{2}\left(\frac{1}{\Delta_{1c}}+\frac{1}{\Delta_{2c}}-\frac{1}{\Sigma_{1c}}-\frac{1}{\Sigma_{2c}}\right). \label{eq:effective_coupling}
	\end{gather}
In the interaction picture, the resulting Hamiltonian $\tilde{\mathcal{H}}_{{qq}}$ in Eq. \eqref{eq:hamiltonian_decoupler_lab} has the same mathematical form as Eq. \eqref{eq:hamiltonian_rot} for the qubit-modulated case.
	
	To activate a parametric interaction, a flux pulse is applied to the tunable coupler, modulating its frequency sinusoidally as $\omega_c(t) = \bar{\omega}_c + \epsilon_{p} \cos(\omega_{p}t + \phi_{p})$, see Fig. \ref{fig:TF_FTF_demo} (c). This modulation makes the effective coupling $\tilde{J}_{12}$ time-dependent. By expanding $\tilde{J}_{12}[\omega_c(t)]$ as a Taylor series in $\omega_c$ around $\bar{\omega}_c$ and applying power-reduction formulae, one obtains \cite{Ma2025}
	 \begin{equation}
	\begin{split}
		\tilde{J}_{12}[\omega_c(t)]=&
		\tilde{J}_{12}(\bar{\omega}_c) + \sum_{n=1}^\infty D_n\delta_{{n\bmod2},0}\binom{n}{\lfloor\frac{n}{2}\rfloor} \\
		&+  \sum_{n=1}^\infty 2D_n\sum_{k=0}^{\lfloor\frac{n-1}{2}\rfloor}\binom{n}{k} \cos [(n-2 k) \omega_{p} t],
	\end{split}
	\label{eq:taylor1}
\end{equation}
where 
    \begin{equation}
  D_n =  \frac{{\epsilon}_{p}^n}{2^nn!} \left.\frac{\partial^n \tilde{J}_{12}}{\partial \omega_c^n}\right|_{\bar{\omega}_c}.
    \end{equation}

	The interaction terms oscillating at harmonics of $\omega_p$ can be used to resonantly drive transitions. For example, the $|100\rangle \leftrightarrow |001\rangle$ transition \cite{McKay2016} is activated when $\Delta_{12} + m\omega_{p} = 0$ for an integer $m$. Here, $\Delta_{12}$ is the Stark-shifted frequency difference between the qubits
	\begin{equation}
			\begin{split}
		\Delta_{12}=
		&\sum_{n=1}^\infty \frac{{\epsilon}_{p}^n}{2^nn!} \left.(\frac{\partial^n {\tilde{\omega}}_1}{\partial \omega_c^n}-\frac{\partial^n {\tilde{\omega}}_2}{\partial \omega_c^n})\right|_{\bar{\omega}_c}\delta_{{n\bmod2},0}\binom{n}{\lfloor\frac{n}{2}\rfloor}\\
		&+\tilde{\omega}_1(\bar{\omega}_c) - \tilde{\omega}_2(\bar{\omega}_c) . \\
			\end{split}
	\end{equation}
	This resonance condition yields a nonzero $m$th-order parametric coupling strength
	 \begin{gather}
	g_{\text{eff}}^{(m)} = \sum_{n=1}^\infty D_n\sum_{k=0}^{\lfloor\frac{n-1}{2}\rfloor}\binom{n}{k}\delta_{ n-2k,|m|},
	\label{eq:coupler_strength}
\end{gather} 
	which is half the coefficient of the $\cos(m\omega_{p}t)$ term in Eq. \eqref{eq:taylor1}. For the fundamental sideband ($m=1$), the dominant contribution typically comes from the $n=1$ term, which to the leading order gives 
	\begin{equation}
		g_{\text{eff}}^{(1)}\approx \epsilon_p \frac{J_{1c}J_{2c}}{4}\left( \frac{1}{\Delta_{1c}^2} + \frac{1}{\Delta_{2c}^2} + \frac{1}{\Sigma_{1c}^2} +\frac{1}{\Sigma_{2c}^2}\right).
	\end{equation}
	This agrees with the standard adiabatic approximation \cite{Roth2017}. Higher-order transitions, such as $|101\rangle \leftrightarrow |002|$ \cite{Huber2025}, $|200\rangle \leftrightarrow |101|$ \cite{Li2022, Ganzhorn2020}, and $|101\rangle \leftrightarrow |000|$ \cite{Roth2017}, can also be activated by matching $m\omega_p$ to the energy difference of the corresponding dressed states (see Appendix \ref{app:higher_levels} for more details).
	
	The SW transformation can be carried out to higher order for increased accuracy. For instance, the third-order correction to the coupling strength is
	 \begin{equation}
	\begin{split}
	\tilde{J}_{12}^{(3)} \approx&  J_{12} +\frac{J_{1c}J_{2c}}{2}\left(\frac{1}{\Delta_{1c}}+\frac{1}{\Delta_{2c}}\right) \\
	&- \frac{J_{12}(J_{1c}^2 +J_{2c}^2)}{2\Delta_{1c}\Delta_{2c}},
		\end{split}
\end{equation}
	after the RWA. For typical circuit parameters, we find this third-order correction to be negligible compared with the second-order result. It is significant to note that the derivation presented thus far is an adiabatic approximation, valid for weak modulations \cite{McKay2016, Roth2017}. A more rigorous analysis for arbitrary drive parameters would require frameworks such as the time-dependent SW perturbation theory \cite{Roth2017, Petrescu2023} or the Floquet transformation \cite{Huber2025}.

	\subsection{Floquet analysis of parametric modulation}

A periodic drive can be used to engineer tunable interactions between multiple circuit elements. Floquet theory provides a powerful and non-perturbative framework for analyzing such systems, which is particularly effective for strong drives where simple perturbation theory may fail (see Appendix \ref{app:floquet_theory} for details). This formalism is broadly applicable to many driven quantum systems, including those implementing microwave-activated gates \cite{Deng2015, Krinner2020, Paolo2022, Nguyen2024, Heya2024, Ding2025}, and is exceptionally well-suited for describing parametric modulation. The interactions engineered by such drives are instances of a general class of phenomena known as sideband transitions, for which the Floquet formalism provides a natural and predictive description \cite{Sameti2019, Petrescu2023, Huber2025, You2025, Huang2025}.

To build intuition for how Floquet theory describes these interactions, it is instructive to first consider the static analog of a frequency collision. In a time-independent system, a frequency collision manifests as an anticrossing between two eigenenergies of the static Hamiltonian. This occurs when two bare states are brought into resonance, and the static coupling between them opens an energy gap. The magnitude of this coupling is determined by half the minimum energy splitting at the resonance point.

\begin{figure}[htbp]
	\centering
	\includegraphics{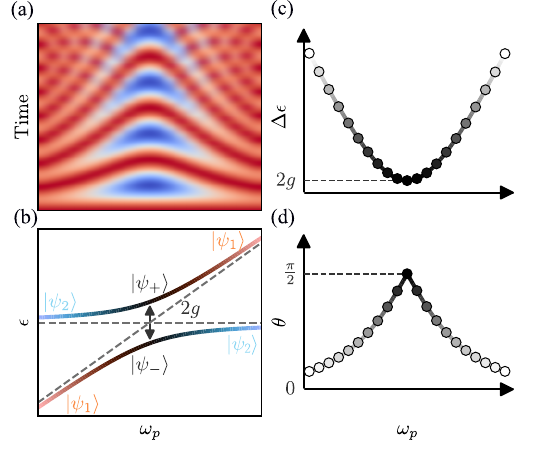}
	\caption{Illustration for Floquet and Schrödinger equation (SE) methods of qubit-modulated parametric coupling in a qubit-qubit system.
		(a) Chevron pattern showing time-dependent population oscillations between states $|10\rangle$ and $|01\rangle$ as a function of the modulation frequency, $\omega_{p}$. The data are obtained from the SE method.
		(b) Schematic of the corresponding anticrossing between the two Floquet states $|\psi_{1,2}\rangle$. The quasienergy splitting between $|\psi_{1,2}\rangle$ varies as the modulation frequency $\omega_p$ increases, and the minimum quasienergy splitting, which occurs at the resonance point, is equal to the parametric coupling strength, $2g$.
		(c) Comparison of the generalized Rabi frequency extracted from the SE method in (a) (circles) and the Floquet quasienergy splitting $\Delta\epsilon$ in (b) (solid line). The excellent agreement validates the Floquet method. The vertex of the parabola corresponds to the on-resonance effective coupling, $2g$.
		(d) The collision angle, $\theta$, as a function of modulation frequency $\omega_{p}$, derived from the coupling strengths and detunings shown in (c). The collision angle is equal to $\pi/2$ at the above resonance point, and the collision angle is close to zero as $\omega_p$ moves away from the resonance point (the strong-dispersive condition).
	}
	\label{fig:avoiding_cross}
\end{figure}

When a system is subjected to a periodic drive, the static description is elevated into the Floquet formalism. The time-dependent Hamiltonian of the finite-dimensional system is mapped onto an equivalent, time-independent but infinite-dimensional Floquet Hamiltonian. In this picture, the static eigenenergies are replaced by the quasienergies, which naturally account for drive-induced effects such as ac Stark and Bloch-Siegert shifts \cite{Bloch1940}. The drive-induced frequency collisions now manifest as anticrossings between these quasienergy levels. The minimum gap at such a resonance point directly corresponds to twice the effective dynamic coupling strength, $2g$, as shown in Fig. \ref{fig:avoiding_cross} (b). More formally, these collisions can be understood as a breakdown of the strong-dispersive condition within the Floquet Hamiltonian itself \cite{Heya2024}. This approach provides a rigorous and predictive tool for mapping out the major collision landscape captured within the chosen truncation, which can then be compared with experimental spectroscopy or the Fourier transform of the system's simulated dynamics \cite{Ma2025}.

	To illustrate this, we analyze the qubit-modulated interaction within the two-qubit subspace $\{|10\rangle, |01\rangle \}$, i.e., the manifold $\mathcal{M}_1$ in Fig. \ref{fig:TF_FTF_demo} (b). The effective Hamiltonian within the two-level subspace spanned by the corresponding Floquet states $|\psi_1\rangle$ and $|\psi_2\rangle$ can be written as 
	\begin{gather}
		H = \Delta \frac{\sigma_z}{2} + 2g \frac{\sigma_x}{2} = \frac{1}{2}
		\begin{pmatrix}
			\Delta & 2g \\
			2g & -\Delta
		\end{pmatrix},
		\label{eq:Flqouet_hami}
	\end{gather} 
	where $\Delta$ and $2g$ represent the detuning and coupling strength between two Floquet states. The Hamiltonian in Eq. \eqref{eq:Flqouet_hami} is equivalent to the $0$th-order Floquet component in the extended space \cite{Son2009}. We can define a collision angle $\theta$ between two Floquet states \cite{Heya2024}, which is the angle between the Hamiltonian vector and the $z$ axis,
	\begin{gather}
		\theta = \arctan\left(\left|\frac{2g}{\Delta}\right|\right)\in (0, \frac{\pi}{2}].
		\label{eq:theta}
	\end{gather}
	The Hamiltonian of Eq. \eqref{eq:Flqouet_hami} can be diagonalized to obtain the eigenvalues $\eta_\pm = \pm\sqrt{\Delta^2 + 4g^2}/2$ and the difference $\Delta \eta= \eta_+-\eta_- =  \sqrt{\Delta^2 + 4g^2} $ corresponds to the quasienergy splitting $\Delta\epsilon$ of identified Floquet states $|\psi_{1,2}\rangle$, as shown in Figs. \ref{fig:avoiding_cross} (b) and \ref{fig:avoiding_cross} (c). 
	
\begin{figure}[h!]
	\centering
	\includegraphics{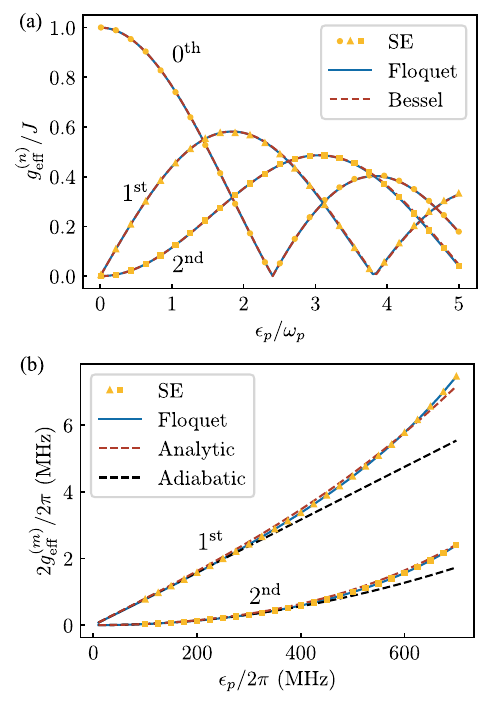}
	\caption{Comparison of parametric coupling strengths for different modulation schemes.
		(a) Parametric coupling strengths for the zeroth, first, and second harmonic orders of the $|10\rangle \leftrightarrow |01\rangle$ transition in a qubit-modulated qubit-qubit system, plotted as a function of the normalized modulation amplitude $\epsilon_p/\omega_p$. Black solid lines, red dashed lines, and gold markers correspond to results from Floquet theory, an analytical model given in Eq. \eqref{eq:qubit_strength}, and the SE method, respectively.
		(b) Parametric coupling strengths for the first and second harmonic orders of the $|100\rangle \leftrightarrow |001\rangle$ transition in a coupler-modulated qubit-coupler-qubit system, as a function of the modulation amplitude $\epsilon_p$. The analytical model (dashed red lines) is derived from Eq. \eqref{eq:coupler_strength}, with the summation truncated at $n=3$ for the first order and $n=4$ for the second. The adiabatic model (dashed black lines) is derived from Eq. \eqref{eq:coupler_strength}, with the summation truncated at $n=1$ for the first order and $n=2$ for the second. The qubit and coupler parameters for numerical simulation and analytic expressions are listed in Table \ref{tab:paras_qq} for (a) and Table \ref{tab:paras_qcq} for (b).
	}
	\label{fig:iswapmefloquetcompare}
\end{figure}
	
	To validate our Floquet analysis, we compare its predictions against full dynamical simulations. We simulate the system by preparing the initial state $|10\rangle$ and evolving it under the Schrödinger equation (SE) for a range of modulation frequencies shown in Fig. \ref{fig:avoiding_cross}(a). The resulting population dynamics within the subspace $\{|10\rangle, |01\rangle\}$ is well-described by a Rabi-driven two-level system with the identical Hamiltonian in Eq. \eqref{eq:Flqouet_hami} and the excited population dynamics follow
	\begin{gather}
		P_{|01\rangle} = \frac{(2g)^2}{(2g)^2 + \Delta^2} \sin^2\left(\frac{\sqrt{(2g)^2 + \Delta^2}}{2} t\right).
		\label{eq:rabi}
	\end{gather}
	We then apply the Floquet numerical method to analyze the qubit-modulated parametric coupling. We calculate all quasienergies of this Hamiltonian with varying-frequency parametric drives and then identify the corresponding quasienergies of Floquet modes (see Appendix \ref{app:identification} for details about state identification). The two identified branches are shown in Fig. \ref{fig:avoiding_cross}(b) and the minimum gap of these branches corresponds to the effective parametric coupling strength $2g$. Figure \ref{fig:avoiding_cross} (c) shows the extracted quasienergy splitting $\Delta\epsilon$ and the corresponding generalized Rabi frequency from the Floquet and SE methods, respectively. Figure \ref{fig:avoiding_cross} (d) shows the corresponding collision angles. The results of these methods demonstrate excellent agreement and validate our Floquet method.  Computationally, the Floquet method is highly efficient, requiring integration over only a single drive period \cite{Petrescu2023}. This is significantly faster than dynamical simulations, which require long evolution times to resolve small coupling strengths. Combined with optimization algorithms, the Floquet method offers a fast and accurate tool for determining coupling strengths and resonance frequencies.

	For a complete comparison, we analyze both qubit-modulated and coupler-modulated parametric transitions within the single-excitation manifold (i.e., $|10\rangle \leftrightarrow |01\rangle$ and $|100\rangle \leftrightarrow |001\rangle$) as a function of (normalized) modulation amplitude (the higher-level parametric couplings are discussed in Appendix \ref{app:higher_levels}). For the qubit-modulated case, we set the modulation frequency $\omega_p$ to satisfy the $n=0,1,2$ order sideband conditions ($\Delta+n\omega_p=0$) and vary the normalized modulation amplitude $\epsilon_p/\omega_{p}$. The resulting coupling strengths are shown in Fig. \ref{fig:iswapmefloquetcompare} (a). The numerical results from both the Floquet and SE methods show excellent agreement with the analytical prediction from Eq. \eqref{eq:qubit_strength}. For the coupler-modulated case, the first- and second-order parametric couplings can be activated through matching the sideband condition $\Delta_{12}+m\omega_p=0$ for $m=1,2$. Figure \ref{fig:iswapmefloquetcompare}(b) shows the coupling strengths as the modulation amplitude $\epsilon_{p}$ increases, and excellent agreement between numerical and analytical results. For analytic results based on Eq. \eqref{eq:coupler_strength}, we truncate the cumulative order to $n=3$ and $n=4$ for the first- and second-order coupling, respectively. For comparison, we also demonstrate the standard adiabatic approximation for weak modulations \cite{McKay2016, Roth2017}, i.e., truncating the cumulative order to $n=1$ and $n=2$ for the first- and second-order coupling, respectively. As expected, for large modulation amplitudes, we observe deviations between the adiabatic approximation and the numerical results. We attribute this discrepancy to the breakdown of the assumptions in the SW transformation in the nonadiabatic regime. This comprehensive comparison demonstrates the effectiveness and accuracy of both our Floquet and analytical methods.

For the numerical simulations, the system is evolved directly under the time-dependent Hamiltonian in the laboratory frame [Eqs. \eqref{eq:hamiltonian_lab} and \eqref{eq:hamiltonian_coupler_lab}] with the explicit modulation terms. Each transmon is truncated to its lowest three energy levels. This exact numerical integration serves as a benchmark, avoiding the RWA, the truncation of the infinite series, and the perturbative truncation errors inherent in the effective Hamiltonian.
	 
	 \section{Frequency collisions in parametrically modulated systems}
	\label{sec:frequency_collisions_in_parametric_modulation}

The ability to precisely activate parametric transitions is the cornerstone of advanced quantum protocols, enabling both the construction of high-fidelity entangled gates and the engineering of analog quantum simulations. Activating one of these target transitions, however, is not a perfectly isolated process. The drive inevitably excites a dense landscape of parasitic sideband transitions that act as error channels. Systematically predicting and navigating this error landscape is therefore a critical task.

The specific structure of frequency collisions is determined by the system's architecture and connectivity. To address this challenge, we develop a complementary approach that combines numerical Floquet simulations with the physical insight of analytical models derived in the preceding section. We employ Floquet theory to numerically map out the complete frequency collision landscape and determine the global maximum collision angle $\max \theta$ across all relevant transitions.  Concurrently, our analytical expressions allow us to identify the dominant error channels and understand their physical origins. The insights from this comprehensive analysis enable us to derive concrete frequency design criteria and propose an optimization algorithm to systematically mitigate the impact of these collisions.

	 \subsection{Qubit-modulated coupling in the qubit-qubit system}
	 \label{sec:Qubit_modulated_coupling_for_the_qubit-qubit system}
	 
The simplest architecture for demonstrating parametric modulation is the directly-coupled two-qubit system. As analytically derived in Sec. \ref{sec:qubit_modulate_parametric_coupling}, parametrically driving such a system induces a rich spectrum of sideband transitions, whose effective coupling strengths and detunings are accurately predicted by Eq. \eqref{eq:Ham8}. Truncating each transmon to its lowest four levels, we identify nine primary transitions of interest, which are listed in Table \ref{tab:TF}. These include three co-rotating and six counterrotating transitions. We note that while state $|3\rangle$ is included here to provide a systematic map of frequency resonance conditions, a quantitative dynamical analysis of transitions involving $|3\rangle$ would require moving beyond the simplified Kerr approximation to include non-conserving terms of the full transmon cosine potential, as discussed in Appendix \ref{app:full_circuit}.

\begin{table}[htbp]
	\centering

\caption{Primary parametric transitions in a qubit-modulated qubit-qubit system. The table lists the most relevant parametrically induced transitions, categorized in the first column as either co-rotating or counterrotating. The second column identifies the specific states involved in each transition. The third column provides the analytical expression for the corresponding time-dependent interaction, expressed as a sum over all harmonic sidebands ($n$). The coefficient of each term in the sum corresponds to the $n$th-order effective coupling strength.}
\begin{ruledtabular}
	\begin{tabular}{ccl}
		Rotating  & Transitions & Couplings ($n\in\mathbb{Z}$) \\ 
		\hline
		Co
		& $|01\rangle\leftrightarrow|10\rangle$ & $J \sum_{n} J_n(\frac{\epsilon_p}{\omega_{p}}) e^{i(\Delta + n\omega_{p}) t} $ \\ 
		& $|11\rangle\leftrightarrow|02\rangle$ & $\sqrt{2}J \sum_{n} J_n(\frac{\epsilon_p}{\omega_{p}}) e^{i(\Delta+\bar\alpha_2 + n\omega_{p} ) t} $  \\ 
		& $|11\rangle\leftrightarrow|20\rangle$ & $\sqrt{2}J \sum_{n} J_n(\frac{\epsilon_p}{\omega_{p}}) e^{i(\Delta-\bar\alpha_1 + n\omega_{p} ) t} $  \\ 
		\hline
		Counter
		& $|00\rangle\leftrightarrow|11\rangle$ & $J \sum_{n} J_n(\frac{\epsilon_p}{\omega_{p}}) e^{i(-\Sigma + n\omega_{p}) t} $ \\ 
		& $|01\rangle\leftrightarrow|12\rangle$ & $\sqrt{2}J \sum_{n} J_n(\frac{\epsilon_p}{\omega_{p}}) e^{i( -\Sigma -\bar\alpha_2 + n\omega_{p}) t} $  \\
		& $|10\rangle\leftrightarrow|21\rangle$ & $\sqrt{2}J \sum_{n} J_n(\frac{\epsilon_p}{\omega_{p}}) e^{i(-\Sigma - \bar\alpha_1 + n\omega_{p}) t} $ \\ 
		& $|02\rangle\leftrightarrow|13\rangle$ & $\sqrt{3}J \sum_{n} J_n(\frac{\epsilon_p}{\omega_{p}}) e^{i( -\Sigma - 2\bar\alpha_2 + n\omega_{p}) t} $ \\
		& $|11\rangle\leftrightarrow|22\rangle$ & $2J \sum_{n} J_n(\frac{\epsilon_p}{\omega_{p}}) e^{i(-\Sigma - \bar\alpha_1 - \bar\alpha_2 + n\omega_{p}) t} $  \\ 
		& $|20\rangle\leftrightarrow|31\rangle$ & $\sqrt{3}J \sum_{n} J_n(\frac{\epsilon_p}{\omega_{p}}) e^{i(-\Sigma - 2\bar\alpha_1 + n\omega_{p}) t} $  \\ 
	\end{tabular}
\end{ruledtabular}
	\label{tab:TF}
\end{table}

\begin{figure}[htbp]
	 	\centering
	 	\includegraphics{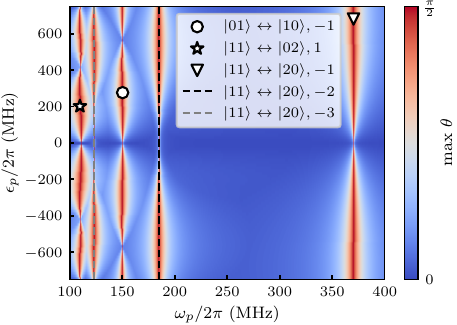}
	 	\caption{Maximum collision angle landscape for qubit-modulated interactions in a qubit-qubit system.
	 		The plot shows the simulated maximum collision angles, $\max \theta$, as a function of the modulation frequency, $\omega_p$, in the range of $100 - 400$ MHz from Floquet theory. Each red branch represents a specific sideband transition, with its prominence determined by the coupling strength and detuning. The markers (circle, star, and triangle) indicate operating points for three target first-order interactions — $|01\rangle\leftrightarrow|10\rangle$, $|11\rangle\leftrightarrow|02\rangle$, and $ |11\rangle\leftrightarrow|20\rangle$, respectively—chosen at modulation amplitudes that maximize their coupling strengths (i.e., $\epsilon_{p}/\omega_p=1.84$). For illustration, the second- (black dashed line) and third-order (gray dashed line) sidebands for the $|11\rangle\leftrightarrow|20\rangle$ transition are explicitly highlighted. The qubit parameters used for the simulation are listed in Table \ref{tab:paras_qq}.
	 		}
	 	\label{fig:thetatf}
	 \end{figure}

For our analysis, we select three first-order sideband transitions ($n = \pm 1$) as representative target interactions: $|01\rangle\leftrightarrow|10\rangle$ at $ \omega_p=\Delta$, $|11\rangle\leftrightarrow|02\rangle$ at $ \omega_p=-\Delta-\alpha_2$, and $|11\rangle\leftrightarrow|20\rangle$ at $\omega_p=\Delta-\alpha_1$ (these settings are to make $\omega_p >0$ for parameters in Table \ref{tab:paras_qq}). Their respective effective strengths are 
\begin{equation}
	\begin{split}
		g_{\text{eff}, |01\rangle\leftrightarrow|10\rangle}^{(-1)} &= JJ_{-1}({\epsilon_{p}}/{\omega_p}),\\
		g_{\text{eff}, |11\rangle\leftrightarrow|02\rangle}^{(1)} &=\sqrt{2} JJ_{1}({\epsilon_{p}}/{\omega_p}),\\
		g_{\text{eff}, |11\rangle\leftrightarrow|20\rangle}^{(-1)} &= \sqrt{2} JJ_{-1}({\epsilon_{p}}/{\omega_p}).
	\end{split}
\end{equation}
To maximize operation speed, we operate at a modulation amplitude $\epsilon_{p}$ that maximizes the magnitude of the first-order Bessel function, corresponding to ${\epsilon_{p}}/{\omega_p}=1.84$ and yielding $|J_{\pm1}({\epsilon_{p}}/{\omega_p})| \approx 0.58$. Figure \ref{fig:thetatf} demonstrates the maximum collision angles of qubit-modulated parametric couplings across a frequency range of $100$ to $400$ MHz. Higher-order collision angles of counterrotating transitions are not shown, as their corresponding coupling strengths are negligible and they are also densely packed.

	At these operating points marked in Fig. \ref{fig:thetatf}, the target operation is perturbed by nonresonant parasitic couplings, which induce two primary error forms: population error, corresponding to population transfer to unwanted states, and phase error, arising from shifts of the computational energy levels. Here, we focus on quantifying the population error under the assumption of an ideal square pulse, neglecting any errors that may arise from the pulse ramps. The magnitude of this error depends on the coupling strength and detuning of each parasitic coupling, which can be described using excited population dynamics described by the Rabi model, as shown in Eq.~\eqref{eq:rabi}. For a target interaction with a strength $g_{\text{t}}$ implemented as a $\pi$ pulse of duration $t = \pi/2g_{\text{t}}$, the total error probability $P_{\text{e}} $ can be bounded by summing the contributions from all unwanted $n$th-order couplings listed in Table \ref{tab:TF}:
	 \begin{equation}
	 	\begin{split}
	P_{\text{e}} 
	&= \sum_{i,n} \frac{(2g_{\text{eff},i}^{(n)})^2}{(2g_{\text{eff},i}^{(n)})^2 + \Delta_{i,n}^2} \sin^2(\frac{\pi\sqrt{(2g_{\text{eff},i}^{(n)})^2 + \Delta_{i,n}^2}}{4g_{t}})\\
&\leq \sum_{i,n} \frac{(2g_{\text{eff},i}^{(n)})^2}{(2g_{\text{eff},i}^{(n)})^2 + \Delta_{i,n}^2}= P_{\text{e}}^{\text{bound}}.
\label{eq:population_errors}
	 	\end{split}
	\end{equation}
	 Here, the index $i$ represents a specific unwanted transition, while $2g_{\text{eff},i}^{(n)}$ and $\Delta_{i,n}$ are its $n$th-order effective coupling strength and corresponding detuning.  $P_{\text{e}}^{\text{bound}}$ defines the worst-case error, i.e., the upper bound of errors, which is suppressed for weaker coupling strengths and larger detunings. This upper bound is a conservative estimate, as the oscillatory nature of the nonresonant evolution can cause the population to return to the initial state at the end of the operation \cite{Barends2019}. We evaluate this conservative bound by treating the individual error channels independently, it is therefore expected to overestimate the actual operation error.
	 
	 \begin{figure*}[htbp]
	\centering
	\includegraphics{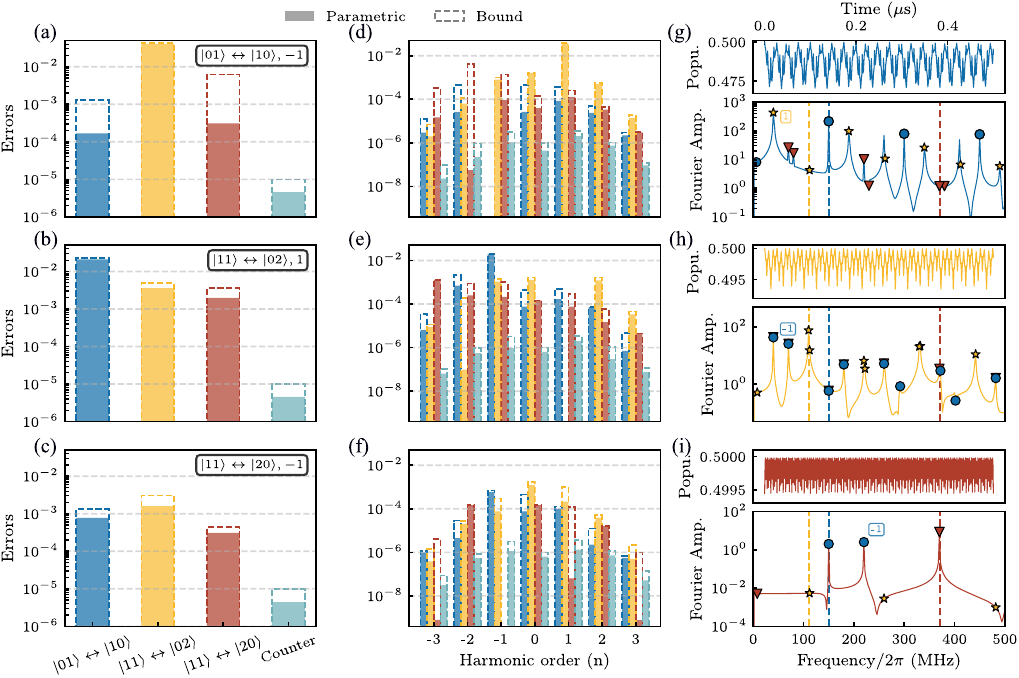}
	\caption{
		Population error analysis and numerical validation for qubit-modulated coupling in the qubit-qubit system.
		(a)-(c) Calculated population errors for three different target first-order interactions: (a) $|01\rangle\leftrightarrow|10\rangle$, 
		(b) $|11\rangle\leftrightarrow|02\rangle$, and (c) $|11\rangle\leftrightarrow|20\rangle$. The operating points are chosen to maximize the respective coupling strengths, as marked in Fig. \ref{fig:thetatf}. The bars show the contributions from parasitic co-rotating transitions---$|01\rangle\leftrightarrow|10\rangle$ (blue), $|11\rangle\leftrightarrow|02\rangle$ (gold), and $|11\rangle\leftrightarrow|20\rangle$ (red)---and the total counterrotating errors (teal). Solid bars represent the calculated error ($P_e$), while dashed borders indicate the upper bound ($P_e^{\text{bound}}$). The results confirm that co-rotating terms are the dominant source of error with negligible counterrotating errors.
		(d)-(f) Error breakdown by harmonic order, $n$, for the corresponding target interactions in (a)-(c). The colored bars identify the source transition for each harmonic's error contribution. The dominant errors originate from low-order harmonics, which have stronger couplings. The harmonic corresponding to the target interaction is omitted from each plot [e.g., $n=-1$ is omitted in (d)].
		(g)-(i) Validation of the error model via direct dynamical simulation. The top panels show the time-domain micromotion of a nontarget state's population over $0.5~\mu\text{s}$ ($100,000$ points), while the bottom panels show the corresponding Fourier transform at the range $[0, 500]$ MHz. Plotted are the populations of  
		(g)~$|11\rangle$ (blue line) when targeting $|01\rangle\leftrightarrow|10\rangle$, (h)~$|01\rangle$ (gold line) when targeting $|11\rangle\leftrightarrow|02\rangle$, and (i)~$|01\rangle$ (red line) when targeting $|11\rangle\leftrightarrow|20\rangle$. The peaks in the Fourier spectrum, which represent the frequencies of the population micromotion, align closely with the theoretically predicted frequencies of parasitic sideband transitions listed in Table \ref{tab:TF} (marked with blue circles, gold stars, and red triangles). The numbers in boxes label the maximum Fourier amplitudes from non-target states, which correspond to the errors in (d)-(f). This confirms that the population errors originate from these unwanted off-resonant couplings. The blue, gold, and red dashed lines in the bottom panels represent the modulation frequency of these three target interactions, respectively. All dynamical simulations were performed in QuTiP~\cite{Lambert2026} using the \texttt{vern9} solver with tolerances of $\text{atol} = \text{rtol} = 10^{-12}$.
	}
	\label{fig:tferrors}
\end{figure*}

 Figures \ref{fig:tferrors}(a)-(c) show the sum theoretical population errors of nonresonant parasitic couplings at the above three operating points. Errors induced by counterrotating terms are substantially smaller than those from co-rotating terms, confirming that the RWA is highly accurate for parametric modulation. Figures \ref{fig:tferrors}(d)-(f) detail the error contributions from different harmonic orders $n\in [-3,3]$. The dominant errors arise from lower-order harmonics, which exhibit stronger couplings at the chosen operating point with ${\epsilon_{p}}/{\omega_p}=1.84$. 

To directly validate the physical assumptions of this theoretical error model, we also perform a full dynamical simulation in the time domain using the SE method. We prepare the initial dressed state $(|01\rangle + |11\rangle)/\sqrt{2}$ and track the subsequent population dynamics of the state $|11\rangle $ (or $|01\rangle $) for each of the three target transitions over $ 0.5$ $\mu\rm{s}$ (assuming $\phi_p=0$). The simulation reveals fast, nonresonant population oscillations, which are termed micromotion. The time trace and corresponding Fourier transforms of this micromotion are shown in Figs. \ref{fig:tferrors}(g)-\ref{fig:tferrors}(i). As seen in the bottom panels, the Fourier spectrum exhibits distinct peaks whose frequencies align closely with the analytical parasitic sideband couplings listed in Table \ref{tab:TF}. This one-to-one correspondence provides compelling evidence that the micromotion in parametrically modulated systems originates precisely from these nonresonant couplings, thereby validating the central hypothesis of our error analysis. We also note that our choice of dressed states as the computational basis prevents such micromotion of the bare states during off-resonant idling, a conclusion supported by our simulations \cite{Galiautdinov2012}.

	Overall, the analysis in Fig. \ref{fig:tferrors} shows that targeting the $ |11\rangle\leftrightarrow|20\rangle$  transition leads to the lowest error. This corresponds to the landscape in Fig. \ref{fig:thetatf}, where fewer undesired collision angles at the operating coefficient $\epsilon_{p}/\omega_{p}$ lead to lower errors. Minor Fourier frequency deviations shown in Figs. \ref{fig:tferrors} (g)-\ref{fig:tferrors}(i) between theory and simulation for higher-energy transitions like $|11\rangle\leftrightarrow|02\rangle$ can be attributed to level shifts from the static coupling $J$.

	Based on this analysis, we can formulate frequency design criteria for minimizing errors, summarized in Table \ref{tab:TF_constraint}. 
	From Sec. \ref{sec:qubit_modulate_parametric_coupling}, an intuitive guideline is to activate the target parametric coupling while suppressing parasitic interactions at the limitation where the modulation frequency is much greater than the corresponding static coupling. Three factors--unwanted coupling strengths, the corresponding detunings, and the duration of the target coupling--determine these nonresonant population errors. Strengths and detunings determine the upper bound of these errors. Unlike the upper-bound $P_e^{\text{bound}}$, calculating the error $P_e$ using the actual interaction duration accounts for the dynamics of population transfer. This yields physically meaningful results even at resonance ($\Delta_{i,n}=0$), where the upper bound incorrectly predicts maximum error for weak couplings. The constraints in Table \ref{tab:TF_constraint} are redundant and we can simplify these constraints according to the realistic requirements and the feature of the Bessel function. For example, the strengths in Eq. \eqref{eq:qubit_strength} are $g_{\text{eff}}^{(n)}= \sqrt{C}JJ_{n}({\epsilon_{p}}/{\omega_{p}}) \lesssim \sqrt{C}J$. So the parametric limitation $ |\omega_p| \gg \sqrt{C}J $ can absorb most constraints from parasitic couplings. Our analysis provides a systemic view to reduce these errors from the unwanted frequency collisions in the parametrically modulated qubit-qubit system.

	\begin{figure*}[htbp]
		\centering
		\includegraphics{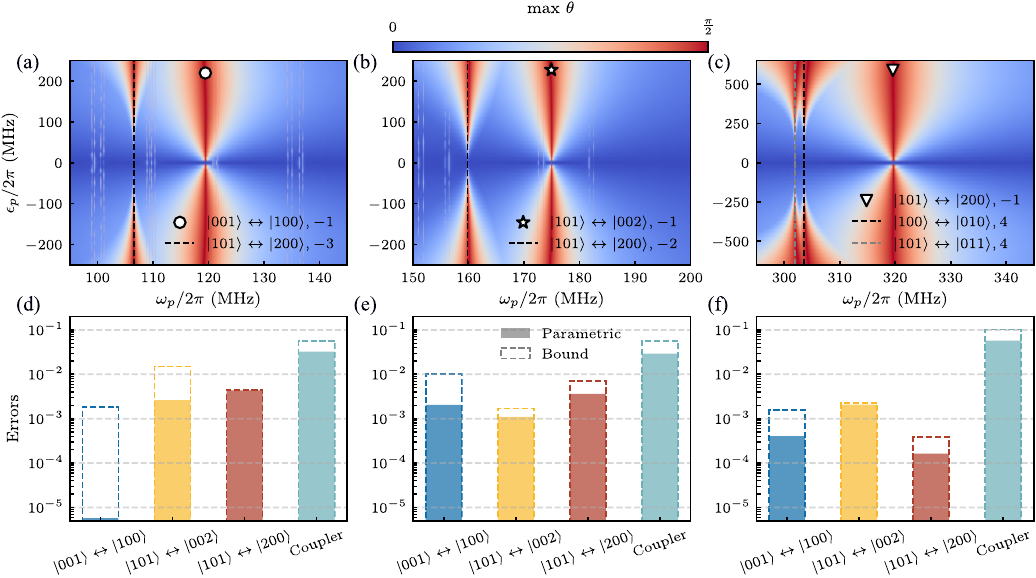}
		
\caption{
	Maximum collision angle landscape and population error analysis for qubit-modulated coupling in the qubit-coupler-qubit system.
	{(a)-(c)} The maximum collision angles, $\max \theta$, are plotted in a $50$ MHz window around the resonant frequency for three target first-order transitions: {(a)} $|001\rangle\leftrightarrow|100\rangle$, {(b)}~$|101\rangle\leftrightarrow|002\rangle$, and {(c)} $|101\rangle\leftrightarrow|200\rangle$. The landscape reveals numerous parasitic sidebands arising from both qubit-qubit and qubit-coupler interactions. Strong parasitic couplings are highlighted with dashed lines and labeled with the sideband order of transitions. The fine and almost invisible branches in (a) and (b) are the collision angles of high-order qubit-coupler parametric couplings. The markers (circle, star, and triangle) indicate the chosen operating points, corresponding to modulation amplitudes of $\epsilon_p/\omega_p=1.84$, $1.3$, and $1.84$, respectively.
	{(d)-(f)} The corresponding population errors calculated at the operating points marked in (a)-(c). The bars detail the error contributions from parasitic co-rotating transitions listed in Table \ref{tab:QCQ}: $|001\rangle\leftrightarrow|100\rangle$ (blue), $|101\rangle\leftrightarrow|002\rangle$ (gold), $|101\rangle\leftrightarrow|200\rangle$ (red), and the sum of directly-driven relevant qubit-coupler transitions (teal) listed in Table \ref{tab:QCQ}. The dashed borders represent the upper error bound. The total error is calculated by summing contributions from up to $\pm 15$ harmonic orders in Eq. \eqref{eq:qubit_strength}, with the results showing that coupler-assisted parasitic couplings are the dominant source of error. Numerical parameters for the system are listed in Table \ref{tab:paras_qcq}.
}

		\label{fig:qcqqthetaerror}
	\end{figure*}

	\begin{figure}[htbp]
	\centering
	\includegraphics{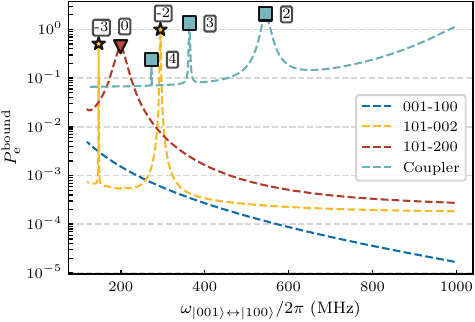}
	\caption{Error competition between qubit- and coupler-mediated parasitic couplings in a qubit-modulated qubit-coupler-qubit system. The plot shows the upper bound of the population error as a function of the resonant modulation frequency $\omega_{|001\rangle\leftrightarrow |100\rangle}$ for the target transition $|001\rangle\leftrightarrow |100\rangle$. The target interaction strength is held fixed at $2g_{\text{eff}}^{(1)} = 3$ MHz. Different dashed lines represent the error contributions from various parasitic channels: qubit-qubit transitions $|001\rangle\leftrightarrow |100\rangle$ (blue, $m \neq 1$), $|101\rangle\leftrightarrow |002\rangle$ (gold), $|101\rangle\leftrightarrow |200\rangle$ (red), and the sum of primary qubit-coupler transitions ($|100\rangle\leftrightarrow |010\rangle$ and $|101\rangle\leftrightarrow |011\rangle$) (teal). Peaks in the plot correspond to resonances with parasitic sidebands, with the harmonic order $m$ indicated in the boxes. The total error bound is calculated by summing contributions up to $|m|= 15$ harmonic orders, revealing the competition between qubit- and coupler-mediated error sources across the frequency range. Numerical parameters are taken from Table \ref{tab:paras_qcq}, with the qubit frequencies $\omega_1$ and $\omega_2$ exchanged to maintain the dispersive regime as $\omega_{|001\rangle\leftrightarrow |100\rangle}$ is varied.}
	\label{fig:errosbndvary}
\end{figure}

	\begin{figure*}[htbp]
		\centering
		\includegraphics{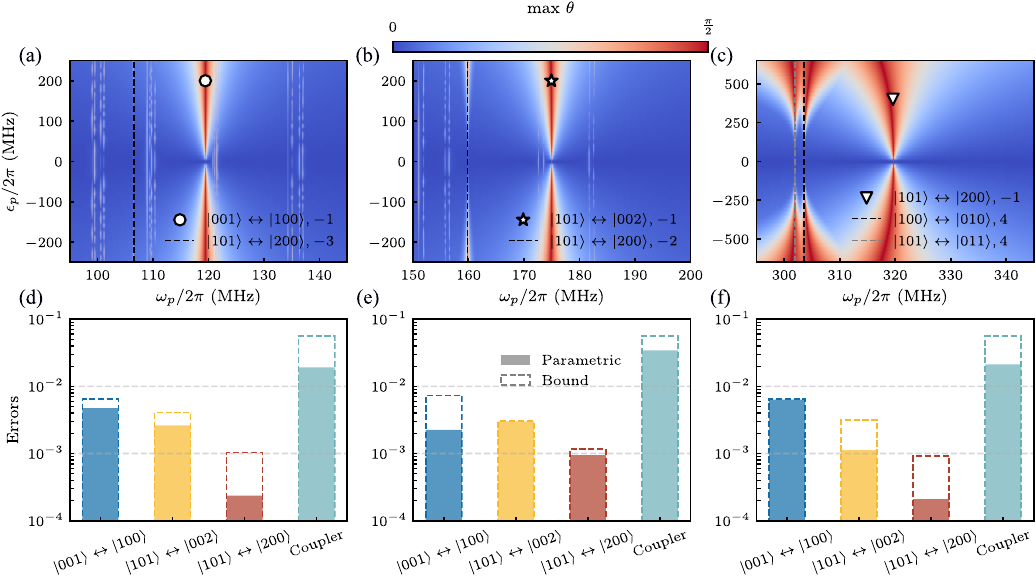}
		\caption{
			Maximum collision angle landscape and population error analysis for coupler-modulated coupling in the qubit-coupler-qubit system.
			{(a)-(c)} The maximum collision angles, $\max \theta$, for three target first-order transitions: {(a)}~$|001\rangle\leftrightarrow|100\rangle$, {(b)} $|101\rangle\leftrightarrow|002\rangle$, and {(c)}~$|101\rangle\leftrightarrow|200\rangle$. The plots are shown in a $50$ MHz window around the respective resonant frequencies. The markers (circle, star, and triangle) indicate the chosen operating points, with modulation amplitudes of $\epsilon_p/2\pi = 200$, $200$, and $400$ MHz, respectively. The complex landscape, featuring numerous parasitic branches, reveals
			qubit-qubit and qubit-coupler interactions listed in Table \ref{tab:QCQ}. Strong interactions cause significant distortions and frequency shifts of the resonance lines, a feature particularly visible in (c).
			{(d)-(f)} The corresponding population error budget calculated at the operating points marked in (a)-(c). The bars detail the error contributions from parasitic co-rotating transitions listed in Table \ref{tab:QCQ} including qubit-qubit couplings [$|001\rangle\leftrightarrow|100\rangle$ (blue), $|101\rangle\leftrightarrow|002\rangle$ (gold), and $|101\rangle\leftrightarrow|200\rangle$ (red)] and the sum of all relevant qubit-coupler transitions (teal). Dashed borders indicate the upper error bound. The total error, summed over $\pm 15$ harmonic orders in Eqs. \eqref{eq:qubit_strength} and \eqref{eq:coupler_strength}, is again dominated by these coupler-assisted parasitic couplings. Numerical parameters for the system are listed in Table \ref{tab:paras_qcq}.
		}
		\label{fig:qcqcthetaerror}
	\end{figure*}
	
\begin{table}[!ht]
	\centering

	\caption{
		Frequency design criteria for qubit-modulated interactions in the qubit-qubit system. Each column is dedicated to a specific target interaction of harmonic order $n'$. The first two rows specify the target transition and its corresponding resonant modulation frequency, $\omega_p$. The third row (``Para. Limit.") presents a general validity condition for the parametric approximation, requiring the modulation frequency to be much larger than the static coupling. The final three rows outline the conditions to suppress the most significant parasitic sideband couplings (where the parasitic order $m \neq n'$). These constraints require the detuning to each parasitic transition to be much greater than its effective coupling strength.}
\begin{ruledtabular}
	\begin{tabular}{cccc}
			& $|01\rangle\overset{n^\prime}\longleftrightarrow|10\rangle$ & $|11\rangle\overset{n^\prime}\longleftrightarrow|02\rangle$ & $|11\rangle\overset{n^\prime}\longleftrightarrow|20\rangle$ \\ 
		\hline

		$\omega_p$& $ |\Delta|/n^\prime$ & $  |\Delta+\bar{\alpha}_2|/n^\prime$ & $  |\Delta-\bar{\alpha}_1|/n^\prime$ \\ 

		Para. Limit. & $\omega_p \gg J$ & $\omega_p \gg  \sqrt{2}J$ & $\omega_p \gg  \sqrt{2}J$ \\ 

		$|01\rangle\leftrightarrow|10\rangle$ &\multicolumn{3}{c}{$|\Delta +m\omega_p| \gg JJ_m(\epsilon_{p}/ \omega_p)$} \\

		$|11\rangle\leftrightarrow|02\rangle$ & \multicolumn{3}{c}{$|\Delta + \bar{\alpha}_2 + m\omega_p| \gg \sqrt{2}JJ_m(\epsilon_{p}/ \omega_p)$} \\

		$|11\rangle\leftrightarrow|20\rangle$ & \multicolumn{3}{c}{$|\Delta - \bar{\alpha}_1 + m\omega_p| \gg \sqrt{2}JJ_m(\epsilon_{p}/ \omega_p)$}  \\

	\end{tabular}
\end{ruledtabular}
	\label{tab:TF_constraint}
\end{table}

	 \subsection{Qubit-modulated coupling in the qubit-coupler-qubit system}
\label{sec:Qubit_modulated_coupling_in_the_qubit_coupler_qubit_system}
To enhance scalability and achieve high-fidelity parametric operations, tunable transmon-type couplers are often introduced to selectively suppress parasitic interactions while enhancing desired ones \cite{Sete2021a, Sete2024, Ma2025}. In the qubit-coupler-qubit architecture shown in Fig. \ref{fig:TF_FTF_demo}(c), the coupler's frequency is tuned to mediate an appropriate effective coupling between the qubits, analogous to the role of a simple coupling capacitor as depicted in Fig. \ref{fig:TF_FTF_demo}(a). While this approach offers greater control, the additional circuit element increases complexity and can open up additional decoherence pathways.

A key advantage of parametric modulation is its ability to mitigate frequency crowding in large-scale processors \cite{Caldwell2018, Reagor2018, Sete2021a}. However, in this architecture, the coupler itself can act as a spectator quantum system \cite{Krauss2025, VallesSanclemente2025}, introducing new potential frequency collisions. This problem is exacerbated by the fact that the qubit-coupler coupling strengths $J_{1c,2c}$ are typically an order of magnitude larger than the direct qubit-qubit coupling $J_{12}$ \cite{Yan2018, Sete2021a}. Consequently, parasitic couplings involving the coupler must be considered, even when the coupler is far detuned from the qubits.

	\begin{table}[!ht]
		\centering
		\caption{
			Classification of parametric transitions in the qubit-coupler-qubit system. The table provides a list of primary parametric transitions, which are categorized based on two criteria. The first column classifies them as either co-rotating or counterrotating. The second column further subdivides them based on the dominant static interaction being parametrically modulated: the direct qubit-qubit coupling ($J_{12}$) or the qubit-coupler couplings ($J_{1c}, J_{2c}$). The third column lists the specific state transitions for each category. The analytical expressions for the effective coupling strengths of these transitions also depend on this classification. Transitions primarily mediated by the direct qubit-qubit coupling ($J_{12}$) are described by the perturbative formulas in Eq. \eqref{eq:effective_coupling} and Appendix \ref{app:higher_levels}. In contrast, transitions mediated by the stronger qubit-coupler couplings are described by simpler expressions analogous to those in the qubit-qubit case of Table~\ref{tab:TF}.}
		\begin{ruledtabular}
		\begin{tabular}{ccc}

			Rotating  & Static Coupling & Transitions  \\ 
			\hline
			Co
			& Qubit  & $|001\rangle\leftrightarrow|100\rangle$  \\ 
			& $\propto J_{12}$ &  $|101\rangle\leftrightarrow|002
			\rangle$  \\ 
			&		    &  $|101\rangle\leftrightarrow|200\rangle$  \\ 
			\cline{2-3} 
			& Coupler &	$|001\rangle\leftrightarrow|010\rangle$ 	\\
			& $\propto J_{1c}, J_{2c}$ & $|100\rangle\leftrightarrow|010\rangle$ 	\\
			& 				& $|101\rangle\leftrightarrow|011\rangle$ 	\\
			& 				& $|101\rangle\leftrightarrow|110\rangle$ 	\\
			&				& $|002\rangle\leftrightarrow|011\rangle$ 	\\
			& 				& $|200\rangle\leftrightarrow|110\rangle$ 	\\
			\hline
			Counter
			& Qubit  & $|000\rangle\leftrightarrow|101\rangle$  \\ 
			& $\propto J_{12}$ &  $|001\rangle\leftrightarrow|102\rangle$  \\ 
			&		    &  $|100\rangle\leftrightarrow|201\rangle$  \\ 
			&		    &  $|101\rangle\leftrightarrow|202\rangle$  \\ 
			&		    &  $|002\rangle\leftrightarrow|103\rangle$  \\ 
			&		    &  $|200\rangle\leftrightarrow|301\rangle$  \\ 
			\cline{2-3} 
			& Coupler &	$|000\rangle\leftrightarrow|011\rangle$ 	\\
			& $\propto J_{1c}, J_{2c}$ & $|000\rangle\leftrightarrow|110\rangle$ 	\\
			& 				& $|001\rangle\leftrightarrow|012\rangle$ 	\\
			&				& $|001\rangle\leftrightarrow|111\rangle$ 	\\
			& 				& $|100\rangle\leftrightarrow|111\rangle$ 	\\
			&				& $|100\rangle\leftrightarrow|210\rangle$ 	\\
			& 				& $|101\rangle\leftrightarrow|112\rangle$ 	\\
			&				& $|101\rangle\leftrightarrow|211\rangle$ 	\\
			& 				& $|002\rangle\leftrightarrow|013\rangle$ 	\\
			&				& $|002\rangle\leftrightarrow|112\rangle$ 	\\
			& 				& $|200\rangle\leftrightarrow|211\rangle$ 	\\
			&				& $|200\rangle\leftrightarrow|310\rangle$ 	\\
		\end{tabular}
	\end{ruledtabular}
		\label{tab:QCQ}
	\end{table}

For our analysis, we adopt the framework from the qubit-qubit case. As confirmed in Sec. \ref{sec:Qubit_modulated_coupling_for_the_qubit-qubit system}, the effects of counterrotating transitions are negligible, so we focus exclusively on co-rotating transitions. The primary parasitic couplings in this system now include not only unwanted qubit-qubit but also qubit-coupler sidebands. We consider all relevant transitions, listed in Table \ref{tab:QCQ}. The effective static coupling strengths of transitions $|001\rangle\leftrightarrow|100\rangle$, $|101\rangle\leftrightarrow|002 \rangle$, and $|101\rangle\leftrightarrow|200 \rangle$ have been discussed in Sec. \ref{sec:coupler_modulated_parametric} and Appendix \ref{app:higher_levels}, resulting in qubit-qubit parametric couplings as discussed in Sec. \ref{sec:qubit_modulate_parametric_coupling}. Those transitions involving coupler excited states can also be considered as qubit-modulated couplings of the qubit-qubit system. Similar to the case in Table \ref{tab:TF}, transitions involving state $|3\rangle$ are included here for kinematic completeness of the resonance map (see Appendix \ref{app:full_circuit} for discussions on exact dynamics).
	
	Following the methodology of Sec. \ref{sec:Qubit_modulated_coupling_for_the_qubit-qubit system}, we analyze the maximum collision angles $\max \theta$ in Eq. \eqref{eq:theta} and population errors $P_\text{e}$ in Eq. \eqref{eq:population_errors} for this system. Three common transitions, $|001\rangle\leftrightarrow|100\rangle$, $|101\rangle\leftrightarrow|002\rangle$, and $|101\rangle\leftrightarrow|200\rangle$ are chosen to demonstrate the maximum collision angles within $50$ MHz around the first-order resonant parametric frequency and analyze the population errors at the operating points shown in Fig. \ref{fig:qcqqthetaerror}. Figures \ref{fig:qcqqthetaerror}(a)-(c) demonstrate the maximum collision angles to reveal the surrounding frequency collisions including other qubit's and coupler's terms, except for those involving coupler states outside the computational subspace, i.e., $|002\rangle\leftrightarrow|011\rangle$ and $|200\rangle\leftrightarrow|110\rangle$. For $|001\rangle\leftrightarrow|100\rangle$ and $|101\rangle\leftrightarrow|002\rangle$ transitions, even if this system owns strong static qubit-coupler coupling strengths, their collision angles are much less pronounced compared with the qubit-qubit coupling strength (i.e., the high order coupling of $|101\rangle\leftrightarrow|200\rangle$) due to the large detuning between the qubit and coupler. For $|101\rangle\leftrightarrow|200\rangle$, it avoids the potential frequency collisions from other qubit-qubit couplings due to its larger first-order resonant frequency, which results in the enhancement of collisions from qubit-coupler parametric couplings. The population errors followed Eq. \eqref{eq:population_errors} at the operating points of these three target interactions are shown in Figs. \ref{fig:qcqqthetaerror}(d)-\ref{fig:qcqqthetaerror}(f). 
	
\begin{table*}[!ht]
	\centering
	\caption{Frequency design criteria for qubit-modulated interactions in the qubit-coupler-qubit system. Each column in the table is dedicated to a specific target $n^\prime$-order interaction. The first three rows are analogous to Table \ref{tab:TF_constraint}. The subsequent rows list the constraints necessary to suppress significant parasitic sidebands (where the parasitic order $m \neq n^\prime$). These constraints are grouped into two types: those mediated by the effective qubit-qubit interaction (e.g., $|001\rangle\leftrightarrow|100\rangle$) and those mediated by the stronger, direct qubit-coupler interactions (e.g., $|100\rangle\leftrightarrow|010\rangle$). Since the drive is applied only to the first qubit, transitions involving the second qubit's coupling to the coupler ($J_{2c}$) are driven indirectly, resulting in a significantly suppressed effective drive amplitude, which is denoted as $\epsilon_p^\prime$.}
	
	\begin{ruledtabular}
		\begin{tabular}{cccc}
			
			& $|001\rangle\overset{n^\prime}\longleftrightarrow|100\rangle$ & $|101\rangle\overset{n^\prime}\longleftrightarrow|002\rangle$ & $|101\rangle\overset{n^\prime}\longleftrightarrow|200\rangle$ \\ 
			\hline
			$\omega_p$& $ |E_{|001\rangle} - E_{|100\rangle}|/n^\prime $ & $|E_{|101\rangle} - E_{|002\rangle}|/n^\prime$  & $ |E_{|101\rangle} - E_{|200\rangle}|/n^\prime$\\ 
			
			Para. Limit. & $\omega_p \gg \tilde{J}_{12}$ & $\omega_p \gg \tilde{J}_{101\leftrightarrow002}$ & $\omega_p \gg  \tilde{J}_{101\leftrightarrow200}$ \\ 
			
			$|001\rangle\leftrightarrow|100\rangle$ & \multicolumn{3}{c}{$|E_{|001\rangle} - E_{|100\rangle} + m\omega_p|\gg \tilde{J}_{12}J_m(\epsilon_{p}/ \omega_p)$} \\ 
			
			$|101\rangle\leftrightarrow|002\rangle$ & \multicolumn{3}{c}{$|E_{|101\rangle} - E_{|002\rangle} + m\omega_p| \gg  \tilde{J}_{101\leftrightarrow002}J_m(\epsilon_{p}/ \omega_p)$}\\ 
			
			$|101\rangle\leftrightarrow|200\rangle$ & \multicolumn{3}{c}{$|E_{|101\rangle} - E_{|200\rangle} + m\omega_p| \gg \tilde{J}_{101\leftrightarrow200}J_m(\epsilon_{p}/ \omega_p)$} \\

			$|001\rangle\leftrightarrow|010\rangle$ & \multicolumn{3}{c}{$|E_{|001\rangle} - E_{|010\rangle} + m\omega_p| \gg J_{2c}J_m(\epsilon_{p}^\prime/ \omega_p)$}\\ 
			
			$|100\rangle\leftrightarrow|010\rangle$ & \multicolumn{3}{c}{$|E_{|100\rangle} - E_{|010\rangle} + m\omega_p| \gg J_{1c}J_m(\epsilon_{p}/ \omega_p)$}\\ 
			
			$|101\rangle\leftrightarrow|011\rangle$ & \multicolumn{3}{c}{$|E_{|101\rangle} - E_{|011\rangle} + m\omega_p| \gg J_{1c}J_m(\epsilon_{p}/ \omega_p)$}\\ 
			
			$|101\rangle\leftrightarrow|110\rangle$ & \multicolumn{3}{c}{$|E_{|101\rangle} - E_{|110\rangle} + m\omega_p| \gg J_{2c}J_m(\epsilon_{p}^\prime/ \omega_p)$}\\ 
		\end{tabular}
	\end{ruledtabular}
	\label{tab:QCQ_Q_constraint}
\end{table*}

The collision landscapes and calculated population errors presented across the multiple panels of Fig. \ref{fig:qcqqthetaerror} reveal a crucial trade-off inherent to the qubit-coupler-qubit architecture under qubit modulation. As observed by comparing the results for different target transitions within Figs. \ref{fig:qcqqthetaerror}(a)-\ref{fig:qcqqthetaerror}(c) and \ref{fig:qcqqthetaerror}(d)-(f), targeting higher-frequency interactions ($|101\rangle\leftrightarrow|200\rangle$) provides isolation from parasitic qubit-qubit sidebands but simultaneously moves the operating point closer to strong qubit-coupler resonances, making them the dominant source of error. Conversely, lower-frequency targets ($|001\rangle\leftrightarrow|100\rangle$ and $|101\rangle\leftrightarrow|002\rangle$) suffer more from qubit-qubit crosstalk but less from coupler-mediated effects. Our analysis further indicates that these coupler-mediated errors primarily originate from low-order sidebands due to the typically large static qubit-coupler coupling strengths $J_{1c,2c}$.

To illustrate this competition more directly for a single case, Fig. \ref{fig:errosbndvary} examines the error budget specifically for the $|001\rangle\leftrightarrow |100\rangle$ target transition (driven via $Q_1$, see constraints in Table \ref{tab:QCQ_Q_constraint}). We plot the upper bound of the population error $P_e^{\text{bound}}$ as its first-order resonant modulation frequency, $\omega_{|001\rangle\leftrightarrow |100\rangle} = E_{|001\rangle} - E_{|100\rangle}$, is varied while keeping the target coupling strength fixed ($E_{|001\rangle} $ and $E_{|100\rangle} $ are the dressed energy levels in Appendix \ref{app:dressed_levels}). As $\omega_{|001\rangle\leftrightarrow |100\rangle}$ increases, the detunings to parasitic transitions involving only qubits (e.g., $|101\rangle\leftrightarrow|002\rangle$) generally increase away from resonance peaks, causing their error contributions to decrease on average. In contrast, the detunings to parasitic transitions involving the coupler (e.g., $|100\rangle\leftrightarrow|010\rangle$) exhibit a different dependence on $\omega_{|001\rangle\leftrightarrow |100\rangle}$, leading to an overall increase in their error contributions across the plotted range. This opposing trend starkly demonstrates the crucial trade-off between minimizing qubit-mediated versus coupler-mediated errors when selecting an operating frequency.

Collectively, the analyses presented in Figs. \ref{fig:qcqqthetaerror} and \ref{fig:errosbndvary} provide complementary perspectives on the significant impact of the spectator coupler. The collision angle landscape, plotted across various potential operating frequencies [Figs. \ref{fig:qcqqthetaerror}(a)-\ref{fig:qcqqthetaerror}(c)], primarily highlights the strengths of individual parasitic resonances, indicating frequency regions where strong unwanted interactions might occur. In contrast, the population error calculations [Figs. \ref{fig:qcqqthetaerror}(d)-\ref{fig:qcqqthetaerror}(f) and Fig. \ref{fig:errosbndvary}] provide a direct estimate of the total performance degradation at specific operating points, reflecting the combined influence of all nearby parasitic couplings. Taken together, these perspectives underscore that, compared with a simple qubit-qubit system, the qubit-coupler-qubit architecture necessitates careful management of an additional, often dominant, set of frequency collisions involving the coupler. The qualitative constraints derived from this analysis are summarized in Table \ref{tab:QCQ_Q_constraint} and inform the optimization problem for designing high-fidelity operations, which we will discuss next.

\begin{table*}[!ht]
	\centering
	\caption{Frequency design criteria for coupler-modulated interactions in the qubit-coupler-qubit system. This table outlines the design rules for implementing three target $n'$th-order interactions via coupler modulation.  The first three rows are analogous to Table \ref{tab:TF_constraint}. The subsequent rows list the constraints required to suppress parasitic sidebands (where the parasitic order $m \neq n'$). The analytical forms of the effective coupling strengths used in these constraints differ based on the interaction type. Transitions between qubit-like states (e.g., $|001\rangle\leftrightarrow|100\rangle$) are governed by the effective strengths $g_{\rm{eff}}^{(m)}$ in Eq. \eqref{eq:coupler_strength}. Parasitic transitions involving the coupler states (e.g., $|001\rangle\leftrightarrow|010\rangle$) are driven more directly and retain their Bessel-function dependence.}
	
	\begin{ruledtabular}
		\begin{tabular}{cccc}
			
			& $|001\rangle\overset{n^\prime}\longleftrightarrow|100\rangle$ & $|101\rangle\overset{n^\prime}\longleftrightarrow|002\rangle$ & $|101\rangle\overset{n^\prime}\longleftrightarrow|200\rangle$ \\ 
			\hline
			$\omega_p$& $ |E_{|001\rangle} - E_{|100\rangle}|/n^\prime $ & $|E_{|101\rangle} - E_{|002\rangle}|/n^\prime$  & $ |E_{|101\rangle} - E_{|200\rangle}|/n^\prime$\\ 
			
			Para. Limit. & $\omega_p \gg \tilde{J}_{12}$ & $\omega_p \gg \tilde{J}_{101\leftrightarrow002}$ & $\omega_p \gg  \tilde{J}_{101\leftrightarrow200}$ \\

			$|001\rangle\leftrightarrow|100\rangle$ & \multicolumn{3}{c}{$|E_{|001\rangle} - E_{|100\rangle} + m\omega_p|\gg g_{\rm{eff}}^{(m)}$}\\ 
			
			$|101\rangle\leftrightarrow|002\rangle$ & \multicolumn{3}{c}{$|E_{|101\rangle} - E_{|002\rangle} + m\omega_p| \gg g_{\rm{eff},101\leftrightarrow002}^{(m)}$}\\ 
			
			$|101\rangle\leftrightarrow|200\rangle$ & \multicolumn{3}{c}{$|E_{|101\rangle} - E_{|200\rangle} + m\omega_p| \gg g_{\rm{eff},101\leftrightarrow200}^{(m)}$} \\

			$|001\rangle\leftrightarrow|010\rangle$ & \multicolumn{3}{c}{$|E_{|001\rangle} - E_{|010\rangle} + m\omega_p| \gg J_{2c}J_m(\epsilon_{p}/ \omega_p)$}\\ 
			
			$|100\rangle\leftrightarrow|010\rangle$ & \multicolumn{3}{c}{$|E_{|100\rangle} - E_{|010\rangle} + m\omega_p| \gg J_{1c}J_m(\epsilon_{p}/ \omega_p)$}\\ 
			
			$|101\rangle\leftrightarrow|011\rangle$ & \multicolumn{3}{c}{$|E_{|101\rangle} - E_{|011\rangle} + m\omega_p| \gg J_{1c}J_m(\epsilon_{p}/ \omega_p)$}\\ 
			
			$|101\rangle\leftrightarrow|110\rangle$ & \multicolumn{3}{c}{$|E_{|101\rangle} - E_{|110\rangle} + m\omega_p| \gg J_{2c}J_m(\epsilon_{p}/ \omega_p)$}\\ 
			
		\end{tabular}
	\end{ruledtabular}
	\label{tab:QCQ_C_constraint}
\end{table*}

	 \subsection{Coupler-modulated coupling in the qubit-coupler-qubit system}
	 An alternative control scheme is coupler-modulated coupling, where the parametric drive is applied to the coupler instead of a qubit. In this configuration, the coupler actively mediates the interaction between the two qubits without directly modulating computational (qubit) levels. While this scheme induces the same set of parametric transitions as in the qubit-modulated scenario (listed in Table \ref{tab:QCQ}), the mechanism is different, leading to distinct effective coupling strengths and detunings.
	 
Following the methodology of Secs. \ref{sec:coupler_modulated_parametric} and \ref{sec:Qubit_modulated_coupling_in_the_qubit_coupler_qubit_system}, we demonstrate the maximum collision angles for the coupler-modulated case as shown in Figs. \ref{fig:qcqcthetaerror}(a)-\ref{fig:qcqcthetaerror}(c). The static Hamiltonian parameters (listed in Table~\ref{tab:paras_qcq}) are identical to those in the qubit-modulated case, but the resulting collision landscapes are qualitatively different. Notably, the collision branches for the $|101\rangle\leftrightarrow|200\rangle$ transition are almost invisible in Figs. \ref{fig:qcqcthetaerror}(a) and \ref{fig:qcqcthetaerror} (b), and branches in Fig. \ref{fig:qcqcthetaerror}(c) exhibit significant distortion, indicative of level repulsion from multiple nearby strong couplings. Overall, coupler modulation results in parametric resonances characterized by markedly reduced magnitudes (smaller collision angles) and sharper, more localized features, contrasting with the stronger and potentially broader resonances typical of the qubit-modulated scenario within the same range of modulation amplitude. This observed feature is consistent with theoretical predictions: coupler-mediated parametric gates are a second-order process, whereas direct qubit modulation is a first-order effect \cite{Krauss2025}. 

We also provide the calculated population errors shown in Figs. \ref{fig:qcqcthetaerror}(d)-\ref{fig:qcqcthetaerror}(f).
The dominant error sources are still from low-order parasitic qubit-coupler sidebands, which means that we should carefully design the coupler parameters to reduce potential collisions like the qubit-modulated scenario. Based on this analysis, we provide modified constraints of qualitative design criteria in Table \ref{tab:QCQ_C_constraint}, which differs from Table \ref{tab:QCQ_Q_constraint} by modifying the qubit-qubit parametric coupling strength and the modulation amplitude related to $J_{2c}$. These constraints can inform an optimization problem for designing high-fidelity operations, which we will discuss next.

	  \subsection{Optimization of frequency allocation}

	The design architecture and desired requirements limit the frequency allocation. The Hamiltonian parameters, such as frequency and anharmonicity, are limited by their inherent features. To realize high-fidelity parametric operations, we prefer a stronger coupling strength and suppressed parasitic couplings. Consequently, the various constraints discussed must be carefully managed to mitigate errors arising from frequency collisions. Combining these constraints with a realistic system model naturally frames the task of frequency allocation as a multi-variable optimization problem. In this section, we synthesize these requirements, considering not only the parasitic couplings central to this work but also other critical factors such as the ZZ coupling \cite{Ganzhorn2020, Mundada2019, Zhao2021, Sung2021, Petrescu2023}.

	  \subsubsection{Constraints}
	  \label{sec:constraints}
	  
The primary constraint, analyzed extensively in this article, is the avoidance of parasitic sideband couplings. As quantified by the population error model given in Eq. \eqref{eq:population_errors} and summarized in Tables \ref{tab:TF_constraint}, \ref{tab:QCQ_Q_constraint}, and \ref{tab:QCQ_C_constraint}, an effective strategy is to minimize the errors arising from any unwanted couplings. Crucially, our comprehensive analysis, encompassing both analytical expressions for coupling strengths and detailed constraint tables, enables us to make informed design choices. By providing a predictive map of target and parasitic interactions as a function of drive parameters, our framework allows for the tailored selection of operating conditions designed to optimize the fidelity of a desired quantum operation while actively suppressing the most detrimental error channels.

While our analysis primarily highlights the dominant low-order transitions, the parametrically driven system also hosts a dense background of numerous higher-order transitions. It is significant to provide a qualitative estimate of their impact. Analytically, the effective coupling strength of an $n$th-order transition generally scales with the Bessel function $|J_n(A)|$, which decays rapidly for large $n$ under typical modulation amplitudes. Numerically, our Floquet analysis captures these numerous transitions, which manifest as the extremely fine, almost invisible branches in Figs. \ref{fig:qcqqthetaerror} and \ref{fig:qcqcthetaerror}. Because bridging a fixed energy detuning with a lower drive frequency requires a larger photon number $n$, this dense background predominantly clusters in the low-frequency regime. Consequently, the collision landscape around the target drive frequencies remains clean and is not overwhelmed, allowing for robust frequency allocation as utilized in our optimization.

	Building upon this observation, a higher modulation frequency $\omega_p$ is generally preferable. As visually confirmed by the maximum collision angles shown in Figs. \ref{fig:thetatf}, \ref{fig:qcqqthetaerror}, and \ref{fig:qcqcthetaerror}, operating at a higher $\omega_p$ provides a larger parameter space for frequency allocation with less severe frequency-crowding effects \cite{Roth2017}, effectively pushing the densely packed high-order parasitic transitions further away. This guideline, along with the requirement that modulation frequency $\omega_p$ is larger than fixed coupling strength, is a direct corollary of the RWA, where faster-oscillating off-resonant terms result in lower errors as discussed in Sec. \ref{sec:qubit_modulate_parametric_coupling}. However, the guideline has a crucial trade-off, which is that higher modulation frequency may bring more collision errors from parasitic qubit-coupler couplings as discussed in Sec. \ref{sec:Qubit_modulated_coupling_in_the_qubit_coupler_qubit_system}. Furthermore, the modulation frequency is constrained by the target coupling strength.

	  \begin{figure}[htbp]
	  	\centering
	  	\includegraphics{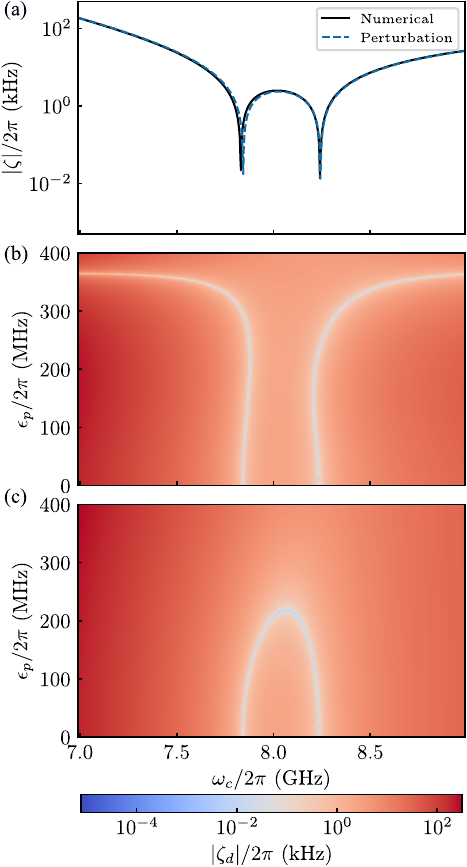}
	  	
		\caption{
			Static and dynamic ZZ coupling in the qubit-coupler-qubit system.
			{(a)} Static ZZ coupling, $\zeta$, as a function of the coupler frequency, $\omega_c$. The black solid line shows the result from exact numerical diagonalization, while the blue dashed line is calculated using perturbation theory up to the fourth order, showing excellent agreement. Numerical parameters for the system are listed in Table \ref{tab:paras_qcq}.
			{(b)-(c)} Dynamic ZZ coupling, $\zeta_d$, for {(b)} qubit-modulated and {(c)} coupler-modulated schemes, plotted as a function of both the coupler frequency, $\omega_c$, and the modulation amplitude, $\epsilon_p$. At zero drive amplitude ($\epsilon_p=0$), the dynamic ZZ correctly reduces to the static values shown in (a). Crucially, the plots reveal contours where the dynamic ZZ coupling can be tuned to zero by co-designing $\omega_c$ and $\epsilon_p$, enabling ZZ-free parametric operations. 
		}
	  	\label{fig:qcqzz}
	  \end{figure}

	  In addition to the above constraints from frequency collisions, one of the major parasitic interactions is ZZ coupling which is mostly caused by repulsion between computational and noncomputational levels. Static and dynamical ZZ couplings have gradually become a major constraint to realizing fast, high-fidelity two-qubit gates and parallel single-qubit gates. Furthermore, the $i$SWAP gate is in particular susceptible to phase errors and ZZ-type crosstalk \cite{Ganzhorn2020}. The static ZZ coupling can be approximated by $\zeta = J^2/(\Delta+\alpha_2) - J^2/(\Delta-\alpha_1) $ for the qubit-qubit system and $\zeta = \zeta^{(2)} +  \zeta^{(3)} + \zeta^{(4)} $ for the qubit-coupler-qubit system where $\zeta^{(n)} $ represents the $n$th-order perturbation result (see Appendix \ref{app:dressed_levels} for more details about derivations). For the qubit-coupler-qubit system, we use exact diagonalization and the fourth-order perturbative theory to obtain the numerical and perturbative static ZZ coupling, respectively, where the theoretical calculation shows good agreement with the numerical simulation, as shown in Fig. \ref{fig:qcqzz}(a). Beyond the static component, parametric modulation introduces a dynamic ZZ coupling. The dynamical ZZ coupling in parametric modulation can be experimentally measured using Ramsey oscillations \cite{Ganzhorn2020} and numerically obtained using the Floquet method \cite{Petrescu2023}. Figures \ref{fig:qcqzz}(b) and \ref{fig:qcqzz}(c) demonstrate, using Floquet numerics, the dynamical ZZ coupling for qubit-modulated and coupler-modulated parametric modulation, respectively, with varying coupler frequency. The dynamical ZZ coupling is defined using the quasienergies \cite{Petrescu2023}
	  \begin{gather}
	  	\zeta_d = \epsilon_{|101\rangle} - \epsilon_{|001\rangle} - \epsilon_{|100\rangle} + \epsilon_{|000\rangle},
	  \end{gather}
	  and the dynamical ZZ coupling reduces to the static value when modulation amplitude $\epsilon_p$ approaches zero. For both qubit-modulated and coupler-modulated schemes, the dynamic ZZ coupling can be tuned to zero by co-designing the coupler frequency $ \omega_c$ and modulation amplitude $\epsilon_{p}$, offering a powerful method for engineering high-fidelity operations. The dynamical ZZ coupling is also dependent on the modulation frequency $\omega_p$ where we keep $\omega_p$ always resonant with the transition $ |100\rangle \leftrightarrow |001\rangle$ for simplicity.

	  In a realistic multi-qubit processor, other constraints must also be considered. Stray couplings to spectator qubits \cite{Krinner2020a, Krauss2025, VallesSanclemente2025} or TLS defects \cite{Mueller2019, Klimov2020} can introduce additional parasitic resonances. During operation execution, the pulse ramp-up may cause transient crossings with unwanted energy levels, leading to population leakage \cite{Sete2024}. Furthermore, the ultimate fidelity is capped by incoherent processes, namely frequency-dependent relaxation and dephasing times \cite{Klimov2024}.

	 Finally, experimental hardware imposes practical constraints. The finite sampling rates of arbitrary waveform generators can introduce signal distortion. Phase noise and clock instability in the control electronics can degrade operational performance, particularly for longer operations \cite{Ganzhorn2020}. These factors must be included in a comprehensive optimization model for predicting achievable fidelities.

	  \subsubsection{Optimization method}
	  The constraints detailed above, while limiting the processor's performance, are predominantly frequency-dependent. This opens the possibility of navigating these limitations by systematically co-designing the system's architecture, chip fabrication, and external flux biasing to allocate operating frequencies into an optimal configuration.
	  
	  To demonstrate this procedure, we consider the example of a scalable architecture using transmon qubits with tunable couplers. Our goal is to find optimal operating frequencies by solving a complex constraint satisfaction problem. Given the large number of coupled, nonlinear constraints, we propose an efficient, two-stage optimization strategy based on satisfiability modulo theories (SMT), a powerful computational tool for solving problems with intricate logical and arithmetic rules (see Appendix \ref{app:smt} for more details about SMT). An extension of SMT is optimization modulo theories (OMT), which finds a solution that also optimizes a specified objective function.

	  First, to ensure the problem is computationally tractable, we formulate an initial model using the system's bare parameters. This is a valid simplification as the qubits and couplers typically operate in the dispersive regime. We then encode all the frequency-dependent constraints from our analysis--including parasitic couplings listed in Tables \ref{tab:TF_constraint}, \ref{tab:QCQ_Q_constraint}, and \ref{tab:QCQ_C_constraint} --into the SMT framework. An SMT solver is then used to find a set of ``satisfiable" bare parameters that meet all specified conditions. In a second stage, these solutions serve as high-quality starting points for a final refinement step. Here, the bare parameters are converted to dressed parameters using perturbation theory or exact diagonalization, allowing for fine-tuning to achieve the precise desired operating points, as outlined in Algorithm \ref{alg:optimization}. Furthermore, optimized results can add new constraints to increase the robustness \cite{Morvan2022}. In the experiment, the measured parameters are dressed levels where we can flexibly solve this constraint optimization problem using SMT \cite{Xu2023, Li2025}.

	  \begin{algorithm}
	  	\caption{Frequency allocation optimization using SMT} \label{alg:optimization}
	  	\KwData{System design targets and component parameter ranges}
	  	\KwResult{Optimized parameter configuration}
	  	Define bare system parameters (frequencies, anharmonicities, couplings) as variables in the SMT solver\;
	  	Encode constraints on static parameters (e.g., qubit frequency ranges, dispersive condition)\;
	  	Encode constraints on dynamic parameters from Tables \ref{tab:TF_constraint}, \ref{tab:QCQ_Q_constraint}, and \ref{tab:QCQ_C_constraint}\;
	  	Encode additional constraints (e.g., minimizing static and dynamic ZZ coupling)\;
	  	Execute SMT or OMT solver to find a set of satisfiable or optimal bare parameters\;
	  	Verify and refine the solution using dressed-level calculations (perturbation theory or numerical exact diagonalization)
	  \end{algorithm}

\subsubsection{A toy model for optimization based on SMT}

The Hamiltonian parameters in Tables \ref{tab:paras_qq} and \ref{tab:paras_qcq} are obtained using the SMT/OMT solver and are employed throughout our numerical simulations. To demonstrate the reproducibility and the detailed logic of our method, we take the qubit-modulated qubit-coupler-qubit system as a toy model.

We utilize pySMT \cite{Gario2015}, a PYTHON library for SMT formulae manipulation and solving, to implement the optimization. This library provides an intuitive and solver-independent interface for our circuit models. Specifically, we map each analytical collision criterion (e.g., those derived in Table \ref{tab:QCQ_Q_constraint}) to a first-order logic formula. The Hamiltonian parameters, including qubit and coupler frequencies and anharmonicities, are defined as real-type symbols.

A key procedure is to incorporate physics-informed constraints to prune the parameter space effectively. For each sideband resonance condition, we impose a constraint $|E_i - E_j + n\omega_p| \ge \kappa \cdot |J_n(A) J_{\text{eff}}|$, where $J_n(A)$ is the $n$th-order Bessel function of the first kind representing the modulation scaling, $J_{\text{eff}}$ is the corresponding effective coupling strength between the energy levels $E_i$ and $E_j$, and $\kappa$ is a safety factor (set to $10$ in this work). This ensures that the solver prioritizes avoiding resonances with larger effective interaction strengths rather than merely searching for arbitrary frequency gaps. Together with hardware-related constraints (e.g., frequency tunability), these constraints are solved using the Z3 solver \cite{Moura2008} to identify the satisfiable or optimal configuration.

To further ensure reproducibility, we provide a concise PYTHON script in Ref. \cite{Ma2026}, demonstrating the automated parameter optimization process.

	\section{Typical applications}
	\label{sec:application_on_lattices}
	
Our systematic analysis of frequency collisions provides a powerful framework for designing and calibrating operations in multi-qubit systems. The two- and three-mode interactions analyzed previously are the fundamental building blocks of large-scale superconducting processors. In this section, we illustrate how our constraint-based optimization method can be applied to a common architectural motif: a quantum lattice for analog simulation. Furthermore, we propose a dynamical ZZ-free parametric $i$SWAP gate based on the above ZZ coupling constraint.

	 \begin{figure}[htbp]
	 	\centering
	 	\includegraphics{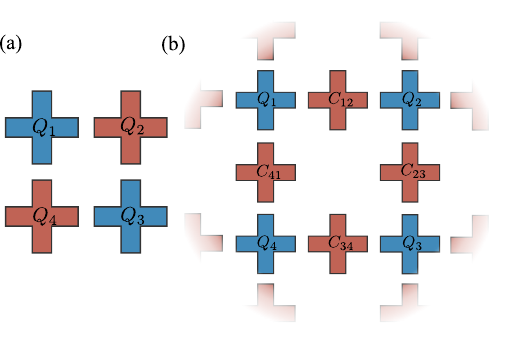}
		\caption{
			Schematic of four-qubit lattice architectures.
			{(a)}~A direct-coupling topology, where four qubits (blue and red cross symbols) are connected via nearest-neighbor capacitive coupling. 
			{(b)}~A coupler-mediated topology, where interactions between the qubits (blue cross symbols) are mediated by tunable couplers (red cross symbols). This arrangement is a common unit cell in scalable quantum processor designs.
		}
	 	
	 	\label{fig:application}
	 \end{figure}

	 \subsection{Analog quantum simulation based on parametric modulation}
	 
A square lattice of four transmon qubits, as depicted in Figs. \ref{fig:application}(a) and \ref{fig:application}(b), is a canonical component for scalable quantum processors. In such an architecture, each qubit is coupled to its nearest neighbors, either directly via capacitors or indirectly via tunable couplers. The system is described by the Hamiltonian:
	 \begin{equation}
	 \begin{split}
		\mathcal{H} &=\sum_{i} \left(\omega_i b_i^{\dagger} b_i+ \frac{\alpha_i}{2} b_i^{\dagger} b_i^{\dagger} b_ib_i\right)  \\
		& +\sum_{\langle i,j \rangle}J_{ij} (b_i + b_i^{\dagger} ) (b_j + b_j^{\dagger} ),
	 \end{split}
	 \end{equation}
 where the index $i$ runs over all quantum elements (qubits and couplers), $\langle i,j \rangle$ denotes nearest-neighbor pairs, and $b_i$ ($b_i^{\dagger}$) is the annihilation (creation) operator for mode $i$. To simulate the desired analog Hamiltonian, a common approach is to engineer a target Hamiltonian within a specific subspace, such as the single-excitation manifold.  This can be achieved using either first-order qubit-modulated \cite{Zheng2022} or coupler-modulated \cite{Zhang2024a, Zhang2025b} schemes, where the modulation parameters (amplitudes, frequencies, and phases) are tuned to realize the desired interactions.
 
 The feasibility of such an analog simulation hinges on avoiding parasitic couplings across the entire device. We can apply our framework to find a viable frequency allocation by making a few simplifying assumptions: we consider only nearest-neighbor couplings, neglect counterrotating terms (a well-justified approximation), and assume a linear flux-frequency response without significant crosstalk.
 
For a qubit-modulated scheme where $Q_1$ and $Q_3$ are parametrically modulated \cite{Zheng2022}, the frequency constraints required to isolate the target interactions are summarized in Table \ref{tab:four_qubits}, following our analysis in Sec. \ref{sec:Qubit_modulated_coupling_for_the_qubit-qubit system}. Alternatively, for a coupler-modulated scheme where drives are applied to all four couplers to form a diamond-like interaction graph \cite{Zhang2024a, Zhang2025b}, the corresponding constraints are given in Table \ref{tab:four_qubits_with_coupler}, following our analysis in Sec. \ref{sec:Qubit_modulated_coupling_in_the_qubit_coupler_qubit_system}.

\begin{table}
	\centering
	\caption{Frequency design criteria for a directly coupled four-qubit analog simulator. This table lists the conditions necessary to suppress parasitic crosstalk in an experiment where qubits $Q_1$ and $Q_3$ are parametrically modulated to simulate a lattice model. The first row (``Para. Limit.") establishes the general condition for the parametric approximation to be valid across all interactions. The subsequent sections (``Modu. $Q_1$" and ``Modu. $Q_3$") specify the constraints needed to isolate the desired interactions by ensuring that all off-resonant parasitic couplings (of order $m$) are suppressed. Here, $E_i$ denotes the dressed frequency of the single-excitation state of qubit $i$.}
		
\begin{ruledtabular}
	\begin{tabular}{cc}
		Para. Limit. & 
		$\omega_{p,12},\ \omega_{p,41},\ \omega_{p,23},\ \omega_{p,34} 
		\gg J_{12},\ J_{41},\ J_{23},\ J_{34}$ 
		\\ \hline
		Modu. $Q_1$ & 
		$\left| E_{4} - E_{1} + m \omega_{p,12} \right|
		\gg J_{41}\, J_{m}\!\left({\epsilon_{p,12}}/{\omega_{p,12}}\right)$
		\\ 

		& $\left| E_{1} - E_{2} + m \omega_{p,41} \right|
		\gg J_{12}\, J_{m}\!\left({\epsilon_{p,41}}/{\omega_{p,41}}\right)$
		\\ \hline
		Modu. $Q_3$ & 
		$\left| E_{3} - E_{4} + m \omega_{p,23} \right|
		\gg J_{23}\, J_{m}\!\left({\epsilon_{p,23}}/{\omega_{p,23}}\right)$
		\\ 

		& $\left| E_{2} - E_{3} + m \omega_{p,34} \right|
		\gg J_{34}\, J_{m}\!\left({\epsilon_{p,34}}/{\omega_{p,34}}\right)$
		\\ 
	\end{tabular}
\end{ruledtabular}
	\label{tab:four_qubits}
\end{table}

\begin{table}
	\centering
	\caption{Frequency design criteria for a coupler-mediated four-qubit analog simulator. This table details the frequency allocation constraints for a lattice where interactions are engineered by applying individual parametric drives to each coupler. The first row (``Para. Limit.") establishes the general validity condition, requiring the drive frequency to greatly exceed the effective qubit-qubit coupling strength, $\tilde{J}_{ij}$. The subsequent rows (``Modu. $C_{12,23,34,41}$") list the specific constraints required when modulating a particular coupler. These conditions ensure that the primary parasitic channels---the direct qubit-coupler sideband interactions---are suppressed, allowing the desired qubit-qubit interaction to dominate. In these expressions, $E_i$ denotes the dressed frequency of the single-excitation state for the corresponding element $i$ (qubit or coupler).}

\begin{ruledtabular}
	\begin{tabular}{cc}
		Para. Limit. & 
		$\omega_{p,12},\ \omega_{p,41},\ \omega_{p,23},\ \omega_{p,34} 
		\gg \tilde{J}_{12},\tilde{J}_{41},\tilde{J}_{23},\tilde{J}_{34}$ 
		\\ \hline
		Modu. $C_{12}$ & 
		$\left| E_{1} - E_{c_{12}} + m \omega_{p,12} \right|
		\gg J_{1,c_{12}}\, J_{m}\!\left({\epsilon_{p,12}}/{\omega_{p,12}}\right)$
		\\ 

		& 
		$\left| E_{c_{12}} - E_{2}  + m \omega_{p,12} \right|
		\gg J_{c_{12},2}\, J_{m}\!\left({\epsilon_{p,12}}/{\omega_{p,12}}\right)$
		\\ \hline
		Modu. $C_{23}$ & 
		$\left| E_{2} - E_{c_{23}} + m \omega_{p,23} \right|
		\gg J_{2,c_{23}}\, J_{m}\!\left({\epsilon_{p,23}}/{\omega_{p,23}}\right)$
		\\ 

		& 
		$\left| E_{c_{23}} - E_{3}  + m \omega_{p,23} \right|
		\gg J_{c_{23},3}\, J_{m}\!\left({\epsilon_{p,23}}/{\omega_{p,23}}\right)$
		\\ \hline
		Modu. $C_{34}$ & 
		$\left| E_{3} - E_{c_{34}} + m \omega_{p,34} \right|
		\gg J_{3,c_{34}}\, J_{m}\!\left({\epsilon_{p,34}}/{\omega_{p,34}}\right)$
		\\ 

		& 
		$\left| E_{c_{34}} - E_{4}  + m \omega_{p,34} \right|
		\gg J_{c_{34},4}\, J_{m}\!\left({\epsilon_{p,34}}/{\omega_{p,34}}\right)$
		\\ \hline
		Modu. $C_{41}$ & 
		$\left| E_{4} - E_{c_{41}} + m \omega_{p,41} \right|
		\gg J_{4,c_{41}}\, J_{m}\!\left({\epsilon_{p,41}}/{\omega_{p,41}}\right)$
		\\ 

		& 
		$\left| E_{c_{41}} - E_{1}  + m \omega_{p,41} \right|
		\gg J_{c_{41},1}\, J_{m}\!\left({\epsilon_{p,41}}/{\omega_{p,41}}\right)$
		\\
	\end{tabular}
\end{ruledtabular}
	\label{tab:four_qubits_with_coupler}
\end{table}

This methodology is readily scalable. For one-dimensional chains or two-dimensional lattices with different connectivity, the constraint set can be systematically constructed following the logic of Tables \ref{tab:four_qubits} and \ref{tab:four_qubits_with_coupler}. The target analog Hamiltonian
determines the choice of computational subspace, which would modify the specific constraints accordingly. By incorporating all relevant constraints into the SMT framework outlined in Algorithm \ref{alg:optimization}, a satisfiable parameter configuration can be found. This approach is particularly well-suited for analog simulation, where the primary goal is often to find any suitable parameter set that reproduces the desired physical phenomenon, rather than a single, globally optimal point. This aligns perfectly with the philosophy of an SMT solver, which is designed to find a satisfiable solution.

	 \subsection{Coupler-assisted dynamical ZZ-free parametric $i$SWAP gate}
		 
	Achieving high-fidelity two-qubit entangling gates is a central challenge in building large-scale quantum computers. While tunable couplers help address scalability, the fidelity of gates like the $i$SWAP is often limited by parasitic ZZ interactions. In a static context, this residual ZZ can be suppressed by carefully tuning the coupler's frequency. However, under parametric modulation, the static ZZ evolves into a dynamic ZZ coupling.
		 
	Our analysis provides a direct pathway to overcome this challenge. As demonstrated in Fig. \ref{fig:qcqzz}, the dynamic ZZ coupling is not just a fixed error but a tunable parameter that can be engineered to zero by co-designing the coupler frequency and the parametric drive. This insight provides a clear recipe for realizing a high-fidelity, dynamically ZZ-free parametric $i$SWAP gate, applicable to both qubit-modulated and coupler-modulated schemes.
	
	Beyond the standard transmon-based coupler, alternative hardware offers further opportunities. For example, using a generalized flux qubit as the coupler can create a system with an intrinsically zero ZZ interaction \cite{Petrescu2023}. Furthermore, this type of coupler leverages three-wave mixing for parametric operations, in contrast to the four-wave mixing used in transmon-based systems. This allows for stronger drives and, consequently, faster gate operations, representing a promising direction for future development.

	 \section{Summary and discussion}
	 \label{sec:summary}

In this article, we have developed a comprehensive framework to systematically analyze and mitigate the frequency collisions in parametrically modulated superconducting quantum circuits. Our approach integrates analytical derivations with Floquet simulations to predict the major landscape captured by the mode of both desired and parasitic sideband interactions. According to this predictive framework, we establish rigorous constraints for high-fidelity operations in both qubit-qubit and qubit-coupler-qubit systems. These physics-informed constraints then serve as the input for a practical, SMT-based optimization algorithm capable of navigating the complex parameter space. Finally, we illustrate the utility of our methodology by applying it to the design of analog quantum simulation and high-fidelity entangling gates.

Our work advances the understanding of time-dependent drives in multi-qubit systems by dissecting the rich landscape of parametric processes that can occur. These processes induce unwanted transitions that constrain the available parameter space for high-fidelity operations. By meticulously modeling these interactions, our framework provides a basis for designing robust control sequences. When combined with advanced optimization strategies, it enables the engineering of precise frequency trajectories that can mitigate errors arising not only from frequency collisions but also from other practical limitations such as TLS defects, stray couplings, and finite coherence times \cite{Klimov2024}.

The framework presented here is general, relying on the fundamental principles of periodic drives rather than a specific qubit implementation. While the specific manifestations of frequency collisions are platform-dependent, our systematic methodology for quantifying them is general and can be readily applied to other platforms where parametric interactions are crucial, such as flux qubits or fluxoniums \cite{Petrescu2023, Ma2024, Zhao2025}. Moreover, the combination of Floquet analysis and constraint-based optimization constitutes a powerful tool for analyzing and engineering other classes of microwave-activated or sideband interactions \cite{Krinner2020, Paolo2022, Heya2024, You2025, Ding2025}.

We acknowledge several avenues for future research that this work enables. Our analysis was primarily based on Hamiltonians with nearest-neighbor couplings. While this is a valid approximation for many architectures, achieving the next echelon of fidelity in large, dense lattices will require accounting for more complex interactions. This challenge can be approached from two directions: from a hardware perspective, by engineering coupling networks that suppress stray interactions \cite{Sete2021, Andersen2025}, and from a software perspective, by incorporating comprehensive, all-to-all coupling models into our constraint-based analysis \cite{Xu2024, Xu2025}. A complete understanding of these many-body effects is essential for mitigating weak, high-order, and long-range parasitic couplings.

Furthermore, our Floquet analysis considered a monochromatic parametric drive without pulse shaping. In practice, gate operations involve ramps, during which the system can transiently cross deleterious resonances \cite{Sete2024}. The adiabaticity of these ramps plays a vital role in suppressing leakage, and their dynamics can be analyzed using the Floquet adiabatic theorem \cite{Weinberg2017}. Designing optimal pulse envelopes to navigate these transient collisions is therefore a crucial next step \cite{Paolo2022, Ding2025}. Extending the analysis to include multichromatic drives, especially in the context of a many-qubit lattice, represents a significant but essential computational challenge for unlocking the full potential of parametric control \cite{Ho1983, Sameti2019, Poertner2020, Gandon2022, Heya2024, BrisenoColunga2025}.

\begin{acknowledgments}
	
	Zhuang Ma would like to thank Baptiste Royer, Camille Le Calonnec, Alberto Mercurio, Kentaro Heya, and Jordan Huang for insightful discussions on Floquet theory. This work was supported by the Innovation Program for Quantum Science and Technology (Grant No. 2021ZD0301702), the NSFC (Grant No. U21A20436), the NSF of Jiangsu Province (Grants No. BE2021015-1, No. BK20232002, and No. BK20233001), the NSF of Shandong Province (Grant No. ZR2023LZH002), and the National Natural Science Foundation of China (Grant No. 12204050).

\end{acknowledgments}

\section*{DATA AVAILABILITY}
The data that support the findings of this article are openly available \cite{Ma2026}.

	\appendix
	\section{Comparison with a full-circuit Hamiltonian}
\label{app:full_circuit}

In the main text, we adopted a simplified model based on the Duffing oscillator and assumed a linear frequency response to external flux. This approach provides an intuitive physical picture of frequency collisions under parametric modulation. In this section, we will discuss the difference between our model and a full-circuit Hamiltonian, which more accurately reflects experimental reality.

A critical simplification in our analytical framework lies in the truncation of the full Josephson cosine potential. The standard Kerr-like model retains only the secular quartic terms (i.e., the Duffing oscillator approximation) to describe the transmon anharmonicity. While this approximation is highly effective for identifying the kinematic resonance conditions of low-lying states, it systematically neglects non-secular and excitation-non-conserving terms. As highlighted in our collision map, when predicting the exact transition matrix elements and dynamical leakage rates involving higher energy levels, specifically the $|3\rangle$ state, these omitted non-secular terms become non-negligible, as they provide additional pathways for multiphoton transitions \cite{Dumas2024}. Therefore, a full-circuit Hamiltonian utilizing the exact cosine potential is required for a rigorous quantitative evaluation of high-excited-state dynamics.

Furthermore, beyond the Hamiltonian truncation, the assumption of a linear flux-to-frequency mapping presents another simplification. In a realistic tunable transmon, the relationship between the applied flux and the qubit frequency is inherently nonlinear, with the exact quantization form determined by the specific circuit geometry and capacitance ratios \cite{You2019, Riwar2022}. This has two significant consequences. First, a single flux drive can be allocated nontrivially across the circuit's elements, potentially creating multiple effective drive operators where only one was intended \cite{Zhao2025}. Second, the nonlinear response to the drive itself induces phenomena not captured by the linear model. Even a monochromatic flux drive can generate higher-order harmonics of the qubit frequency, and the time-averaged frequency can experience a significant drive-power-dependent shift \cite{Rosen2024, Ma2025, Huber2025}.

The full-circuit model also becomes crucial when considering strong drives, which are desirable for fast gate operations. In a linear model, a strong drive implies a large frequency excursion, causing the modulated element's frequency to transiently cross the transition frequencies of other elements, leading to parasitic resonances. Realistic circuit designs mitigate this by employing asymmetric Josephson junctions in the superconducting quantum interference device geometry, which bounds the frequency excursion to avoid this resonance \cite{Didier2018, Li2022}. Furthermore, the strength of coupler-modulated interactions depends directly on the coupler's flux responsivity (i.e., the gradient $d\omega_c/d\Phi$) at the bias point \cite{Roth2017}. A full-circuit model allows this responsivity to be engineered for optimal coupling strength. However, strong drives are not a panacea, and they can also induce unwanted higher-order effects, ionization, and even lead to chaos, which degrades gate fidelity \cite{Lagemann2022, Petrescu2023, Dumas2024, Huber2025, Zhao2025, Xia2025, Baskov2025, Xiao2025}.

Ultimately, the phenomena captured by a full-circuit Hamiltonian introduce a richer, more physically accurate set of constraints into the optimization problem. Incorporating these realistic constraints is essential for designing high-performance circuits tailored for specific parametric processes. While adding complexity, these detailed models concurrently reveal new knobs for control and co-design opportunities. By integrating Floquet theory with such realistic circuit models \cite{Petrescu2023, Paolo2022, Ding2025}, one can create a unifying framework to systematically understand and engineer strong-drive phenomena, paving the way for faster and higher-fidelity operations on superconducting processors.

\section{Analysis of higher-energy-level transitions}
\label{app:higher_levels}

\begin{figure*}[htbp]
	\centering
	\includegraphics{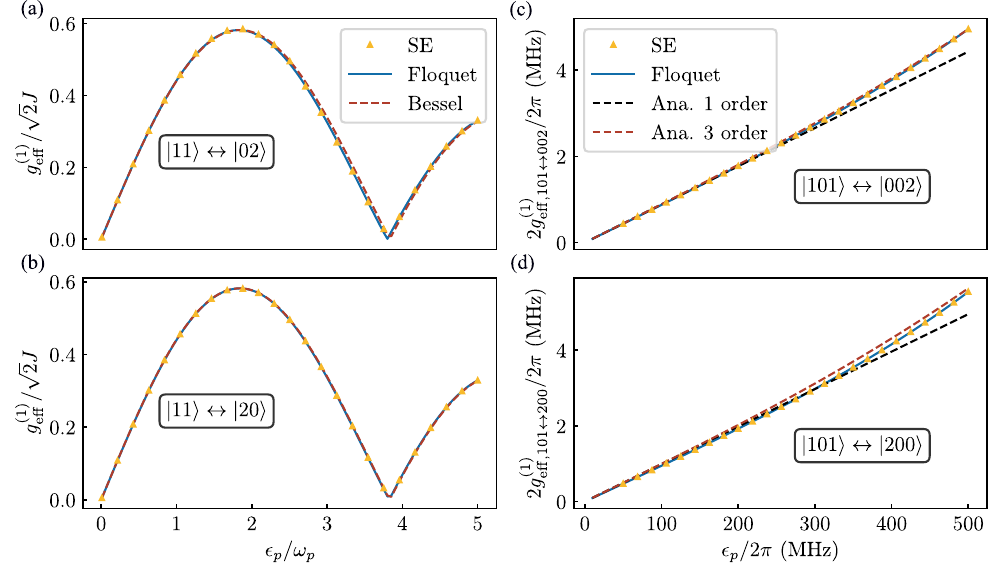}
	\caption{
		Comparison of parametric coupling strengths for higher-energy-level transitions across different schemes.
		{(a)-(b)} Qubit-modulated coupling strengths for the transitions $|11\rangle \leftrightarrow |02\rangle$ and $|11\rangle \leftrightarrow |20\rangle$ as a function of the normalized modulation amplitude $\epsilon_p/\omega_p$. Results from Floquet theory (solid lines), an analytical Bessel-function model (dashed lines), and SE simulations (triangle markers) are compared, showing excellent agreement.
		{(c)-(d)} Coupler-modulated coupling strengths for the transitions $|101\rangle \leftrightarrow |002\rangle$ and $|101\rangle \leftrightarrow |200\rangle$ as a function of the modulation amplitude $\epsilon_p$. The comparison includes two analytical models derived from Eq. \eqref{eq:coupler_strength}: a first-order model (black) where the summation is truncated at $n=1$, and a third-order model (red) where the summation is truncated at $n=3$. The third-order model shows significantly better agreement with the Floquet (solid lines) and SE simulation (triangle markers) results, validating the perturbative approach.
	}
	\label{fig:tfftfczcompare}
\end{figure*}

While the main text focuses on parametric processes within the primary computational subspaces (e.g., $|01\rangle\leftrightarrow|10\rangle$ for the qubit-qubit system and $|001\rangle\leftrightarrow|100\rangle$ for the qubit-coupler-qubit system), a complete error analysis requires considering transitions involving higher energy levels. Parasitic couplings originating from the two-excitation manifold, such as $|11\rangle \leftrightarrow |02\rangle$ and $|11\rangle \leftrightarrow |20\rangle$ in the two-qubit case, or $|101\rangle \leftrightarrow |002\rangle$ and $|101\rangle \leftrightarrow |200\rangle$ in the qubit-coupler-qubit case, are often significant error channels and must be accurately modeled. This appendix provides the detailed analytical formulas used to calculate the coupling strengths for these various higher-energy-level transitions and compares them with numerical simulations, as shown in Fig. \ref{fig:tfftfczcompare}.

For the higher-energy-level transitions in the qubit-qubit system, the coupling strengths are described by the same analytical form as Eq. \eqref{eq:qubit_strength}. The results of this model are plotted as the dashed lines in Fig. \ref{fig:tfftfczcompare}(a) and \ref{fig:tfftfczcompare}(b).

For the qubit-coupler-qubit system, the calculation is more involved. The static Hamiltonian is first approximately diagonalized using the SW transformation \cite{Sete2021a}. This transformation yields the effective static coupling strengths for the transitions $|101\rangle\leftrightarrow|002\rangle$ and $|101\rangle\leftrightarrow|200\rangle$ as:
\begin{gather}
	\begin{split}
	\tilde{J}_{101\leftrightarrow002} =& \sqrt{2}J_{12} +\frac{J_{1c}J_{2c}}{\sqrt{2}}\\
&\times \left(\frac{1}{\Delta_{1c}}+\frac{1}{\Delta_{2c} + \alpha_{2}}-\frac{1}{\Sigma_{1c}}-\frac{1}{\Sigma_{2c}+ \alpha_{2}}\right),\\
\tilde{J}_{101\leftrightarrow200} = &\sqrt{2}J_{12} +\frac{J_{1c}J_{2c}}{\sqrt{2}}\\
&\times \left(\frac{1}{\Delta_{1c}+ \alpha_{1}}+\frac{1}{\Delta_{2c} }-\frac{1}{\Sigma_{1c}+ \alpha_{1}}-\frac{1}{\Sigma_{2c}}\right).
	\end{split}
\end{gather}
So we can replace $\tilde{J}_{12}$ in the formula \eqref{eq:coupler_strength} with $\tilde{J}_{101\leftrightarrow002}$ or $\tilde{J}_{101\leftrightarrow200}$ to obtain the parametric coupling strength $g_{\rm{eff},101\leftrightarrow002}^{(m)}$ or $g_{\rm{eff},101\leftrightarrow200}^{(m)}$ shown in Fig. \ref{fig:tfftfczcompare}(c) and \ref{fig:tfftfczcompare}(d).

	\section{Floquet theory}
	\label{app:floquet_theory}

	Floquet theory provides the structural form for solutions to linear ordinary differential equations with periodic coefficients. In quantum mechanics, it is an essential tool for analyzing the dynamics of periodically driven systems. Analogous to Bloch's theorem in condensed matter physics, Floquet theory has found wide application across physics and has recently proven exceptionally valuable in quantum technology \cite{Grifoni1998, Eckardt2017, Oka2019}, with applications such as Floquet engineering \cite{Kyriienko2018, Sameti2019} and time crystals \cite{Mi2021}, Floquet protection \cite{Huang2021, Huang2022}, gate calibration \cite{Arute2020} and Floquet codes \cite{Hastings2021}.

\subsection{The Floquet theorem}

The dynamics of the state $|\Psi(t)\rangle$ of a periodically driven quantum system are governed by the time-dependent Schrödinger equation:
	\begin{gather}
	i \hbar \frac{\partial}{\partial t} |\Psi(t)\rangle = H(t) |\Psi(t)\rangle \label{eq:Schrodinger}
	\end{gather}
where the Hamiltonian is periodic in time, $H(t+T) = H(t)$, for a system with a $d$-dimensional Hilbert space. The Floquet theorem \cite{Floquet1883} states that there exists a complete set of solutions, $|\Psi_{\alpha}(t)\rangle$, such that any arbitrary solution $ |\Psi(t) \rangle$ can be expressed as their superposition:
	\begin{gather}
	|\Psi(t) \rangle=\sum_{\alpha=1}^{d} c_{\alpha} |\Psi_{\alpha}(t)\rangle. \label{eq:floquet_state}
\end{gather}
The coefficients $ c_{\alpha} $ are determined by the initial state $|\Psi(0)\rangle=\sum_{\alpha=1 }^{d} c_{\alpha} |\Psi_{\alpha}(0)\rangle$. 
These basis solutions, known as the Floquet states, have the characteristic form
\begin{gather}
	|\Psi_{\alpha}(t)\rangle =e^{-\frac{i}{\hbar} \epsilon_{\alpha} t }|\Phi_{\alpha}(t)\rangle. \label{eq:floquet_mode}
\end{gather}
Here, $\epsilon_{\alpha}$ is the quasienergy associated with the Floquet state, and $|\Phi_{\alpha}(t)\rangle$ is the corresponding Floquet mode, which is periodic with the same period as the Hamiltonian, $|\Phi_{\alpha}(t)\rangle = |\Phi_{\alpha}(t+T)\rangle$. The index $\alpha$ labels the $d$ orthonormal Floquet modes which are complete and orthogonal
\begin{equation}
	\begin{split}
	\sum_\alpha |\Phi_{\alpha}(t)\rangle \langle \Phi_{\alpha}(t)| &= I,\\
\langle \Phi_{\alpha}(t)|\Phi_{\alpha^\prime}(t)\rangle &= \delta_{\alpha \alpha^\prime}.
	\end{split}
\end{equation}

By substituting Eq. \eqref{eq:floquet_mode} into the Schrödinger equation in Eq. \eqref{eq:Schrodinger}, we obtain an eigenvalue equation for the quasienergies:
	\begin{gather}
	\left(H(t)-i \hbar \frac{\partial}{\partial t} \right)|\Phi_{\alpha}(t)\rangle=\epsilon_{\alpha} |\Phi_{\alpha}(t)\rangle. \label{eq:eigenvalue_equation}
\end{gather}
The operator $H_F = H(t)-i \hbar \frac{\partial}{\partial t} $ is often referred to as the Floquet Hamiltonian. This Eq. \eqref{eq:eigenvalue_equation} forms the basis for finding the quasienergies and Floquet modes numerically or analytically \cite{Creffield2003, Lambert2026}.

\subsection{Time-domain approach: The propagator method}

From the view of the evolution operator (propagator), the evolution of Floquet states can be defined as
	\begin{gather}
	U(T+t,t) |\Psi(t)\rangle = |\Psi(T+t)\rangle.
\end{gather}
A natural method for finding the quasienergies and Floquet modes is to analyze the system's evolution operator over one full period, $U(T,0)$. This operator, or propagator, describes the system's dynamics from $t=0$ to $t=T$:
\begin{gather}
	U(T,0) = \mathcal{T} \exp \left(-\frac{i}{\hbar} \int_0^T H(\tau) d\tau \right),
\end{gather}
where $\mathcal{T}$ is the time-ordering operator \cite{Creffield2003}. Combining the relation $|\Psi_\alpha(T)\rangle = U(T,0)|\Psi_\alpha(0)\rangle$ with Eq. \eqref{eq:floquet_mode} yields an eigenvalue equation for the propagator:
	\begin{gather}
	U(T,0)|\Phi_\alpha(0)\rangle = e^{-\frac{i}{\hbar}\epsilon_\alpha T} |\Phi_\alpha(0)\rangle.
\end{gather}
This shows that the initial Floquet modes, $|\Phi_\alpha(0)\rangle$, are the eigenstates of the one-period evolution operator $U(T,0)$. By numerically diagonalizing $U(T,0)$, one can find its eigenvalues $ E_\alpha = \exp(-i\epsilon_\alpha T/\hbar)$
and solve for the quasienergies $ \epsilon_\alpha = i \hbar \log(E_\alpha)/T$. Once the initial modes $|\Phi_\alpha(0)\rangle$ are known, the modes at any other time can be found via evolution:
\begin{gather}
	|\Phi_\alpha(t)\rangle = e^{i\epsilon_\alpha t / \hbar} U(t,0)|\Phi_\alpha(0)\rangle.
\end{gather}

\subsection{Frequency-domain approach: The Sambe space formalism}

Since both the Hamiltonian $H(t)$ and the Floquet modes $|\Phi_\alpha(t)\rangle$ are periodic, they can be expanded as a Fourier series with the drive frequency $\omega = 2\pi/T$:
\begin{equation}
	\begin{split}
	H(t) &= \sum_{n=-\infty}^{\infty} H^n e^{-i n\omega t},\\
	|\Phi_\alpha(t)\rangle &= \sum_{n=-\infty}^{\infty} |\phi_\alpha^n\rangle e^{-i n\omega t}, \label{eq:Fourier_expand}
	\end{split}
\end{equation}
where $H^n$ and $ |\phi_\alpha^n\rangle $ are respective Fourier components. This transforms the finite-dimensional, time-dependent problem into an infinite-dimensional, time-independent matrix eigenvalue problem \cite{Shirley1965, Chu2004, Son2009}.

This infinite-dimensional space is known as the extended space or Sambe space \cite{Sambe1973}. A set of basis states in this space can be written as $|\alpha, n\rangle = |\alpha\rangle\otimes|n\rangle $, where $|\alpha \rangle $ is a basis state of the original system and $|n\rangle$ represents the Fourier (or ``photon") index. Substituting the Fourier series Eq. \eqref{eq:Fourier_expand} into Eq. \eqref{eq:eigenvalue_equation} yields its frequency-domain representation:
\begin{gather}
	\sum_{n^\prime} (H^{(n-n^\prime)} - n\hbar \omega\delta_{nn^\prime})  |\phi_\alpha^{n^\prime} \rangle= \epsilon_\alpha |\phi_\alpha^n\rangle. \label{eq:eigenvalue_equation_frequency}
\end{gather}
This can be expressed as a single eigenvalue equation $	H_F |\varphi_\alpha\rangle \rangle =  \epsilon_\alpha |\varphi_\alpha \rangle\rangle$, where $H_F$ is the time-independent Floquet Hamiltonian matrix with elements \cite{Son2009}:
\begin{gather}
	\langle\alpha n| H_{F}|\alpha^\prime n^\prime\rangle=H_{\alpha \alpha^\prime}^{n-n^\prime}-n \hbar \omega \delta_{\alpha \alpha^\prime} \delta_{n n^\prime},\label{eq:Floquqet_matrix2}
\end{gather}
and $|\varphi_\alpha \rangle\rangle$ is the corresponding eigenvector.
This infinite matrix is truncated by restricting the range of the Fourier index $n$, providing a powerful method to analyze the periodically driven quantum system 
\subsection{Properties}

\textbf{Periodicity.} The quasienergy spectrum is periodic. While a Floquet state is unique, its corresponding quasienergy is defined only up to integer multiples of $\hbar\omega$. Shifting the quasienergy by $m\hbar\omega$ is equivalent to relabeling the Fourier components, leaving the physical state unchanged:
	\begin{equation}
		\begin{split}
				|\Psi_{\alpha}(t)\rangle &=e^{-\frac{i}{\hbar} (\epsilon_{\alpha}+m\hbar \omega)t}  
				  \sum_{n=-\infty}^{\infty} |\phi_\alpha^{n+m}\rangle e^{-i n\omega t}.
		\end{split}
\end{equation}
Thus, the quasienergies are defined modulo $\hbar \omega$, i.e., $\epsilon_{\alpha,m} = \epsilon_{\alpha} +m\hbar \omega$ for $ m\in \mathbb{Z}$.
Analogous to Bloch's theorem in solid-state physics, this allows the quasienergy spectrum to be folded into a first Floquet-Brillouin zone, typically 
$[-\hbar \omega/2,\hbar \omega/2]$.

\textbf{Gauge choice.} The specific form of the Floquet Hamiltonian depends on the chosen gauge (or frame). Different gauges are related by a periodic unitary transformation, $U(t)$, such that $H_F^\prime = U^\dagger(t) H_F U(t)$. The physical quasienergy spectrum is gauge-invariant and the choice of gauge is typically a matter of practical convenience.

\section{Identification of dressed states}
\label{app:identification}

Our analysis focuses on the computational states of the system, which, in the presence of static interactions or dynamic drives, are more accurately described as dressed states. The Floquet states resulting from a periodic drive are a specific class of such dressed states \cite{Huang2021}. When we numerically solve for the eigenstates of a static or Floquet Hamiltonian, the solver typically returns the states sorted by their corresponding energy eigenvalues or quasienergies \cite{Lambert2026}. This numerical ordering often breaks the correspondence with the physical labels of the computational states when parameters are swept across regions with level crossings or anti-crossings. Therefore, a robust method is required to correctly identify and track the physical dressed states.

In the weakly interacting (dispersive) regime of the static Hamiltonian, or the weak-drive limit of the dynamic Hamiltonian, this identification is straightforward: the dressed states are well-approximated by the bare states of the uncoupled components and can be labeled by finding the bare state with the maximum overlap. However, this simple approach fails in the strongly hybridized regime, where multiple mode frequencies are close to resonance, or when strong drives create highly entangled Floquet states. To address this challenge, we employ a recursive tracking method inspired by the principles of adiabatic evolution \cite{Shillito2022, Dumas2024}.

For the static Hamiltonian, we begin the identification process in a known dispersive regime, where the parameters are set such that all components are far detuned from one another. In this limit, the label of each numerically calculated eigenstate is unambiguously determined by finding the bare state with which it has the maximum overlap. We then incrementally change a system parameter (e.g., a coupler frequency) in small steps. At each new step, we identify the new set of dressed states by assigning each one the label of the state from the previous step with which it has the maximum overlap. By repeating this recursive process, we can reliably track each dressed state and its corresponding eigenenergy from the simple dispersive regime into the complex, strongly hybridized regime.

We adopt this recursive method to identify the Floquet modes of the dynamic Hamiltonian. The process begins at zero drive amplitude ($\epsilon_p=0$), where the Floquet modes are identical to the static dressed states of the system. We first identify these static states using the method described above. These labels then serve as the starting point for the dynamic tracking. We incrementally increase the drive amplitude $\epsilon_p$ in small steps. At each step, we identify the new Floquet modes by finding the mode from the previous amplitude step that has the maximum overlap. This allows us to track the states as they become progressively more dressed by the drive. The same recursive procedure is also applied when sweeping other parameters, such as the drive frequency $\omega_p$.

This recursive tracking method is robust enough to resolve the identification of states even through complex regions of narrowly avoided crossings. By ensuring the correct state identification, we can accurately calculate all quantities of interest, such as the effective parametric coupling strengths and parasitic ZZ interactions, which is a prerequisite for the entire analysis presented in this article.	
	
	\section{Dressed energy levels of the qubit-coupler-qubit system}
	\label{app:dressed_levels}
	
	This appendix provides the analytical expressions for the dressed energy levels of the qubit-coupler-qubit system, calculated using perturbation theory up to the fourth order \cite{Krishnan1978, Zhao2021}. These perturbative results are used in the main text to inform the frequency allocation and to derive the static ZZ interaction. To avoid excessively lengthy formulas, the expressions presented here are derived from a system Hamiltonian where the counterrotating terms have been neglected (i.e., under the RWA). The full, unabridged expressions can be readily derived using a symbolic computation library.

The total eigenenergy for a target bare state $s = |Q_1CQ_2\rangle$ is expressed as a perturbative expansion:  $E_s = E_s^{(0)}+ E_s^{(2)}+  E_s^{(3)} +  E_s^{(4)}$, where $E_s ^{(0)}$ is the unperturbed energy and $E_s^{(n)}(n\neq 0)$ denotes the $n$th-order energy correction with
\begin{equation}
	\begin{split}
		E_s^{(2)} &= \sum_{j \neq s} \frac{|V_{sj}|^2}{E_{sj}}, \\
		E_s^{(3)} &= \sum_{j,k \neq s} \frac{V_{sj}V_{jk}V_{ks}}{E_{sj}E_{sk}}, \\
		E_s^{(4)} &=  \sum_{j,k,l \neq s} \frac{V_{sj}V_{jk}V_{kl}V_{ls}}{E_{sj}E_{sk}E_{sl}} -  \sum_{j,k \neq s} \frac{|V_{sj}|^2|V_{sk}|^2}{E_{sj}^2E_{sk}}.
	\end{split}
\label{eq:perturbative}
\end{equation}
Here, the subscripts $ s,j,k,l $ run over all bare states in the system. We denote the energy denominator as $E_{sj} = E_s^{(0)} - E_j^{(0)} $ and $V_{sj} = \langle s|\mathcal{V}_{qcq}| j \rangle $, where $\mathcal{V}_{qcq}$ is the coupling Hamiltonian from Eq. \eqref{eq:hamiltonian_coupler_lab}. The coupling Hamiltonian $\mathcal{V}_{qcq}$ has zero diagonal elements (e.g., $V_{ss}=0$). Thus the first-order energy correction $E_s^{(1)}$ vanishes.

 The perturbative results for the involved dressed states in the main text can be derived using the SymPy symbolic computation library \cite{Meurer2017}, which are given as:
\begin{equation}
	E_{|000\rangle}^{(0)} = E_{|000\rangle}^{(2)} = E_{|000\rangle}^{(3)} = E_{|000\rangle}^{(4)} = 0,
\end{equation}

\begin{equation}
	\begin{split}
		E_{|001\rangle}^{(0)} = &\omega_2,\\
		E_{|001\rangle}^{(2)} = &\frac{J_{2c}^{2}}{\Delta_{2c}} - \frac{J_{12}^{2}}{\Delta_{12}},\\
		E_{|001\rangle}^{(3)} = & - \frac{2 J_{12} J_{1c} J_{2c}}{\Delta_{12} \Delta_{2c}},\\
		E_{|001\rangle}^{(4)} =& J_{12}^{2} J_{2c}^{2} \left(\frac{1}{\Delta_{12} \Delta_{2c}^{2}} - \frac{1}{\Delta_{12}^{2} \Delta_{2c}}\right) - \frac{J_{2c}^{4}}{\Delta_{2c}^{3}} \\
		&- \frac{J_{1c}^{2} J_{2c}^{2}}{\Delta_{12} \Delta_{2c}^{2}} + \frac{J_{12}^{2} J_{1c}^{2}}{\Delta_{12}^{2} \Delta_{2c}} + \frac{J_{12}^{4}}{\Delta_{12}^{3}},\\
	\end{split}
\end{equation}

\begin{equation}
	\begin{split}
		E_{|010\rangle}^{(0)} =& \omega_c,\\
		E_{|010\rangle}^{(2)} = &- \frac{J_{2c}^{2}}{\Delta_{2c}} - \frac{J_{1c}^{2}}{\Delta_{1c}},\\
		E_{|010\rangle}^{(3)} = &\frac{2 J_{12} J_{1c} J_{2c}}{\Delta_{1c} \Delta_{2c}},\\
		E_{|010\rangle}^{(4)} =&J_{1c}^{2} J_{2c}^{2} \left(\frac{1}{\Delta_{1c} \Delta_{2c}^{2}} + \frac{1}{\Delta_{1c}^{2} \Delta_{2c}}\right) + \frac{J_{2c}^{4}}{\Delta_{2c}^{3}} \\
		&- \frac{J_{12}^{2} J_{2c}^{2}}{\Delta_{1c} \Delta_{2c}^{2}} - \frac{J_{12}^{2} J_{1c}^{2}}{\Delta_{1c}^{2} \Delta_{2c}} + \frac{J_{1c}^{4}}{\Delta_{1c}^{3}},\\
	\end{split}
\end{equation}

\begin{equation}
	\begin{split}
		E_{|100\rangle}^{(0)} =& \omega_1,\\
		E_{|100\rangle}^{(2)} = &\frac{J_{1c}^{2}}{\Delta_{1c}} + \frac{J_{12}^{2}}{\Delta_{12}},\\
		E_{|100\rangle}^{(3)} = &\frac{2 J_{12} J_{1c} J_{2c}}{\Delta_{12} \Delta_{1c}},\\
		E_{|100\rangle}^{(4)} =&- J_{12}^{2} J_{1c}^{2} \left(\frac{1}{\Delta_{12} \Delta_{1c}^{2}} + \frac{1}{\Delta_{12}^{2} \Delta_{1c}}\right) - \frac{J_{1c}^{4}}{\Delta_{1c}^{3}} \\
		&+ \frac{J_{1c}^{2} J_{2c}^{2}}{\Delta_{12} \Delta_{1c}^{2}} + \frac{J_{12}^{2} J_{2c}^{2}}{\Delta_{12}^{2} \Delta_{1c}} - \frac{J_{12}^{4}}{\Delta_{12}^{3}},\\
	\end{split}
\end{equation}

\begin{widetext}
	\begin{equation}
		\begin{split}
			E_{|002\rangle}^{(0)} = &2\omega_2+\alpha_2, 
			E_{|002\rangle}^{(2)} = \frac{2 J_{12}^{2}}{- \Delta_{12} + \alpha_{2}} + \frac{2 J_{2c}^{2}}{\Delta_{2c} + \alpha_{2}},
			E_{|002\rangle}^{(3)} = \frac{4 J_{12} J_{1c} J_{2c}}{\left(- \Delta_{12} + \alpha_{2}\right) \left(\Delta_{2c} + \alpha_{2}\right)},\\
			E_{|002\rangle}^{(4)} =&\frac{4 J_{12}^{4} \left(\Delta_{12} + \alpha_{1}\right)}{\left(\Delta_{12} - \alpha_{2}\right)^{3} \left(2 \Delta_{12} + \alpha_{1} - \alpha_{2}\right)} + \frac{2 J_{12}^{2} J_{1c}^{2}}{\left(\Delta_{12} - \alpha_{2}\right)^{2} \left(\Delta_{2c} + \alpha_{2}\right)} - \frac{2 J_{1c}^{2} J_{2c}^{2}}{\left(\Delta_{12} - \alpha_{2}\right) \left(\Delta_{2c} + \alpha_{2}\right)^{2}}\\
			&- \frac{4 J_{2c}^{4} \left(\Delta_{2c} - \alpha_{c}\right)}{\left(\Delta_{2c} + \alpha_{2}\right)^{3} \left(2 \Delta_{2c} + \alpha_{2} - \alpha_{c}\right)}
		 	- \frac{2 J_{12}^{2} J_{2c}^{2} \left(\Delta_{12} - \Delta_{2c} - 2 \alpha_{2}\right) \left(\Delta_{12} - \Delta_{2c} - 2 \Sigma_{1c} + 4 \omega_{2}\right)}{\left(\Delta_{12} - \alpha_{2}\right)^{2} \left(\Delta_{2c} + \alpha_{2}\right)^{2} \left(\Sigma_{1c} - \alpha_{2} - 2 \omega_{2}\right)},\\
		\end{split}
	\end{equation}
\end{widetext}

\begin{widetext}
	\begin{equation}
		\begin{split}
			E_{|200\rangle}^{(0)} = & 2 \omega_{1} + \alpha_{1},
			E_{|200\rangle}^{(2)} = \frac{2 J_{12}^{2}}{\Delta_{12} + \alpha_{1}} + \frac{2 J_{1c}^{2}}{\Delta_{1c} + \alpha_{1}},
			E_{|200\rangle}^{(3)} = \frac{4 J_{12} J_{1c} J_{2c}}{\left(\Delta_{12} + \alpha_{1}\right) \left(\Delta_{1c} + \alpha_{1}\right)},\\
			E_{|200\rangle}^{(4)}  =&\frac{4 J_{12}^{4} \left(- \Delta_{12} + \alpha_{2}\right)}{\left(\Delta_{12} + \alpha_{1}\right)^{3} \left(2 \Delta_{12} + \alpha_{1} - \alpha_{2}\right)} + \frac{2 J_{12}^{2} J_{1c}^{2} \left(\Delta_{12} + \Delta_{1c} + 2 \alpha_{1}\right) \left(\Delta_{12} + \Delta_{1c} + 2 \Sigma_{2c} - 4 \omega_{1}\right)}{\left(\Delta_{12} + \alpha_{1}\right)^{2} \left(\Delta_{1c} + \alpha_{1}\right)^{2} \left(- \Sigma_{2c} + \alpha_{1} + 2 \omega_{1}\right)} \\
			&+ \frac{2 J_{12}^{2} J_{2c}^{2}}{\left(\Delta_{12} + \alpha_{1}\right)^{2} \left(\Delta_{1c} + \alpha_{1}\right)} + \frac{4 J_{1c}^{4} \left(- \Delta_{1c} + \alpha_{c}\right)}{\left(\Delta_{1c} + \alpha_{1}\right)^{3} \left(2 \Delta_{1c} + \alpha_{1} - \alpha_{c}\right)} + \frac{2 J_{1c}^{2} J_{2c}^{2}}{\left(\Delta_{12} + \alpha_{1}\right) \left(\Delta_{1c} + \alpha_{1}\right)^{2}},\\
		\end{split}
	\end{equation}
\end{widetext}

\begin{widetext}
	\begin{equation}
		\begin{split}
			E_{|011\rangle}^{(0)} =&   \Sigma_{2c},\\
			E_{|011\rangle}^{(2)} =&   2 J_{2c}^{2} \left(\frac{1}{\Delta_{2c} - \alpha_{c}} + \frac{1}{- \Delta_{2c} - \alpha_{2}}\right) - \frac{J_{1c}^{2}}{\Delta_{1c}} - \frac{J_{12}^{2}}{\Delta_{12}},\\
			E_{|011\rangle}^{(3)} = &2 J_{12} J_{1c} J_{2c} \left(\frac{2}{\Delta_{1c} \left(\Delta_{2c} + \alpha_{2}\right)} + \frac{2}{\Delta_{12} \left(- \Delta_{2c} + \alpha_{c}\right)} + \frac{1}{\Delta_{12} \Delta_{1c}}\right),\\
			E_{|011\rangle}^{(4)} =&J_{12}^{2} J_{1c}^{2} \left(- \frac{2}{\Delta_{1c}^{2} \left(- \Sigma_{2c} + \alpha_{1} + 2 \omega_{1}\right)} - \frac{2}{\Delta_{1c}^{2} \left(\Delta_{2c} + \alpha_{2}\right)} - \frac{4}{\Delta_{12} \Delta_{1c} \left(- \Sigma_{2c} + \alpha_{1} + 2 \omega_{1}\right)} + \frac{1}{\Delta_{12} \Delta_{1c}^{2}} \right.\\
			&\left.- \frac{2}{\Delta_{12}^{2} \left(- \Sigma_{2c} + \alpha_{1} + 2 \omega_{1}\right)} - \frac{2}{\Delta_{12}^{2} \left(- \Delta_{2c} + \alpha_{c}\right)} + \frac{1}{\Delta_{12}^{2} \Delta_{1c}}\right) + J_{12}^{2} J_{2c}^{2} \left(- \frac{4}{\Delta_{1c} \left(\Delta_{2c} + \alpha_{2}\right)^{2}} \right.\\
			&+ \frac{2}{\Delta_{12} \left(\Delta_{2c} + \alpha_{2}\right)^{2}} + \frac{2}{\Delta_{12} \left(- \Delta_{2c} + \alpha_{c}\right)^{2}} - \frac{4}{\Delta_{12} \Delta_{1c} \left(\Delta_{2c} + \alpha_{2}\right)} + \frac{2}{\Delta_{12}^{2} \left(\Delta_{2c} + \alpha_{2}\right)} + \frac{2}{\Delta_{12}^{2} \left(- \Delta_{2c} + \alpha_{c}\right)}  \\
			&\left.- \frac{1}{\Delta_{12}^{2} \Delta_{1c}}\right)+ J_{1c}^{2} J_{2c}^{2} \left(\frac{2}{\Delta_{1c} \left(\Delta_{2c} + \alpha_{2}\right)^{2}} + \frac{2}{\Delta_{1c} \left(- \Delta_{2c} + \alpha_{c}\right)^{2}} + \frac{2}{\Delta_{1c}^{2} \left(\Delta_{2c} + \alpha_{2}\right)} + \frac{2}{\Delta_{1c}^{2} \left(- \Delta_{2c} + \alpha_{c}\right)}  \right. \\
			&\left. - \frac{4}{\Delta_{12} \left(- \Delta_{2c} + \alpha_{c}\right)^{2}} -\frac{4}{\Delta_{12} \Delta_{1c} \left(- \Delta_{2c} + \alpha_{c}\right)} - \frac{1}{\Delta_{12} \Delta_{1c}^{2}}\right) + 4 J_{2c}^{4} \left(\frac{1}{\left(\Delta_{2c} + \alpha_{2}\right)^{3}} + \frac{1}{\left(- \Delta_{2c} + \alpha_{c}\right) \left(\Delta_{2c} + \alpha_{2}\right)^{2}} \right. \\
			&\left.  + \frac{1}{\left(- \Delta_{2c} + \alpha_{c}\right)^{2} \left(\Delta_{2c} + \alpha_{2}\right)} + \frac{1}{\left(- \Delta_{2c} + \alpha_{c}\right)^{3}}\right) + \frac{J_{1c}^{4}}{\Delta_{1c}^{3}} + \frac{J_{12}^{4}}{\Delta_{12}^{3}},\\
		\end{split}
	\end{equation}
\end{widetext}

\begin{widetext}
	\begin{equation}
		\begin{split}
			E_{|101\rangle}^{(0)} =&  \Sigma_{12},\\
			E_{|101\rangle}^{(2)} =& 2 J_{12}^{2} \left(\frac{1}{\Delta_{12} - \alpha_{2}} - \frac{1}{\Delta_{12} + \alpha_{1}}\right) + \frac{J_{2c}^{2}}{\Delta_{2c}} + \frac{J_{1c}^{2}}{\Delta_{1c}},\\
			E_{|101\rangle}^{(3)} = &2 J_{12} J_{1c} J_{2c} \left(- \frac{2}{\Delta_{2c} \left(\Delta_{12} + \alpha_{1}\right)} + \frac{2}{\Delta_{1c} \left(\Delta_{12} - \alpha_{2}\right)} + \frac{1}{\Delta_{1c} \Delta_{2c}}\right),\\
			E_{|101\rangle}^{(4)} =&4 J_{12}^{4} \left(- \frac{1}{\left(\Delta_{12} - \alpha_{2}\right)^{3}} + \frac{1}{\left(\Delta_{12} + \alpha_{1}\right) \left(\Delta_{12} - \alpha_{2}\right)^{2}} - \frac{1}{\left(\Delta_{12} + \alpha_{1}\right)^{2} \left(\Delta_{12} - \alpha_{2}\right)} + \frac{1}{\left(\Delta_{12} + \alpha_{1}\right)^{3}}\right) \\
			&+ J_{12}^{2} J_{1c}^{2} \left(\frac{4}{\Delta_{2c} \left(\Delta_{12} + \alpha_{1}\right)^{2}} - \frac{2}{\Delta_{1c} \left(\Delta_{12} - \alpha_{2}\right)^{2}} - \frac{2}{\Delta_{1c} \left(\Delta_{12} + \alpha_{1}\right)^{2}} - \frac{4}{\Delta_{1c} \Delta_{2c} \left(\Delta_{12} + \alpha_{1}\right)} - \frac{2}{\Delta_{1c}^{2} \left(\Delta_{12} - \alpha_{2}\right)} \right.\\
			&\left.+ \frac{2}{\Delta_{1c}^{2} \left(\Delta_{12} + \alpha_{1}\right)} + \frac{1}{\Delta_{1c}^{2} \Delta_{2c}}\right) + J_{12}^{2} J_{2c}^{2} \left(- \frac{2}{\Delta_{2c} \left(\Delta_{12} - \alpha_{2}\right)^{2}} - \frac{2}{\Delta_{2c} \left(\Delta_{12} + \alpha_{1}\right)^{2}} - \frac{2}{\Delta_{2c}^{2} \left(\Delta_{12} - \alpha_{2}\right)} \right.\\ 
			&\left.+ \frac{2}{\Delta_{2c}^{2} \left(\Delta_{12} + \alpha_{1}\right)} + \frac{4}{\Delta_{1c} \left(\Delta_{12} - \alpha_{2}\right)^{2}} + \frac{4}{\Delta_{1c} \Delta_{2c} \left(\Delta_{12} - \alpha_{2}\right)} + \frac{1}{\Delta_{1c} \Delta_{2c}^{2}}\right) + J_{1c}^{2} J_{2c}^{2} \left(\frac{2}{\Delta_{2c}^{2} \left(\Delta_{1c} + \Delta_{2c} - \alpha_{c}\right)} \right.\\ 
			&- \frac{2}{\Delta_{2c}^{2} \left(\Delta_{12} + \alpha_{1}\right)} + \frac{4}{\Delta_{1c} \Delta_{2c} \left(\Delta_{1c} + \Delta_{2c} - \alpha_{c}\right)} - \frac{1}{\Delta_{1c} \Delta_{2c}^{2}} + \frac{2}{\Delta_{1c}^{2} \left(\Delta_{1c} + \Delta_{2c} - \alpha_{c}\right)} \\
			&\left.+ \frac{2}{\Delta_{1c}^{2} \left(\Delta_{12} - \alpha_{2}\right)} - \frac{1}{\Delta_{1c}^{2} \Delta_{2c}}\right) - \frac{J_{2c}^{4}}{\Delta_{2c}^{3}} - \frac{J_{1c}^{4}}{\Delta_{1c}^{3}},\\
		\end{split}
	\end{equation}
\end{widetext}

\begin{widetext}
	\begin{equation}
		\begin{split}
			E_{|110\rangle}^{(0)} = & \Sigma_{1c},\\
			E_{|110\rangle}^{(2)} =&  2 J_{1c}^{2} \left(\frac{1}{\Delta_{1c} - \alpha_{c}} - \frac{1}{\Delta_{1c} + \alpha_{1}}\right) - \frac{J_{2c}^{2}}{\Delta_{2c}} + \frac{J_{12}^{2}}{\Delta_{12}},\\
			E_{|110\rangle}^{(3)} =& 2 J_{12} J_{1c} J_{2c} \left(\frac{2}{\Delta_{2c} \left(\Delta_{1c} + \alpha_{1}\right)} + \frac{2}{\Delta_{12} \left(\Delta_{1c} - \alpha_{c}\right)} - \frac{1}{\Delta_{12} \Delta_{2c}}\right),\\
			E_{|110\rangle}^{(4)} =&J_{12}^{2} J_{1c}^{2} \left(- \frac{4}{\Delta_{2c} \left(\Delta_{1c} + \alpha_{1}\right)^{2}} - \frac{2}{\Delta_{12} \left(\Delta_{1c} - \alpha_{c}\right)^{2}} - \frac{2}{\Delta_{12} \left(\Delta_{1c} + \alpha_{1}\right)^{2}} + \frac{4}{\Delta_{12} \Delta_{2c} \left(\Delta_{1c} + \alpha_{1}\right)} - \frac{2}{\Delta_{12}^{2} \left(\Delta_{1c} - \alpha_{c}\right)} \right.\\ 
			&\left.+ \frac{2}{\Delta_{12}^{2} \left(\Delta_{1c} + \alpha_{1}\right)} - \frac{1}{\Delta_{12}^{2} \Delta_{2c}}\right) + J_{12}^{2} J_{2c}^{2} \left(\frac{2}{\Delta_{2c}^{2} \left(\Sigma_{1c} - \alpha_{2} - 2 \omega_{2}\right)} - \frac{2}{\Delta_{2c}^{2} \left(\Delta_{1c} + \alpha_{1}\right)} + \frac{4}{\Delta_{12} \Delta_{2c} \left(- \Sigma_{1c} + \alpha_{2} + 2 \omega_{2}\right)} \right.\\ 
			&\left.- \frac{1}{\Delta_{12} \Delta_{2c}^{2}} + \frac{2}{\Delta_{12}^{2} \left(\Sigma_{1c} - \alpha_{2} - 2 \omega_{2}\right)} + \frac{2}{\Delta_{12}^{2} \left(\Delta_{1c} - \alpha_{c}\right)} + \frac{1}{\Delta_{12}^{2} \Delta_{2c}}\right) + 4 J_{1c}^{4} \left(- \frac{1}{\left(\Delta_{1c} - \alpha_{c}\right)^{3}}+ \frac{1}{\left(\Delta_{1c} + \alpha_{1}\right)^{3}} \right.\\
			&\left.+ \frac{1}{\left(\Delta_{1c} + \alpha_{1}\right) \left(\Delta_{1c} - \alpha_{c}\right)^{2}} - \frac{1}{\left(\Delta_{1c} + \alpha_{1}\right)^{2} \left(\Delta_{1c} - \alpha_{c}\right)} \right) + J_{1c}^{2} J_{2c}^{2} \left(\frac{2}{\Delta_{2c} \left(\Delta_{1c} - \alpha_{c}\right)^{2}} + \frac{2}{\Delta_{2c} \left(\Delta_{1c} + \alpha_{1}\right)^{2}} \right.\\
			&\left. - \frac{2}{\Delta_{2c}^{2} \left(\Delta_{1c} - \alpha_{c}\right)} + \frac{2}{\Delta_{2c}^{2} \left(\Delta_{1c} + \alpha_{1}\right)} + \frac{4}{\Delta_{12} \left(\Delta_{1c} - \alpha_{c}\right)^{2}} - \frac{4}{\Delta_{12} \Delta_{2c} \left(\Delta_{1c} - \alpha_{c}\right)} + \frac{1}{\Delta_{12} \Delta_{2c}^{2}}\right) 
			+ \frac{J_{2c}^{4}}{\Delta_{2c}^{3}} - \frac{J_{12}^{4}}{\Delta_{12}^{3}}.\\
		\end{split}
	\end{equation}
\end{widetext}

The above expressions for the dressed energies can be adopted to refine the optimized solution in Algorithm \ref{alg:optimization} and derive an analytical expression for the static ZZ coupling, which is defined as $\zeta = E_{|101\rangle} + E_{|000\rangle} - E_{|001\rangle} - E_{|100\rangle}$. Summing the contributions up to the fourth order gives $\zeta = \zeta^{(2)} + \zeta^{(3)} + \zeta^{(4)}$, yielding after RWA:
\begin{equation}
	\begin{split}
	 \zeta^{(2)}   =& 2 J_{12}^{2} \left(\frac{1}{\Delta_{12} - \alpha_{2}} - \frac{1}{\Delta_{12} + \alpha_{1}}\right), \\ 
	 \zeta^{(3)}   =& 2 J_{12} J_{1c} J_{2c} \left(- \frac{2}{\Delta_{2c} \left(\Delta_{12} + \alpha_{1}\right)} + \frac{1}{\Delta_{1c} \Delta_{2c}}\right.\\ 
	 &\left. + \frac{2}{\Delta_{1c} \left(\Delta_{12} - \alpha_{2}\right)}  + \frac{1}{\Delta_{12} \Delta_{2c}} - \frac{1}{\Delta_{12} \Delta_{1c}}\right),\\
	 \zeta^{(4)}  \approx& J_{1c}^{2} J_{2c}^{2} \left(\frac{2}{\Delta_{2c}^{2} \left(\Delta_{1c} + \Delta_{2c} - \alpha_{c}\right)}- \frac{1}{\Delta_{1c} \Delta_{2c}^{2}} \right.\\ 
	 &\left. - \frac{2}{\Delta_{2c}^{2} \left(\Delta_{12} + \alpha_{1}\right)} + \frac{4}{\Delta_{1c} \Delta_{2c} \left(\Delta_{1c} + \Delta_{2c} - \alpha_{c}\right)} \right.\\ 
	 &\left. + \frac{2}{\Delta_{1c}^{2} \left(\Delta_{1c} + \Delta_{2c} - \alpha_{c}\right)} + \frac{2}{\Delta_{1c}^{2} \left(\Delta_{12} - \alpha_{2}\right)} \right.\\ 
	 &\left.- \frac{1}{\Delta_{1c}^{2} \Delta_{2c}} + \frac{1}{\Delta_{12} \Delta_{2c}^{2}} - \frac{1}{\Delta_{12} \Delta_{1c}^{2}}\right).
	\end{split}
\label{eq:qcq_zz}
\end{equation}

For the expression of $\zeta^{(4)}$ shown in Eq. \eqref{eq:qcq_zz}, we omit contributing terms including $\mathcal{O}(J_{12}^{4})$, $\mathcal{O}(J_{12}^{2} J_{1c}^{2})$, and $\mathcal{O}(J_{12}^{2} J_{2c}^{2})  $ as these are negligible in the typical regime where the direct qubit-qubit coupling is much weaker than the qubit-coupler couplings ($J_{1c,2c} \gg J_{12}$) in our system. It is significant to note that for the plot in Fig. \ref{fig:qcqzz}, we used the full symbolic expression without the RWA to ensure an accurate comparison with numerical simulations and experimental results \cite{Sung2021}.

\section{Satisfiability Modulo Theory}
\label{app:smt}

SMT is a decision problem that determines whether a complex mathematical formula is satisfiable. It generalizes the well-known Boolean satisfiability (SAT) problem by enriching Boolean logic with theories from first-order logic, such as the theories of real numbers, integers, and various data structures \cite{Biere2021}. An SMT solver, therefore, can handle problems with intricate logical and arithmetic rules far beyond the scope of a simple SAT solver. A powerful extension of SMT is OMT, which moves beyond a simple ``satisfiable" or ``unsatisfiable" answer. An OMT solver searches for a solution that not only satisfies all given constraints but also optimizes (minimizes or maximizes) a specified objective function. Crucially, OMT solvers are designed to find a globally optimal solution, in contrast to many local optimization methods.

The task of frequency allocation in a multi-qubit processor is a natural fit for the SMT/OMT framework. The design criteria, such as the numerous constraints detailed in Sec. \ref{sec:constraints}, can be directly encoded as a set of logical and arithmetic formulas. The SMT solver then determines if a valid set of system parameters (frequencies, anharmonicities, etc.) exists that simultaneously satisfies all of these conditions. This approach offers several key advantages. First, its expressiveness natively supports the nonlinear constraints that arise from the physics of superconducting circuits. Second, the OMT extension provides the crucial capability of finding a globally optimal set of parameters, for instance, one that maximizes the detuning from the most dangerous parasitic transition. Finally, modern solvers, such as the Z3 solver \cite{Moura2008}, ensure efficiency by employing powerful techniques like logical deduction and constraint propagation to efficiently prune vast, unsolvable regions of the parameter space.

However, the primary challenge of using SMT/OMT is its computational complexity, which typically scales exponentially with the number of variables (e.g., qubits). To make the problem tractable for larger systems, careful model simplification is essential before mapping the physics onto the solver's constraints. To accelerate the search for a solution, several strategies can be employed. First, it is often beneficial to reformulate the constraints to avoid nonlinearities where possible, such as multiplication and division between variables. Second, for systems with local interactions, one can exploit the problem's structure by first solving for a smaller unit cell of qubits and then extending the solution to a larger lattice.

	\bibliography{collision_parametric_floquet_cleaned}

@Article{Mueller2019,
  author    = {Müller, Clemens and Cole, Jared H and Lisenfeld, Jürgen},
  journal   = {Rep. Prog. Phys.},
  title     = {Towards understanding two-level-systems in amorphous solids: insights from quantum circuits},
  year      = {2019},
  issn      = {1361-6633},
  month     = oct,
  number    = {12},
  pages     = {124501},
  volume    = {82},
  doi       = {10.1088/1361-6633/ab3a7e},
  fjournal  = {Reports on Progress in Physics},
  publisher = {IOP Publishing},
}

@Article{Gandon2022,
  author    = {Gandon, Anthony and Le Calonnec, Camille and Shillito, Ross and Petrescu, Alexandru and Blais, Alexandre},
  journal   = {Phys. Rev. Appl.},
  title     = {Engineering, Control, and Longitudinal Readout of Floquet Qubits},
  year      = {2022},
  issn      = {2331-7019},
  month     = jun,
  number    = {6},
  pages     = {064006},
  volume    = {17},
  doi       = {10.1103/physrevapplied.17.064006},
  fjournal  = {Physical Review Applied},
  publisher = {American Physical Society (APS)},
}

@Article{Wang2024b,
  author    = {Wang, Bi-Ying and Liu, Wuxin and Chen, Xiangyu and Xu, Shu and Cui, Jiangyu and Yung, Man-Hong},
  journal   = {Sci. China Phys. Mech. Astron.},
  title     = {Optimizing frequency allocation for superconducting quantum processors with frequency-tunable qubits},
  year      = {2024},
  issn      = {1869-1927},
  month     = dec,
  number    = {2},
  pages     = {220312},
  volume    = {68},
  doi       = {10.1007/s11433-024-2527-3},
  publisher = {Springer Science and Business Media LLC},
}

@Article{Kyriienko2018,
  author    = {Kyriienko, Oleksandr and Sørensen, Anders S.},
  journal   = {Phys. Rev. Appl.},
  title     = {Floquet Quantum Simulation with Superconducting Qubits},
  year      = {2018},
  issn      = {2331-7019},
  month     = jun,
  number    = {6},
  pages     = {064029},
  volume    = {9},
  doi       = {10.1103/physrevapplied.9.064029},
  fjournal  = {Physical Review Applied},
  publisher = {American Physical Society (APS)},
}

@Book{Biere2021,
  author    = {Armin Biere and Marijn J. H. Heule and Hans Maaren and Toby Walsh},
  publisher = {IOS Press},
  title     = {Handbook of satisfiability},
  year      = {2021},
  address   = {Amsterdam},
  edition   = {Second edition},
  isbn      = {9781643681610},
  number    = {volume 336},
  series    = {Frontiers in artificial intelligence and applications},
  pagetotal = {11465},
  ppn_gvk   = {1775115682},
  subtitle  = {Part 1/part 2},
}

@Article{Sete2021a,
  author    = {Sete, Eyob A. and Didier, Nicolas and Chen, Angela Q. and Kulshreshtha, Shobhan and Manenti, Riccardo and Poletto, Stefano},
  journal   = {Phys. Rev. Appl.},
  title     = {Parametric-Resonance Entangling Gates with a Tunable Coupler},
  year      = {2021},
  issn      = {2331-7019},
  month     = aug,
  number    = {2},
  pages     = {024050},
  volume    = {16},
  doi       = {10.1103/physrevapplied.16.024050},
  fjournal  = {Physical Review Applied},
  publisher = {American Physical Society (APS)},
}

@Article{Son2009,
  author    = {Son, Sang-Kil and Han, Siyuan and Chu, Shih-I},
  journal   = {Phys. Rev. A},
  title     = {Floquet formulation for the investigation of multiphoton quantum interference in a superconducting qubit driven by a strong ac field},
  year      = {2009},
  issn      = {1094-1622},
  month     = mar,
  number    = {3},
  pages     = {032301},
  volume    = {79},
  doi       = {10.1103/physreva.79.032301},
  fjournal  = {Physical Review A},
  publisher = {American Physical Society (APS)},
}

@Article{Chow2011,
  author    = {Chow, Jerry M. and Córcoles, A. D. and Gambetta, Jay M. and Rigetti, Chad and Johnson, B. R. and Smolin, John A. and Rozen, J. R. and Keefe, George A. and Rothwell, Mary B. and Ketchen, Mark B. and Steffen, M.},
  journal   = {Phys. Rev. Lett.},
  title     = {Simple All-Microwave Entangling Gate for Fixed-Frequency Superconducting Qubits},
  year      = {2011},
  issn      = {1079-7114},
  month     = aug,
  number    = {8},
  pages     = {080502},
  volume    = {107},
  doi       = {10.1103/physrevlett.107.080502},
  fjournal  = {Physical Review Letters},
  publisher = {American Physical Society (APS)},
}

@Article{Hertzberg2021,
  author    = {Hertzberg, Jared B. and Zhang, Eric J. and Rosenblatt, Sami and Magesan, Easwar and Smolin, John A. and Yau, Jeng-Bang and Adiga, Vivekananda P. and Sandberg, Martin and Brink, Markus and Chow, Jerry M. and Orcutt, Jason S.},
  journal   = {npj Quantum Inf.},
  title     = {Laser-annealing Josephson junctions for yielding scaled-up superconducting quantum processors},
  year      = {2021},
  issn      = {2056-6387},
  month     = aug,
  volume    = {7},
  doi       = {10.1038/s41534-021-00464-5},
  eid       = {129},
  fjournal  = {npj Quantum Information},
  publisher = {Springer Science and Business Media LLC},
}

@Article{Xiao2025,
  author    = {Xiao, Xu and Venkatraman, Jayameenakshi and Corti\~nas, Rodrigo G. and Chowdhury, Shoumik and Devoret, Michel H.},
  journal   = {Phys. Rev. Appl.},
  title     = {Diagrammatic method to compute the effective Hamiltonian of a driven nonlinear oscillator},
  year      = {2025},
  month     = {Oct},
  pages     = {044021},
  volume    = {24},
  doi       = {10.1103/d6j2-lsf7},
  fjournal  = {Physical Review Applied},
  issue     = {4},
  numpages  = {72},
  publisher = {American Physical Society},
  url       = {https://link.aps.org/doi/10.1103/d6j2-lsf7},
}

@Article{Lu2025,
  author    = {Lu, Bin-Han and Li, Qing-Song and Wang, Peng and Chen, Zhao-Yun and Wu, Yu-Chun and Guo, Guo-Ping},
  journal   = {Chin. Phys. Lett.},
  title     = {Neural Network-Based Frequency Optimization for Superconducting Quantum Chips},
  year      = {2025},
  issn      = {1741-3540},
  month     = mar,
  number    = {3},
  pages     = {030204},
  volume    = {42},
  doi       = {10.1088/0256-307x/42/3/030204},
  fjournal  = {Chinese Physics Letters},
  publisher = {IOP Publishing},
}

@Article{Chu2004,
  author    = {Chu, Shih-I and Telnov, Dmitry A.},
  journal   = {Phys. Rep.},
  title     = {Beyond the Floquet theorem: generalized Floquet formalisms and quasienergy methods for atomic and molecular multiphoton processes in intense laser fields},
  year      = {2004},
  issn      = {0370-1573},
  month     = feb,
  number    = {1–2},
  pages     = {1--131},
  volume    = {390},
  doi       = {10.1016/j.physrep.2003.10.001},
  fjournal  = {Physics Reports},
  publisher = {Elsevier BV},
}

@Article{Abrams2020,
  author    = {Abrams, Deanna M. and Didier, Nicolas and Johnson, Blake R. and Silva, Marcus P. da and Ryan, Colm A.},
  journal   = {Nat. Electron.},
  title     = {Implementation of XY entangling gates with a single calibrated pulse},
  year      = {2020},
  issn      = {2520-1131},
  month     = nov,
  number    = {12},
  pages     = {744--750},
  volume    = {3},
  doi       = {10.1038/s41928-020-00498-1},
  fjournal  = {Nature Electronics},
  publisher = {Springer Science and Business Media LLC},
}

@Article{Zheng2022,
  author    = {Zheng, Wen and Xu, Jianwen and Ma, Zhuang and Li, Yong and Dong, Yuqian and Zhang, Yu and Wang, Xiaohan and Sun, Guozhu and Wu, Peiheng and Zhao, Jie and Li, Shaoxiong and Lan, Dong and Tan, Xinsheng and Yu, Yang},
  journal   = {Chin. Phys. Lett.},
  title     = {Measuring Quantum Geometric Tensor of Non-Abelian System in Superconducting Circuits},
  year      = {2022},
  issn      = {1741-3540},
  month     = sep,
  number    = {10},
  pages     = {100202},
  volume    = {39},
  doi       = {10.1088/0256-307x/39/10/100202},
  fjournal  = {Chinese Physics Letters},
  publisher = {IOP Publishing},
}

@Article{Grifoni1998,
  author    = {Grifoni, Milena and Hänggi, Peter},
  journal   = {Phys. Rep.},
  title     = {Driven quantum tunneling},
  year      = {1998},
  issn      = {0370-1573},
  month     = oct,
  number    = {5–6},
  pages     = {229--354},
  volume    = {304},
  doi       = {10.1016/s0370-1573(98)00022-2},
  fjournal  = {Physics Reports},
  publisher = {Elsevier BV},
}

@Article{Shillito2022,
  author    = {Shillito, Ross and Petrescu, Alexandru and Cohen, Joachim and Beall, Jackson and Hauru, Markus and Ganahl, Martin and Lewis, Adam G.M. and Vidal, Guifre and Blais, Alexandre},
  journal   = {Phys. Rev. Appl.},
  title     = {Dynamics of Transmon Ionization},
  year      = {2022},
  issn      = {2331-7019},
  month     = sep,
  number    = {3},
  pages     = {034031},
  volume    = {18},
  doi       = {10.1103/physrevapplied.18.034031},
  fjournal  = {Physical Review Applied},
  publisher = {American Physical Society (APS)},
}

@Misc{VallesSanclemente2025,
  author        = {S. Vallés-Sanclemente and T. H. F. Vroomans and T. R. van Abswoude and F. Brulleman and T. Stavenga and S. L. M. van der Meer and Y. Xin and A. Lawrence and V. Singh and M. A. Rol and L. DiCarlo},
  title         = {Optimizing the frequency positioning of tunable couplers in a circuit QED processor to mitigate spectator effects on quantum operations},
  year          = {2025},
  archiveprefix = {arXiv},
  eprint        = {2503.13225},
  primaryclass  = {quant-ph},
  url           = {https://arxiv.org/abs/2503.13225},
}

@Article{Dumas2024,
  author    = {Dumas, Marie Frédérique and Groleau-Paré, Benjamin and McDonald, Alexander and Muñoz-Arias, Manuel H. and Lledó, Cristóbal and D’Anjou, Benjamin and Blais, Alexandre},
  journal   = {Phys. Rev. X},
  title     = {Measurement-Induced Transmon Ionization},
  year      = {2024},
  issn      = {2160-3308},
  month     = oct,
  number    = {4},
  pages     = {041023},
  volume    = {14},
  doi       = {10.1103/physrevx.14.041023},
  fjournal  = {Physical Review X},
  publisher = {American Physical Society (APS)},
}

@InProceedings{Smith2022,
  author    = {Smith, Kaitlin N. and Ravi, Gokul Subramanian and Baker, Jonathan M. and Chong, Frederic T.},
  booktitle = {2022 55th IEEE/ACM International Symposium on Microarchitecture (MICRO)},
  title     = {Scaling Superconducting Quantum Computers with Chiplet Architectures},
  year      = {2022},
  pages     = {1092-1109},
  doi       = {10.1109/MICRO56248.2022.00078},
}

@Misc{Baskov2025,
  author        = {Roman Baskov and Daniel K. Weiss and Steven M. Girvin},
  title         = {Exact amplitudes of parametric processes in driven Josephson circuits},
  year          = {2025},
  archiveprefix = {arXiv},
  eprint        = {2501.07784},
  primaryclass  = {quant-ph},
  url           = {https://arxiv.org/abs/2501.07784},
}

@Article{Rosen2024,
  author    = {Rosen, Ilan T. and Muschinske, Sarah and Barrett, Cora N. and Chatterjee, Arkya and Hays, Max and DeMarco, Michael A. and Karamlou, Amir H. and Rower, David A. and Das, Rabindra and Kim, David K. and Niedzielski, Bethany M. and Schuldt, Meghan and Serniak, Kyle and Schwartz, Mollie E. and Yoder, Jonilyn L. and Grover, Jeffrey A. and Oliver, William D.},
  journal   = {Nat. Phys.},
  title     = {A synthetic magnetic vector potential in a 2D superconducting qubit array},
  year      = {2024},
  issn      = {1745-2481},
  month     = oct,
  number    = {12},
  pages     = {1881--1887},
  volume    = {20},
  doi       = {10.1038/s41567-024-02661-3},
  fjournal  = {Nature Physics},
  publisher = {Springer Science and Business Media LLC},
}

@Article{Hour2024,
  author    = {Hour, Leanghok and Heng, Sengthai and Heng, Sovanmonynuth and Go, Myeongseong and Han, Youngsun},
  journal   = {Quantum Sci. Technol.},
  title     = {Context-aware coupler reconfiguration for tunable coupler-based superconducting quantum computers},
  year      = {2024},
  issn      = {2058-9565},
  month     = oct,
  number    = {1},
  pages     = {015016},
  volume    = {10},
  doi       = {10.1088/2058-9565/ad8510},
  fjournal  = {Quantum Science and Technology},
  publisher = {IOP Publishing},
}

@Misc{Xu2025,
  author        = {Xuexin Xu and Kuljeet Kaur and Chloé Vignes and Mohammad H. Ansari and John M. Martinis},
  title         = {Surface-Code Hardware Hamiltonian},
  year          = {2025},
  archiveprefix = {arXiv},
  eprint        = {2507.06201},
  primaryclass  = {quant-ph},
  url           = {https://arxiv.org/abs/2507.06201},
}

@Article{Sambe1973,
  author    = {Sambe, Hideo},
  journal   = {Phys. Rev. A},
  title     = {Steady States and Quasienergies of a Quantum-Mechanical System in an Oscillating Field},
  year      = {1973},
  issn      = {0556-2791},
  month     = jun,
  number    = {6},
  pages     = {2203--2213},
  volume    = {7},
  doi       = {10.1103/physreva.7.2203},
  fjournal  = {Physical Review A},
  publisher = {American Physical Society (APS)},
}

@Misc{Krauss2025,
  author        = {Matthias G. Krauss and Christiane P. Koch},
  title         = {The perfect entangler spectrum as a tool to analyze crosstalk},
  year          = {2025},
  archiveprefix = {arXiv},
  eprint        = {2506.03137},
  primaryclass  = {quant-ph},
  url           = {https://arxiv.org/abs/2506.03137},
}

@Article{Osman2023,
  author    = {Osman, Amr and Fernández-Pendás, Jorge and Warren, Christopher and Kosen, Sandoko and Scigliuzzo, Marco and Frisk Kockum, Anton and Tancredi, Giovanna and Fadavi Roudsari, Anita and Bylander, Jonas},
  journal   = {Phys. Rev. Res.},
  title     = {Mitigation of frequency collisions in superconducting quantum processors},
  year      = {2023},
  issn      = {2643-1564},
  month     = oct,
  number    = {4},
  pages     = {043001},
  volume    = {5},
  doi       = {10.1103/physrevresearch.5.043001},
  fjournal  = {Physical Review Research},
  publisher = {American Physical Society (APS)},
}

@Article{Xu2023,
  author    = {Xu, Shibo and Sun, Zheng-Zhi and Wang, Ke and Xiang, Liang and Bao, Zehang and Zhu, Zitian and Shen, Fanhao and Song, Zixuan and Zhang, Pengfei and Ren, Wenhui and Zhang, Xu and Dong, Hang and Deng, Jinfeng and Chen, Jiachen and Wu, Yaozu and Tan, Ziqi and Gao, Yu and Jin, Feitong and Zhu, Xuhao and Zhang, Chuanyu and Wang, Ning and Zou, Yiren and Zhong, Jiarun and Zhang, Aosai and Li, Weikang and Jiang, Wenjie and Yu, Li-Wei and Yao, Yunyan and Wang, Zhen and Li, Hekang and Guo, Qiujiang and Song, Chao and Wang, H. and Deng, Dong-Ling},
  journal   = {Chin. Phys. Lett.},
  title     = {Digital Simulation of Projective Non-Abelian Anyons with 68 Superconducting Qubits},
  year      = {2023},
  issn      = {1741-3540},
  month     = jun,
  number    = {6},
  pages     = {060301},
  volume    = {40},
  doi       = {10.1088/0256-307x/40/6/060301},
  fjournal  = {Chinese Physics Letters},
  publisher = {IOP Publishing},
}

@Article{Xu2024,
  author    = {Xu, Xuexin and Manabputra and Vignes, Chloé and Ansari, Mohammad H. and Martinis, John M.},
  journal   = {Phys. Rev. Appl.},
  title     = {Lattice Hamiltonians and stray interactions within quantum processors},
  year      = {2024},
  issn      = {2331-7019},
  month     = dec,
  number    = {6},
  pages     = {064030},
  volume    = {22},
  doi       = {10.1103/physrevapplied.22.064030},
  fjournal  = {Physical Review Applied},
  publisher = {American Physical Society (APS)},
}

@Article{Bloch1940,
  author    = {Bloch, F. and Siegert, A.},
  journal   = {Phys. Rev.},
  title     = {Magnetic Resonance for Nonrotating Fields},
  year      = {1940},
  issn      = {0031-899X},
  month     = mar,
  number    = {6},
  pages     = {522--527},
  volume    = {57},
  doi       = {10.1103/physrev.57.522},
  fjournal  = {Physical Review},
  publisher = {American Physical Society (APS)},
}

@Article{Huber2025,
  author    = {Huber, G.B.P. and Roy, F.A. and Koch, L. and Tsitsilin, I. and Schirk, J. and Glaser, N.J. and Bruckmoser, N. and Schweizer, C. and Romeiro, J. and Krylov, G. and Singh, M. and Haslbeck, F.X. and Knudsen, M. and Marx, A. and Pfeiffer, F. and Schneider, C. and Wallner, F. and Bunch, D. and Richard, L. and Södergren, L. and Liegener, K. and Werninghaus, M. and Filipp, S.},
  journal   = {PRX Quantum},
  title     = {Parametric {Multielement} {Coupling} {Architecture} for {Coherent} and {Dissipative} {Control} of {Superconducting} {Qubits}},
  year      = {2025},
  month     = jul,
  number    = {3},
  pages     = {030313},
  volume    = {6},
  doi       = {10.1103/9shv-l4cx},
  publisher = {American Physical Society},
  url       = {https://link.aps.org/doi/10.1103/9shv-l4cx},
  urldate   = {2025-09-06},
}

@Article{Roth2017,
  author    = {Roth, Marco and Ganzhorn, Marc and Moll, Nikolaj and Filipp, Stefan and Salis, Gian and Schmidt, Sebastian},
  journal   = {Phys. Rev. A},
  title     = {Analysis of a parametrically driven exchange-type gate and a two-photon excitation gate between superconducting qubits},
  year      = {2017},
  issn      = {2469-9934},
  month     = dec,
  number    = {6},
  pages     = {062323},
  volume    = {96},
  doi       = {10.1103/physreva.96.062323},
  fjournal  = {Physical Review A},
  publisher = {American Physical Society (APS)},
}

@Article{Zhang2025a,
  author    = {Zhang, Zewen and Gokhale, Pranav and Larson, Jeffrey M.},
  journal   = {Phys. Rev. A},
  title     = {Efficient frequency allocation for superconducting quantum processors using improved optimization techniques},
  year      = {2025},
  month     = {Jan},
  pages     = {012619},
  volume    = {111},
  doi       = {10.1103/PhysRevA.111.012619},
  fjournal  = {Physical Review A},
  issue     = {1},
  numpages  = {9},
  publisher = {American Physical Society},
  url       = {https://link.aps.org/doi/10.1103/PhysRevA.111.012619},
}

@Article{Zhang2025b,
  author    = {Zhang, Yu and Xu, Jianwen and Ma, Zhuang and Zheng, Wen and Lan, Dong and Tan, Xinsheng and Yu, Yang},
  journal   = {Commun. Phys.},
  title     = {Experimental simulation of Dirac equation in superconducting qubits},
  year      = {2025},
  issn      = {2399-3650},
  month     = jun,
  volume    = {8},
  doi       = {10.1038/s42005-025-02112-2},
  eid       = {248},
  publisher = {Springer Science and Business Media LLC},
}

@InBook{Moura2008,
  author    = {de Moura, Leonardo and Bjørner, Nikolaj},
  pages     = {337--340},
  publisher = {Springer Berlin Heidelberg},
  title     = {Z3: An Efficient SMT Solver},
  year      = {2008},
  isbn      = {9783540788003},
  booktitle = {Tools and Algorithms for the Construction and Analysis of Systems},
  doi       = {10.1007/978-3-540-78800-3_24},
  issn      = {1611-3349},
}

@Article{Ganzhorn2020,
  author    = {Ganzhorn, M. and Salis, G. and Egger, D. J. and Fuhrer, A. and Mergenthaler, M. and Müller, C. and Müller, P. and Paredes, S. and Pechal, M. and Werninghaus, M. and Filipp, S.},
  journal   = {Phys. Rev. Res.},
  title     = {Benchmarking the noise sensitivity of different parametric two-qubit gates in a single superconducting quantum computing platform},
  year      = {2020},
  issn      = {2643-1564},
  month     = sep,
  number    = {3},
  pages     = {033447},
  volume    = {2},
  doi       = {10.1103/physrevresearch.2.033447},
  fjournal  = {Physical Review Research},
  publisher = {American Physical Society (APS)},
}

@Article{Oka2019,
  author    = {Oka, Takashi and Kitamura, Sota},
  journal   = {Annu. Rev. Condens. Matter Phys.},
  title     = {Floquet Engineering of Quantum Materials},
  year      = {2019},
  issn      = {1947-5462},
  month     = mar,
  number    = {1},
  pages     = {387--408},
  volume    = {10},
  doi       = {10.1146/annurev-conmatphys-031218-013423},
  fjournal  = {Annual Review of Condensed Matter Physics},
  publisher = {Annual Reviews},
}

@Article{Weinberg2017,
  author    = {Weinberg, Phillip and Bukov, Marin and D’Alessio, Luca and Polkovnikov, Anatoli and Vajna, Szabolcs and Kolodrubetz, Michael},
  journal   = {Phys. Rep.},
  title     = {Adiabatic perturbation theory and geometry of periodically-driven systems},
  year      = {2017},
  issn      = {0370-1573},
  month     = may,
  pages     = {1--35},
  volume    = {688},
  doi       = {10.1016/j.physrep.2017.05.003},
  fjournal  = {Physics Reports},
  publisher = {Elsevier BV},
}

@Article{Floquet1883,
  author    = {Floquet, G.},
  journal   = {Annales scientifiques de l’École normale supérieure},
  title     = {Sur les équations différentielles linéaires à coefficients périodiques},
  year      = {1883},
  issn      = {1873-2151},
  pages     = {47--88},
  volume    = {12},
  doi       = {10.24033/asens.220},
  publisher = {Societe Mathematique de France},
}

@Article{Ma2023,
  author    = {Ma, Zhuang and Xu, Jianwen and Chen, Tao and Zhang, Yu and Zheng, Wen and Li, Shaoxiong and Lan, Dong and Xue, Zheng-Yuan and Tan, Xinsheng and Yu, Yang},
  journal   = {Phys. Rev. Appl.},
  title     = {Noncyclic nonadiabatic geometric quantum gates in a superconducting circuit},
  year      = {2023},
  issn      = {2331-7019},
  month     = nov,
  number    = {5},
  pages     = {054047},
  volume    = {20},
  doi       = {10.1103/physrevapplied.20.054047},
  fjournal  = {Physical Review Applied},
  publisher = {American Physical Society (APS)},
}

@Article{Morvan2022,
  author    = {Morvan, Alexis and Chen, Larry and Larson, Jeffrey M. and Santiago, David I. and Siddiqi, Irfan},
  journal   = {Phys. Rev. Res.},
  title     = {Optimizing frequency allocation for fixed-frequency superconducting quantum processors},
  year      = {2022},
  issn      = {2643-1564},
  month     = apr,
  number    = {2},
  pages     = {023079},
  volume    = {4},
  doi       = {10.1103/physrevresearch.4.023079},
  fjournal  = {Physical Review Research},
  publisher = {American Physical Society (APS)},
}

@Misc{Ma2025,
  author        = {Zhuang Ma and Xianke Li and Hongyi Shi and Ruonan Guo and Jianwen Xu and Xinsheng Tan and Yang Yu},
  title         = {Parametric Phase Modulation in Superconducting Circuits},
  year          = {2025},
  archiveprefix = {arXiv},
  eprint        = {2510.20192},
  primaryclass  = {quant-ph},
  url           = {https://arxiv.org/abs/2510.20192},
}

@Article{Ma2024,
  author    = {Ma, Xizheng and Zhang, Gengyan and Wu, Feng and Bao, Feng and Chang, Xu and Chen, Jianjun and Deng, Hao and Gao, Ran and Gao, Xun and Hu, Lijuan and Ji, Honghong and Ku, Hsiang-Sheng and Lu, Kannan and Ma, Lu and Mao, Liyong and Song, Zhijun and Sun, Hantao and Tang, Chengchun and Wang, Fei and Wang, Hongcheng and Wang, Tenghui and Xia, Tian and Ying, Make and Zhan, Huijuan and Zhou, Tao and Zhu, Mengyu and Zhu, Qingbin and Shi, Yaoyun and Zhao, Hui-Hai and Deng, Chunqing},
  journal   = {Phys. Rev. Lett.},
  title     = {Native Approach to Controlled- Z Gates in Inductively Coupled Fluxonium Qubits},
  year      = {2024},
  issn      = {1079-7114},
  month     = feb,
  number    = {6},
  pages     = {060602},
  volume    = {132},
  doi       = {10.1103/physrevlett.132.060602},
  fjournal  = {Physical Review Letters},
  publisher = {American Physical Society (APS)},
}

@Article{Shirley1965,
  author    = {Shirley, Jon H.},
  journal   = {Phys. Rev.},
  title     = {Solution of the Schrödinger Equation with a Hamiltonian Periodic in Time},
  year      = {1965},
  issn      = {0031-899X},
  month     = may,
  number    = {4B},
  pages     = {B979--B987},
  volume    = {138},
  doi       = {10.1103/physrev.138.b979},
  fjournal  = {Physical Review},
  publisher = {American Physical Society (APS)},
}

@Article{Zhang2024a,
  author    = {Zhang, Yu and Zhu, Yan-Qing and Xu, Jianwen and Zheng, Wen and Lan, Dong and Palumbo, Giandomenico and Goldman, Nathan and Zhu, Shi-Liang and Tan, Xinsheng and Wang, Z.D. and Yu, Yang},
  journal   = {Phys. Rev. Appl.},
  title     = {Exploring parity magnetic effects through quantum simulation with superconducting qubits},
  year      = {2024},
  issn      = {2331-7019},
  month     = mar,
  number    = {3},
  pages     = {034052},
  volume    = {21},
  doi       = {10.1103/physrevapplied.21.034052},
  fjournal  = {Physical Review Applied},
  publisher = {American Physical Society (APS)},
}

@Article{Lagemann2022,
  author    = {Lagemann, H. and Willsch, D. and Willsch, M. and Jin, F. and De Raedt, H. and Michielsen, K.},
  journal   = {Phys. Rev. A},
  title     = {Numerical analysis of effective models for flux-tunable transmon systems},
  year      = {2022},
  issn      = {2469-9934},
  month     = aug,
  number    = {2},
  pages     = {022615},
  volume    = {106},
  doi       = {10.1103/physreva.106.022615},
  fjournal  = {Physical Review A},
  publisher = {American Physical Society (APS)},
}

@Article{Krinner2020a,
  author    = {Krinner, S. and Lazar, S. and Remm, A. and Andersen, C.K. and Lacroix, N. and Norris, G.J. and Hellings, C. and Gabureac, M. and Eichler, C. and Wallraff, A.},
  journal   = {Phys. Rev. Appl.},
  title     = {Benchmarking Coherent Errors in Controlled-Phase Gates due to Spectator Qubits},
  year      = {2020},
  issn      = {2331-7019},
  month     = aug,
  number    = {2},
  pages     = {024042},
  volume    = {14},
  doi       = {10.1103/physrevapplied.14.024042},
  fjournal  = {Physical Review Applied},
  publisher = {American Physical Society (APS)},
}

@Article{Heya2024,
  author    = {Heya, Kentaro and Malekakhlagh, Moein and Merkel, Seth and Kanazawa, Naoki and Pritchett, Emily},
  journal   = {Phys. Rev. Appl.},
  title     = {Floquet analysis of frequency collisions},
  year      = {2024},
  issn      = {2331-7019},
  month     = feb,
  number    = {2},
  pages     = {024035},
  volume    = {21},
  doi       = {10.1103/physrevapplied.21.024035},
  fjournal  = {Physical Review Applied},
  publisher = {American Physical Society (APS)},
}

@Article{Creffield2003,
  author    = {Creffield, C.E.},
  journal   = {Phys. Rev. B},
  title     = {Location of crossings in the Floquet spectrum of a driven two-level system},
  year      = {2003},
  issn      = {1095-3795},
  month     = apr,
  number    = {16},
  pages     = {165301},
  volume    = {67},
  doi       = {10.1103/physrevb.67.165301},
  fjournal  = {Physical Review B},
  publisher = {American Physical Society (APS)},
}

@Article{Andersen2025,
  author    = {Andersen, T. I. and Astrakhantsev, N. and Karamlou, A. H. et al.},
  journal   = {Nature},
  title     = {Thermalization and criticality on an analogue–digital quantum simulator},
  year      = {2025},
  issn      = {1476-4687},
  month     = feb,
  number    = {8049},
  pages     = {79--85},
  volume    = {638},
  doi       = {10.1038/s41586-024-08460-3},
  publisher = {Springer Science and Business Media LLC},
}

@Article{You2025,
  author    = {You, Xinyuan and Li, Andy C.Y. and Roy, Tanay and Zhu, Shaojiang and Romanenko, Alexander and Grassellino, Anna and Lu, Yao and Chakram, Srivatsan},
  journal   = {Phys. Rev. Appl.},
  title     = {Floquet-engineered fast snap gates in weakly coupled circuit-QED systems},
  year      = {2025},
  month     = {Sep},
  pages     = {034072},
  volume    = {24},
  doi       = {10.1103/smcc-t465},
  fjournal  = {Physical Review Applied},
  issue     = {3},
  numpages  = {20},
  publisher = {American Physical Society},
  url       = {https://link.aps.org/doi/10.1103/smcc-t465},
}

@Article{Ai2025,
  author    = {Ai, Hao and Liu, Yu-xi},
  journal   = {Phys. Rev. Lett.},
  title     = {Scalable Parameter Design for Superconducting Quantum Circuits with Graph Neural Networks},
  year      = {2025},
  month     = {Jul},
  pages     = {040601},
  volume    = {135},
  doi       = {10.1103/yr9d-7z8k},
  fjournal  = {Physical Review Letters},
  issue     = {4},
  numpages  = {7},
  publisher = {American Physical Society},
  url       = {https://link.aps.org/doi/10.1103/yr9d-7z8k},
}

@Article{Nguyen2024,
  author    = {Nguyen, Long B. and Kim, Yosep and Hashim, Akel and Goss, Noah and Marinelli, Brian and Bhandari, Bibek and Das, Debmalya and Naik, Ravi K. and Kreikebaum, John Mark and Jordan, Andrew N. and Santiago, David I. and Siddiqi, Irfan},
  journal   = {Nat. Phys.},
  title     = {Programmable Heisenberg interactions between Floquet qubits},
  year      = {2024},
  issn      = {1745-2481},
  month     = jan,
  number    = {2},
  pages     = {240--246},
  volume    = {20},
  doi       = {10.1038/s41567-023-02326-7},
  fjournal  = {Nature Physics},
  publisher = {Springer Science and Business Media LLC},
}

@Article{Krinner2020,
  author    = {Krinner, S. and Kurpiers, P. and Royer, B. and Magnard, P. and Tsitsilin, I. and Besse, J.-C. and Remm, A. and Blais, A. and Wallraff, A.},
  journal   = {Phys. Rev. Appl.},
  title     = {Demonstration of an All-Microwave Controlled-Phase Gate between Far-Detuned Qubits},
  year      = {2020},
  issn      = {2331-7019},
  month     = oct,
  number    = {4},
  pages     = {044039},
  volume    = {14},
  doi       = {10.1103/physrevapplied.14.044039},
  fjournal  = {Physical Review Applied},
  publisher = {American Physical Society (APS)},
}

@Article{Sung2021,
  author    = {Sung, Youngkyu and Ding, Leon and Braumüller, Jochen and Vepsäläinen, Antti and Kannan, Bharath and Kjaergaard, Morten and Greene, Ami and Samach, Gabriel O. and McNally, Chris and Kim, David and Melville, Alexander and Niedzielski, Bethany M. and Schwartz, Mollie E. and Yoder, Jonilyn L. and Orlando, Terry P. and Gustavsson, Simon and Oliver, William D.},
  journal   = {Phys. Rev. X},
  title     = {Realization of High-Fidelity CZ and ZZ -Free iSWAP Gates with a Tunable Coupler},
  year      = {2021},
  issn      = {2160-3308},
  month     = jun,
  number    = {2},
  pages     = {021058},
  volume    = {11},
  doi       = {10.1103/physrevx.11.021058},
  fjournal  = {Physical Review X},
  publisher = {American Physical Society (APS)},
}

@Article{Sete2024,
  author    = {Sete, Eyob A. and Tripathi, Vinay and Valery, Joseph A. and Lidar, Daniel and Mutus, Josh Y.},
  journal   = {Phys. Rev. Appl.},
  title     = {Error budget of a parametric resonance entangling gate with a tunable coupler},
  year      = {2024},
  issn      = {2331-7019},
  month     = jul,
  number    = {1},
  pages     = {014059},
  volume    = {22},
  doi       = {10.1103/physrevapplied.22.014059},
  fjournal  = {Physical Review Applied},
  publisher = {American Physical Society (APS)},
}

@Article{Malekakhlagh2020,
  author    = {Malekakhlagh, Moein and Magesan, Easwar and McKay, David C.},
  journal   = {Phys. Rev. A},
  title     = {First-principles analysis of cross-resonance gate operation},
  year      = {2020},
  issn      = {2469-9934},
  month     = oct,
  number    = {4},
  pages     = {042605},
  volume    = {102},
  doi       = {10.1103/physreva.102.042605},
  fjournal  = {Physical Review A},
  publisher = {American Physical Society (APS)},
}

@Article{Sete2021,
  author    = {Sete, Eyob A. and Chen, Angela Q. and Manenti, Riccardo and Kulshreshtha, Shobhan and Poletto, Stefano},
  journal   = {Phys. Rev. Appl.},
  title     = {Floating Tunable Coupler for Scalable Quantum Computing Architectures},
  year      = {2021},
  issn      = {2331-7019},
  month     = jun,
  number    = {6},
  pages     = {064063},
  volume    = {15},
  doi       = {10.1103/physrevapplied.15.064063},
  fjournal  = {Physical Review Applied},
  publisher = {American Physical Society (APS)},
}

@Article{Zhao2023,
  author    = {Zhao, Peng},
  journal   = {Phys. Rev. Appl.},
  title     = {Mitigation of quantum crosstalk in cross-resonance-based qubit architectures},
  year      = {2023},
  issn      = {2331-7019},
  month     = nov,
  number    = {5},
  pages     = {054033},
  volume    = {20},
  doi       = {10.1103/physrevapplied.20.054033},
  fjournal  = {Physical Review Applied},
  publisher = {American Physical Society (APS)},
}

@Article{Rasmussen2021,
  author    = {Rasmussen, S.E. and Christensen, K.S. and Pedersen, S.P. and Kristensen, L.B. and Bækkegaard, T. and Loft, N.J.S. and Zinner, N.T.},
  journal   = {PRX Quantum},
  title     = {Superconducting Circuit Companion—an Introduction with Worked Examples},
  year      = {2021},
  issn      = {2691-3399},
  month     = dec,
  number    = {4},
  pages     = {040204},
  volume    = {2},
  doi       = {10.1103/prxquantum.2.040204},
  publisher = {American Physical Society (APS)},
}

@Misc{Li2025,
  author        = {Tian-Ming Li and Zheng-Hang Sun and Yun-Hao Shi and Zhen-Ting Bao and Yong-Yi Wang and Jia-Chi Zhang and Yu Liu and Cheng-Lin Deng and Yi-Han Yu and Zheng-He Liu and Chi-Tong Chen and Li Li and Hao Li and Hao-Tian Liu and Si-Yun Zhou and Zhen-Yu Peng and Yan-Jun Liu and Ziting Wang and Yue-Shan Xu and Kui Zhao and Yang He and Da'er Feng and Jia-Cheng Song and Cai-Ping Fang and Junrui Deng and Mingyu Xu and Yu-Tao Chen and Bozhen zhou and Gui-Han Liang and Zhong-Cheng Xiang and Guangming Xue and Dongning Zheng and Kaixuan Huang and Zheng-An Wang and Haifeng Yu and Piotr Sierant and Kai Xu and Heng Fan},
  title         = {Many-body delocalization with a two-dimensional 70-qubit superconducting quantum simulator},
  year          = {2025},
  archiveprefix = {arXiv},
  eprint        = {2507.16882},
  primaryclass  = {quant-ph},
  url           = {https://arxiv.org/abs/2507.16882},
}

@Article{Zhao2021,
  author    = {Zhao, Peng and Lan, Dong and Xu, Peng and Xue, Guangming and Blank, Mace and Tan, Xinsheng and Yu, Haifeng and Yu, Yang},
  journal   = {Phys. Rev. Appl.},
  title     = {Suppression of Static ZZ Interaction in an All-Transmon Quantum Processor},
  year      = {2021},
  issn      = {2331-7019},
  month     = aug,
  number    = {2},
  pages     = {024037},
  volume    = {16},
  doi       = {10.1103/physrevapplied.16.024037},
  fjournal  = {Physical Review Applied},
  publisher = {American Physical Society (APS)},
}

@Article{Li2022,
  author    = {Li, Shaowei and Fan, Daojin and Gong, Ming and Ye, Yangsen and Chen, Xiawei and Wu, Yulin and Guan, Huijie and Deng, Hui and Rong, Hao and Huang, He-Liang and Zha, Chen and Yan, Kai and Guo, Shaojun and Qian, Haoran and Zhang, Haibin and Chen, Fusheng and Zhu, Qingling and Zhao, Youwei and Wang, Shiyu and Ying, Chong and Cao, Sirui and Yu, Jiale and Liang, Futian and Xu, Yu and Lin, Jin and Guo, Cheng and Sun, Lihua and Li, Na and Han, Lianchen and Peng, Cheng-Zhi and Zhu, Xiaobo and Pan, Jian-Wei},
  journal   = {Chin. Phys. Lett.},
  title     = {Realization of Fast All-Microwave Controlled-Z Gates with a Tunable Coupler},
  year      = {2022},
  issn      = {1741-3540},
  month     = feb,
  number    = {3},
  pages     = {030302},
  volume    = {39},
  doi       = {10.1088/0256-307x/39/3/030302},
  fjournal  = {Chinese Physics Letters},
  publisher = {IOP Publishing},
}

@Article{Li2021,
  author   = {Li, Danyu and Zheng, Wen and Chu, Ji and Yang, Xiaopei and Song, Shuqing and Han, Zhikun and Dong, Yuqian and Wang, Zhimin and Yu, Xiangmin and Lan, Dong and Zhao, Jie and Li, Shaoxiong and Tan, Xinsheng and Yu, Yang},
  journal  = {Appl. Phys. Lett.},
  title    = {Coherent state transfer between superconducting qubits via stimulated Raman adiabatic passage},
  year     = {2021},
  issn     = {0003-6951},
  month    = {03},
  number   = {10},
  pages    = {104003},
  volume   = {118},
  doi      = {10.1063/5.0040079},
  fjournal = {Applied Physics Letters},
  url      = {https://doi.org/10.1063/5.0040079},
}

@Misc{Zhao2025,
  author        = {Peng Zhao and Peng Xu and Zheng-Yuan Xue},
  title         = {Fast entangling gates on fluxoniums via parametric modulation of plasmon interaction},
  year          = {2025},
  archiveprefix = {arXiv},
  eprint        = {2509.04762},
  primaryclass  = {quant-ph},
  url           = {https://arxiv.org/abs/2509.04762},
}

@Article{Meurer2017,
  author    = {Meurer, Aaron and Smith, Christopher P. and Paprocki, Mateusz and Certik, Ond{\v{r}}ej and Kirpichev, Sergey B. and Rocklin, Matthew and Kumar, AMiT and Ivanov, Sergiu and Moore, Jason K. and Singh, Sartaj and Rathnayake, Thilina and Vig, Sean and Granger, Brian E. and Muller, Richard P. and Bonazzi, Francesco and Gupta, Harsh and Vats, Shivam and Johansson, Fredrik and Pedregosa, Fabian and Curry, Matthew J. and Terrel, Andy R. and Roucka, {\v{S}}t{\v{e}}p{\'a}n and Saboo, Ashutosh and Fernando, Isuru and Kulal, Sumith and Cimrman, Robert and Scopatz, Anthony},
  journal   = {PeerJ Comput. Sci.},
  title     = {SymPy: Symbolic computing in Python},
  year      = {2017},
  issn      = {2376-5992},
  month     = {January},
  pages     = {e103},
  volume    = {3},
  doi       = {10.7717/peerj-cs.103},
  fjournal  = {PeerJ Computer Science},
  publisher = {PeerJ},
}

@Article{You2019,
  author    = {You, Xinyuan and Sauls, J. A. and Koch, Jens},
  journal   = {Phys. Rev. B},
  title     = {Circuit quantization in the presence of time-dependent external flux},
  year      = {2019},
  issn      = {2469-9969},
  month     = may,
  number    = {17},
  pages     = {174512},
  volume    = {99},
  doi       = {10.1103/physrevb.99.174512},
  fjournal  = {Physical Review B},
  publisher = {American Physical Society (APS)},
}

@Misc{BrisenoColunga2025,
  author        = {D. Dominic Briseño-Colunga and Bibek Bhandari and Debmalya Das and Long B. Nguyen and Yosep Kim and David I. Santiago and Irfan Siddiqi and Andrew N. Jordan and Justin Dressel},
  title         = {Dynamical Sweet and Sour Regions in Bichromatically Driven Floquet Qubits},
  year          = {2025},
  archiveprefix = {arXiv},
  eprint        = {2505.22606},
  primaryclass  = {quant-ph},
  url           = {https://arxiv.org/abs/2505.22606},
}

@Article{Yan2018,
  author    = {Yan, Fei and Krantz, Philip and Sung, Youngkyu and Kjaergaard, Morten and Campbell, Daniel L. and Orlando, Terry P. and Gustavsson, Simon and Oliver, William D.},
  journal   = {Phys. Rev. Appl.},
  title     = {Tunable Coupling Scheme for Implementing High-Fidelity Two-Qubit Gates},
  year      = {2018},
  issn      = {2331-7019},
  month     = nov,
  number    = {5},
  pages     = {054062},
  volume    = {10},
  doi       = {10.1103/physrevapplied.10.054062},
  fjournal  = {Physical Review Applied},
  publisher = {American Physical Society (APS)},
}

@Article{Hastings2021,
  author    = {Hastings, Matthew B. and Haah, Jeongwan},
  journal   = {Quantum},
  title     = {Dynamically Generated Logical Qubits},
  year      = {2021},
  issn      = {2521-327X},
  month     = oct,
  pages     = {564},
  volume    = {5},
  doi       = {10.22331/q-2021-10-19-564},
  publisher = {Verein zur Forderung des Open Access Publizierens in den Quantenwissenschaften},
}

@Article{Riwar2022,
  author    = {Riwar, R.-P. and DiVincenzo, D. P.},
  journal   = {npj Quantum Inf.},
  title     = {Circuit quantization with time-dependent magnetic fields for realistic geometries},
  year      = {2022},
  issn      = {2056-6387},
  month     = mar,
  volume    = {8},
  doi       = {10.1038/s41534-022-00539-x},
  eid       = {36},
  fjournal  = {npj Quantum Information},
  publisher = {Springer Science and Business Media LLC},
}

@Article{Koch2007,
  author    = {Koch, Jens and Yu, Terri M. and Gambetta, Jay and Houck, A. A. and Schuster, D. I. and Majer, J. and Blais, Alexandre and Devoret, M. H. and Girvin, S. M. and Schoelkopf, R. J.},
  journal   = {Phys. Rev. A},
  title     = {Charge-insensitive qubit design derived from the Cooper pair box},
  year      = {2007},
  issn      = {1094-1622},
  month     = oct,
  number    = {4},
  pages     = {042319},
  volume    = {76},
  doi       = {10.1103/physreva.76.042319},
  fjournal  = {Physical Review A},
  publisher = {American Physical Society (APS)},
}

@Article{Sameti2019,
  author    = {Sameti, Mahdi and Hartmann, Michael J.},
  journal   = {Phys. Rev. A},
  title     = {Floquet engineering in superconducting circuits: From arbitrary spin-spin interactions to the Kitaev honeycomb model},
  year      = {2019},
  issn      = {2469-9934},
  month     = jan,
  number    = {1},
  pages     = {012333},
  volume    = {99},
  doi       = {10.1103/physreva.99.012333},
  fjournal  = {Physical Review A},
  publisher = {American Physical Society (APS)},
}

@Misc{Kelly2018,
  author        = {Julian Kelly and Peter O'Malley and Matthew Neeley and Hartmut Neven and John M. Martinis},
  title         = {Physical qubit calibration on a directed acyclic graph},
  year          = {2018},
  archiveprefix = {arXiv},
  eprint        = {1803.03226},
  primaryclass  = {quant-ph},
  url           = {https://arxiv.org/abs/1803.03226},
}

@Article{Paraoanu2006,
  author    = {Paraoanu, G. S.},
  journal   = {Phys. Rev. B},
  title     = {Microwave-induced coupling of superconducting qubits},
  year      = {2006},
  issn      = {1550-235X},
  month     = oct,
  number    = {14},
  pages     = {140504},
  volume    = {74},
  doi       = {10.1103/physrevb.74.140504},
  fjournal  = {Physical Review B},
  publisher = {American Physical Society (APS)},
}

@Article{Galiautdinov2012,
  author    = {Galiautdinov, Andrei and Korotkov, Alexander N. and Martinis, John M.},
  journal   = {Phys. Rev. A},
  title     = {Resonator–zero-qubit architecture for superconducting qubits},
  year      = {2012},
  issn      = {1094-1622},
  month     = apr,
  number    = {4},
  pages     = {042321},
  volume    = {85},
  doi       = {10.1103/physreva.85.042321},
  fjournal  = {Physical Review A},
  publisher = {American Physical Society (APS)},
}

@Article{Petrescu2023,
  author    = {Petrescu, Alexandru and Le Calonnec, Camille and Leroux, Catherine and Di Paolo, Agustin and Mundada, Pranav and Sussman, Sara and Vrajitoarea, Andrei and Houck, Andrew A. and Blais, Alexandre},
  journal   = {Phys. Rev. Appl.},
  title     = {Accurate Methods for the Analysis of Strong-Drive Effects in Parametric Gates},
  year      = {2023},
  issn      = {2331-7019},
  month     = apr,
  number    = {4},
  pages     = {044003},
  volume    = {19},
  doi       = {10.1103/physrevapplied.19.044003},
  fjournal  = {Physical Review Applied},
  publisher = {American Physical Society (APS)},
}

@Article{Klimov2024,
  author    = {Klimov, Paul V. and Bengtsson, Andreas and Quintana, Chris and Bourassa, Alexandre and Hong, Sabrina and Dunsworth, Andrew and Satzinger, Kevin J. and Livingston, William P. and Sivak, Volodymyr and Niu, Murphy Yuezhen and Andersen, Trond I. and Zhang, Yaxing and Chik, Desmond and Chen, Zijun and Neill, Charles and Erickson, Catherine and Grajales Dau, Alejandro and Megrant, Anthony and Roushan, Pedram and Korotkov, Alexander N. and Kelly, Julian and Smelyanskiy, Vadim and Chen, Yu and Neven, Hartmut},
  journal   = {Nat. Commun.},
  title     = {Optimizing quantum gates towards the scale of logical qubits},
  year      = {2024},
  issn      = {2041-1723},
  month     = mar,
  volume    = {15},
  doi       = {10.1038/s41467-024-46623-y},
  eid       = {2442},
  fjournal  = {Nature Communications},
  publisher = {Springer Science and Business Media LLC},
}

@Article{Barends2019,
  author    = {Barends, R. and Quintana, C. M. and Petukhov, A. G. and Chen, Yu and Kafri, D. and Kechedzhi, K. and Collins, R. and Naaman, O. and Boixo, S. and Arute, F. and Arya, K. and Buell, D. and Burkett, B. and Chen, Z. and Chiaro, B. and others},
  journal   = {Phys. Rev. Lett.},
  title     = {Diabatic Gates for Frequency-Tunable Superconducting Qubits},
  year      = {2019},
  issn      = {1079-7114},
  month     = nov,
  number    = {21},
  pages     = {210501},
  volume    = {123},
  doi       = {10.1103/physrevlett.123.210501},
  fjournal  = {Physical Review Letters},
  publisher = {American Physical Society (APS)},
}

@Article{Chu2020,
  author    = {Chu, Ji and Li, Danyu and Yang, Xiaopei and Song, Shuqing and Han, Zhikun and Yang, Zhen and Dong, Yuqian and Zheng, Wen and Wang, Zhimin and Yu, Xiangmin and Lan, Dong and Tan, Xinsheng and Yu, Yang},
  journal   = {Phys. Rev. Appl.},
  title     = {Realization of Superadiabatic Two-Qubit Gates Using Parametric Modulation in Superconducting Circuits},
  year      = {2020},
  issn      = {2331-7019},
  month     = jun,
  number    = {6},
  pages     = {064012},
  volume    = {13},
  doi       = {10.1103/physrevapplied.13.064012},
  fjournal  = {Physical Review Applied},
  publisher = {American Physical Society (APS)},
}

@Misc{Klimov2020,
  author        = {Paul V. Klimov and Julian Kelly and John M. Martinis and Hartmut Neven},
  title         = {The Snake Optimizer for Learning Quantum Processor Control Parameters},
  year          = {2020},
  archiveprefix = {arXiv},
  eprint        = {2006.04594},
  primaryclass  = {quant-ph},
  url           = {https://arxiv.org/abs/2006.04594},
}

@Misc{Xia2025,
  author        = {Mingkang Xia and Cristóbal Lledó and Matthew Capocci and Jacob Repicky and Benjamin D'Anjou and Ian Mondragon-Shem and Ryan Kaufman and Jens Koch and Alexandre Blais and Michael Hatridge},
  title         = {Exceeding the Parametric Drive Strength Threshold in Nonlinear Circuits},
  year          = {2025},
  archiveprefix = {arXiv},
  eprint        = {2506.03456},
  primaryclass  = {quant-ph},
  url           = {https://arxiv.org/abs/2506.03456},
}

@Article{Eckardt2017,
  author    = {Eckardt, André},
  journal   = {Rev. Mod. Phys.},
  title     = {Colloquium: Atomic quantum gases in periodically driven optical lattices},
  year      = {2017},
  issn      = {1539-0756},
  month     = mar,
  number    = {1},
  pages     = {011004},
  volume    = {89},
  doi       = {10.1103/revmodphys.89.011004},
  fjournal  = {Reviews of Modern Physics},
  publisher = {American Physical Society (APS)},
}

@Article{Didier2018,
  author    = {Didier, Nicolas and Sete, Eyob A. and da Silva, Marcus P. and Rigetti, Chad},
  journal   = {Phys. Rev. A},
  title     = {Analytical modeling of parametrically modulated transmon qubits},
  year      = {2018},
  issn      = {2469-9934},
  month     = feb,
  number    = {2},
  pages     = {022330},
  volume    = {97},
  doi       = {10.1103/physreva.97.022330},
  fjournal  = {Physical Review A},
  publisher = {American Physical Society (APS)},
}

@Article{McKay2016,
  author    = {McKay, David C. and Filipp, Stefan and Mezzacapo, Antonio and Magesan, Easwar and Chow, Jerry M. and Gambetta, Jay M.},
  journal   = {Phys. Rev. Appl.},
  title     = {Universal Gate for Fixed-Frequency Qubits via a Tunable Bus},
  year      = {2016},
  issn      = {2331-7019},
  month     = dec,
  number    = {6},
  pages     = {064007},
  volume    = {6},
  doi       = {10.1103/physrevapplied.6.064007},
  fjournal  = {Physical Review Applied},
  publisher = {American Physical Society (APS)},
}

@Misc{Huang2025,
  author        = {Jordan Huang and Thomas J. DiNapoli and Gavin Rockwood and Ming Yuan and Prathyankara Narasimhan and Eesh Gupta and Mustafa Bal and Francesco Crisa and Sabrina Garattoni and Yao Lu and Liang Jiang and Srivatsan Chakram},
  title         = {Fast Sideband Control of a Weakly Coupled Multimode Bosonic Memory},
  year          = {2025},
  archiveprefix = {arXiv},
  eprint        = {2503.10623},
  primaryclass  = {quant-ph},
  url           = {https://arxiv.org/abs/2503.10623},
}

@Article{Rigetti2010,
  author    = {Rigetti, Chad and Devoret, Michel},
  journal   = {Phys. Rev. B},
  title     = {Fully microwave-tunable universal gates in superconducting qubits with linear couplings and fixed transition frequencies},
  year      = {2010},
  issn      = {1550-235X},
  month     = apr,
  number    = {13},
  pages     = {134507},
  volume    = {81},
  doi       = {10.1103/physrevb.81.134507},
  fjournal  = {Physical Review B},
  publisher = {American Physical Society (APS)},
}

@Article{Huang2022,
  author    = {Huang, Ziwen and You, Xinyuan and Alyanak, Ugur and Romanenko, Alexander and Grassellino, Anna and Zhu, Shaojiang},
  journal   = {Phys. Rev. Appl.},
  title     = {High-Order Qubit Dephasing at Sweet Spots by Non-Gaussian Fluctuators: Symmetry Breaking and Floquet Protection},
  year      = {2022},
  issn      = {2331-7019},
  month     = dec,
  number    = {6},
  pages     = {l061001},
  volume    = {18},
  doi       = {10.1103/physrevapplied.18.l061001},
  fjournal  = {Physical Review Applied},
  publisher = {American Physical Society (APS)},
}

@Article{Huang2021,
  author    = {Huang, Ziwen and Mundada, Pranav S. and Gyenis, András and Schuster, David I. and Houck, Andrew A. and Koch, Jens},
  journal   = {Phys. Rev. Appl.},
  title     = {Engineering Dynamical Sweet Spots to Protect Qubits from 1/f Noise},
  year      = {2021},
  issn      = {2331-7019},
  month     = mar,
  number    = {3},
  pages     = {034065},
  volume    = {15},
  doi       = {10.1103/physrevapplied.15.034065},
  fjournal  = {Physical Review Applied},
  publisher = {American Physical Society (APS)},
}

@Article{Strauch2003,
  author    = {Strauch, Frederick W. and Johnson, Philip R. and Dragt, Alex J. and Lobb, C. J. and Anderson, J. R. and Wellstood, F. C.},
  journal   = {Phys. Rev. Lett.},
  title     = {Quantum Logic Gates for Coupled Superconducting Phase Qubits},
  year      = {2003},
  issn      = {1079-7114},
  month     = oct,
  number    = {16},
  pages     = {167005},
  volume    = {91},
  doi       = {10.1103/physrevlett.91.167005},
  fjournal  = {Physical Review Letters},
  publisher = {American Physical Society (APS)},
}

@Article{Ho1983,
  author    = {Ho, Tak-San and Chu, Shih-I and Tietz, James V.},
  journal   = {Chem. Phys. Lett.},
  title     = {Semiclassical many-mode floquet theory},
  year      = {1983},
  issn      = {0009-2614},
  month     = apr,
  number    = {4},
  pages     = {464--471},
  volume    = {96},
  doi       = {10.1016/0009-2614(83)80732-5},
  fjournal  = {Chemical Physics Letters},
  publisher = {Elsevier BV},
}

@Article{Bravyi2011,
  author    = {Bravyi, Sergey and DiVincenzo, David P. and Loss, Daniel},
  journal   = {Ann. Phys-new. York.},
  title     = {Schrieffer–Wolff transformation for quantum many-body systems},
  year      = {2011},
  issn      = {0003-4916},
  month     = oct,
  number    = {10},
  pages     = {2793--2826},
  volume    = {326},
  doi       = {10.1016/j.aop.2011.06.004},
  fjournal  = {Annals of Physics},
  publisher = {Elsevier BV},
}

@Article{Deng2015,
  author    = {Deng, Chunqing and Orgiazzi, Jean-Luc and Shen, Feiruo and Ashhab, Sahel and Lupascu, Adrian},
  journal   = {Phys. Rev. Lett.},
  title     = {Observation of Floquet States in a Strongly Driven Artificial Atom},
  year      = {2015},
  issn      = {1079-7114},
  month     = sep,
  number    = {13},
  pages     = {133601},
  volume    = {115},
  doi       = {10.1103/physrevlett.115.133601},
  fjournal  = {Physical Review Letters},
  publisher = {American Physical Society (APS)},
}

@Article{Poertner2020,
  author    = {Poertner, A. N. and Martin, J. D. D.},
  journal   = {Phys. Rev. A},
  title     = {Validity of many-mode Floquet theory with commensurate frequencies},
  year      = {2020},
  issn      = {2469-9934},
  month     = mar,
  number    = {3},
  pages     = {032116},
  volume    = {101},
  doi       = {10.1103/physreva.101.032116},
  fjournal  = {Physical Review A},
  publisher = {American Physical Society (APS)},
}

@InProceedings{Zhang2025,
  author    = {Zhang, Junyao and Wang, Hanrui and Ding, Qi and Gu, Jiaqi and Assouly, Reouven and Oliver, William and Han, Song and Brown, Kenneth and Li, Hai and Chen, Yiran},
  booktitle = {Proceedings of the 52nd Annual International Symposium on Computer Architecture},
  title     = {QPlacer: Frequency-Aware Component Placement for Superconducting Quantum Computers},
  year      = {2025},
  address   = {New York, NY, USA},
  pages     = {1554–1567},
  publisher = {Association for Computing Machinery},
  series    = {ISCA '25},
  doi       = {10.1145/3695053.3730994},
  isbn      = {9798400712616},
  numpages  = {14},
  url       = {https://doi.org/10.1145/3695053.3730994},
}

@Misc{Paolo2022,
  author        = {Agustin Di Paolo and Catherine Leroux and Thomas M. Hazard and Kyle Serniak and Simon Gustavsson and Alexandre Blais and William D. Oliver},
  title         = {Extensible circuit-QED architecture via amplitude- and frequency-variable microwaves},
  year          = {2022},
  archiveprefix = {arXiv},
  eprint        = {2204.08098},
  primaryclass  = {quant-ph},
  url           = {https://arxiv.org/abs/2204.08098},
}

@Article{Groot2010,
  author    = {de Groot, P. C. and Lisenfeld, J. and Schouten, R. N. and Ashhab, S. and Lupaşcu, A. and Harmans, C. J. P. M. and Mooij, J. E.},
  journal   = {Nat. Phys.},
  title     = {Selective darkening of degenerate transitions demonstrated with two superconducting quantum bits},
  year      = {2010},
  issn      = {1745-2481},
  month     = aug,
  number    = {10},
  pages     = {763--766},
  volume    = {6},
  doi       = {10.1038/nphys1733},
  fjournal  = {Nature Physics},
  publisher = {Springer Science and Business Media LLC},
}

@Article{Krishnan1978,
  author    = {Krishnan, R. and Pople, J. A.},
  journal   = {Int. J. Quantum Chem.},
  title     = {Approximate fourth‐order perturbation theory of the electron correlation energy},
  year      = {1978},
  issn      = {1097-461X},
  month     = jul,
  number    = {1},
  pages     = {91--100},
  volume    = {14},
  doi       = {10.1002/qua.560140109},
  fjournal  = {International Journal of Quantum Chemistry},
  publisher = {Wiley},
}

@InProceedings{Brink2018,
  author    = {Brink, Markus and Chow, Jerry M. and Hertzberg, Jared and Magesan, Easwar and Rosenblatt, Sami},
  booktitle = {2018 IEEE International Electron Devices Meeting (IEDM)},
  title     = {Device challenges for near term superconducting quantum processors: frequency collisions},
  pages     = {6.1.1--6.1.3},
  publisher = {IEEE},
  date      = {2018-12},
  doi       = {10.1109/iedm.2018.8614500},
}

@Article{Pappas2024,
  author    = {Pappas, David P. and Field, Mark and Kopas, Cameron J. and Howard, Joel A. and Wang, Xiqiao and Lachman, Ella and Oh, Jinsu and Zhou, Lin and Gold, Alysson and Stiehl, Gregory M. and Yadavalli, Kameshwar and Sete, Eyob A. and Bestwick, Andrew and Kramer, Matthew J. and Mutus, Josh Y.},
  journal   = {Commun. Mater.},
  title     = {Alternating-bias assisted annealing of amorphous oxide tunnel junctions},
  year      = {2024},
  issn      = {2662-4443},
  month     = aug,
  volume    = {5},
  doi       = {10.1038/s43246-024-00596-z},
  eid       = {150},
  publisher = {Springer Science and Business Media LLC},
}

@Article{Mundada2019,
  author    = {Mundada, Pranav and Zhang, Gengyan and Hazard, Thomas and Houck, Andrew},
  journal   = {Phys. Rev. Appl.},
  title     = {Suppression of Qubit Crosstalk in a Tunable Coupling Superconducting Circuit},
  year      = {2019},
  issn      = {2331-7019},
  month     = nov,
  number    = {5},
  pages     = {054023},
  volume    = {12},
  doi       = {10.1103/physrevapplied.12.054023},
  fjournal  = {Physical Review Applied},
  publisher = {American Physical Society (APS)},
}

@Misc{Liu2025,
  author        = {Zheng-He Liu and Yu Liu and Gui-Han Liang and Cheng-Lin Deng and Keyang Chen and Yun-Hao Shi and Tian-Ming Li and Lv Zhang and Bing-Jie Chen and Cai-Ping Fang and Da'er Feng and Xu-Yang Gu and Yang He and Kaixuan Huang and Hao Li and Hao-Tian Liu and Li Li and Zheng-Yang Mei and Zhen-Yu Peng and Jia-Cheng Song and Ming-Chuan Wang and Shuai-Li Wang and Ziting Wang and Yongxi Xiao and Minke Xu and Yue-Shan Xu and Yu Yan and Yi-Han Yu and Wei-Ping Yuan and Jia-Chi Zhang and Jun-Jie Zhao and Kui Zhao and Si-Yun Zhou and Zheng-An Wang and Xiaohui Song and Ye Tian and Florian Mintert and Johannes Knolle and Roderich Moessner and Yu-Ran Zhang and Pan Zhang and Zhongcheng Xiang and Dongning Zheng and Kai Xu and Hongzheng Zhao and Heng Fan},
  title         = {Prethermalization by Random Multipolar Driving on a 78-Qubit Superconducting Processor},
  year          = {2025},
  archiveprefix = {arXiv},
  eprint        = {2503.21553},
  primaryclass  = {quant-ph},
  url           = {https://arxiv.org/abs/2503.21553},
}

@Article{Mi2021,
  author    = {Mi, Xiao and Ippoliti, Matteo and Quintana, Chris and Greene, Ami and Chen, Zijun and Gross, Jonathan and Arute, Frank and Arya, Kunal and Atalaya, Juan and Babbush, Ryan and Bardin, Joseph C. and Basso, Joao and Bengtsson, Andreas and Bilmes, Alexander and Bourassa, Alexandre and others},
  journal   = {Nature},
  title     = {Time-crystalline eigenstate order on a quantum processor},
  year      = {2021},
  issn      = {1476-4687},
  month     = nov,
  number    = {7894},
  pages     = {531--536},
  volume    = {601},
  doi       = {10.1038/s41586-021-04257-w},
  publisher = {Springer Science and Business Media LLC},
}

@Article{Caldwell2018,
  author    = {Caldwell, S. A. and Didier, N. and Ryan, C. A. and Sete, E. A. and Hudson, A. and Karalekas, P. and Manenti, R. and da Silva, M. P. and Sinclair, R. and Acala, E. and Alidoust, N. and Angeles, J. and Bestwick, A. and Block, M. and Bloom, B. and others},
  journal   = {Phys. Rev. Appl.},
  title     = {Parametrically Activated Entangling Gates Using Transmon Qubits},
  year      = {2018},
  issn      = {2331-7019},
  month     = sep,
  number    = {3},
  pages     = {034050},
  volume    = {10},
  doi       = {10.1103/physrevapplied.10.034050},
  fjournal  = {Physical Review Applied},
  publisher = {American Physical Society (APS)},
}

@Article{Reagor2018,
  author    = {Reagor, Matthew and Osborn, Christopher B. and Tezak, Nikolas and Staley, Alexa and Prawiroatmodjo, Guenevere and Scheer, Michael and Alidoust, Nasser and Sete, Eyob A. and Didier, Nicolas and da Silva, Marcus P. and Acala, Ezer and Angeles, Joel and Bestwick, Andrew and Block, Maxwell and Bloom, Benjamin and others},
  journal   = {Sci. Adv.},
  title     = {Demonstration of universal parametric entangling gates on a multi-qubit lattice},
  year      = {2018},
  issn      = {2375-2548},
  month     = feb,
  number    = {2},
  pages     = {eaao3603},
  volume    = {4},
  doi       = {10.1126/sciadv.aao3603},
  fjournal  = {Science Advances},
  publisher = {American Association for the Advancement of Science (AAAS)},
}

@Misc{Arute2020,
  author        = {Frank Arute and Kunal Arya and Ryan Babbush and Dave Bacon and Joseph C. Bardin and Rami Barends and Andreas Bengtsson and Sergio Boixo and Michael Broughton and Bob B. Buckley and David A. Buell and Brian Burkett and Nicholas Bushnell and Yu Chen and Zijun Chen and others},
  title         = {Observation of separated dynamics of charge and spin in the Fermi-Hubbard model},
  year          = {2020},
  archiveprefix = {arXiv},
  eprint        = {2010.07965},
  primaryclass  = {quant-ph},
  url           = {https://arxiv.org/abs/2010.07965},
}

@InProceedings{Ding2020,
  author    = {Ding, Yongshan and Gokhale, Pranav and Lin, Sophia Fuhui and Rines, Richard and Propson, Thomas and Chong, Frederic T.},
  booktitle = {2020 53rd Annual IEEE/ACM International Symposium on Microarchitecture (MICRO)},
  title     = {Systematic Crosstalk Mitigation for Superconducting Qubits via Frequency-Aware Compilation},
  year      = {2020},
  pages     = {201-214},
  doi       = {10.1109/MICRO50266.2020.00028},
}

@Article{Acharya2025,
  author    = {Acharya, Rajeev and Abanin, Dmitry A. and Aghababaie-Beni, Laleh and Aleiner, Igor and Andersen, Trond I. and Ansmann, Markus and Arute, Frank and Arya, Kunal and Asfaw, Abraham and Astrakhantsev, Nikita and Atalaya, Juan and Babbush, Ryan and Bacon, Dave and Ballard, Brian and Bardin, Joseph C. and others},
  journal   = {Nature},
  title     = {Quantum error correction below the surface code threshold},
  year      = {2025},
  issn      = {1476-4687},
  month     = dec,
  number    = {8052},
  pages     = {920--926},
  volume    = {638},
  doi       = {10.1038/s41586-024-08449-y},
  publisher = {Springer Science and Business Media LLC},
}

@Misc{Ding2025,
  author        = {Qi Ding and Shoumik Chowdhury and Agustin Di Paolo and Réouven Assouly and Alan V. Oppenheim and Jeffrey A. Grover and William D. Oliver},
  title         = {Frequency- and Amplitude-Modulated Gates for Universal Quantum Control},
  year          = {2025},
  archiveprefix = {arXiv},
  eprint        = {2511.03164},
  primaryclass  = {quant-ph},
  url           = {https://arxiv.org/abs/2511.03164},
}

@Article{Kim2022,
  author    = {Kim, Hyunseong and Jünger, Christian and Morvan, Alexis and Barnard, Edward S. and Livingston, William P. and Altoé, M. Virginia P. and Kim, Yosep and Song, Chengyu and Chen, Larry and Kreikebaum, John Mark and Ogletree, D. Frank and Santiago, David I. and Siddiqi, Irfan},
  journal   = {Appl. Phys. Lett.},
  title     = {Effects of laser-annealing on fixed-frequency superconducting qubits},
  year      = {2022},
  issn      = {1077-3118},
  month     = oct,
  number    = {14},
  volume    = {121},
  doi       = {10.1063/5.0102092},
  fjournal  = {Applied Physics Letters},
  publisher = {AIP Publishing},
}

@Article{Lambert2026,
  author   = {Neill Lambert and Eric Giguère and Paul Menczel and Boxi Li and Patrick Hopf and Gerardo Suárez and Marc Gali and Jake Lishman and Rushiraj Gadhvi and Rochisha Agarwal and Asier Galicia and Nathan Shammah and Paul Nation and J.R. Johansson and Shahnawaz Ahmed and Simon Cross and Alexander Pitchford and Franco Nori},
  journal  = {Phys. Rep.},
  title    = {QuTiP 5: The Quantum Toolbox in Python},
  year     = {2026},
  issn     = {0370-1573},
  note     = {QuTiP 5: The Quantum Toolbox in Python},
  pages    = {1-62},
  volume   = {1153},
  doi      = {https://doi.org/10.1016/j.physrep.2025.10.001},
  fjournal = {Physics Reports},
  url      = {https://www.sciencedirect.com/science/article/pii/S0370157325002704},
}

@InProceedings{Gario2015,
  author    = {Gario, Marco and Micheli, Andrea},
  booktitle = {SMT Workshop 2015},
  title     = {PySMT: a solver-agnostic library for fast prototyping of SMT-based algorithms},
  year      = {2015},
}

@Dataset{Ma2026,
  author    = {Ma, Zhuang},
  doi       = {10.5281/zenodo.18787593},
  month     = feb,
  publisher = {Zenodo},
  title     = {Data and Figures in "Ma, Z., et. al. Analysis of Frequency Collisions in Parametrically Modulated Superconducting Circuits. arXiv.2511.05031"},
  url       = {https://doi.org/10.5281/zenodo.18787593},
  year      = {2026},
}

\end{document}